\documentclass[twocolumn]{aastex62}

\usepackage{diagbox}

\newcommand{\kms}{\,km\,s$^{-1}$}

\newcommand{\mum}{\,$\mu$m}

\newcommand{\Msun}{$M_{\odot}$}

\newcommand{\rev}{ }
\newcommand{\newrev}{ }

\newcommand{\LiI}{Li\,I\,$\lambda$6708 \AA}

\newcommand{\HII}{\mbox{H\,\textsc{ii}}}%
\def\arcsec{\hbox{$^{\hbox{\rlap{\hbox{\lower4pt\hbox{$\,\prime\prime$}}}\hbox{$\frown$}}}$}}
          
\graphicspath{{./}{figures/}}

\shorttitle{}
\shortauthors{}

\begin{document}

\title{The First Extensive Spectroscopic Study of Young Stars in the North America and Pelican Nebulae Region}

\author{Min Fang}
\affiliation{1Department of Astronomy, California Institute of Technology, Pasadena CA 91125}
\author{Lynne A. Hillenbrand}
\affiliation{1Department of Astronomy, California Institute of Technology, Pasadena CA 91125}
\author{Jinyoung Serena Kim}
\affiliation{Department of Astronomy, University of Arizona, 933 North Cherry Avenue, Tucson, AZ 85721, USA}
\author{Krzysztof Findeisen}
\affiliation{University of Washington, Department of Astronomy, Seattle, WA, 98195, U.S.A}
\author{Gregory J. Herczeg}
\affiliation{Kavli Institute for Astronomy and Astrophysics, Peking University, Yiheyuan 5, Haidian Qu, 100871 Beijing, China}
\author{John M. Carpenter }
\affiliation{Joint ALMA Observatory, Alonso de Cordova 3107 Vitacura, Santiago, Chile}
\author{Luisa M. Rebull}
\affiliation{Infrared Science Archive (IRSA), IPAC, 1200 E. California Blvd., California Institute of Technology, Pasadena, CA 91125, USA}
\author{Hongchi Wang}
\affiliation{Purple Mountain Observatory and Key Laboratory of Radio Astronomy, Chinese Academy of Sciences, 2 West Beijing Road, 210008 Nanjing, China}

\begin{abstract}
We present a spectroscopic survey of over 3400 potential members in the North America and Pelican nebulae (NAP) using several {\newrev low}-resolution ($R\approx$ 1300-2000) spectrographs: Palomar/Norris,  WIYN/HYDRA, Keck/DEIMOS, and MMT/Hectospec. We {\newrev identify} 580 young stars as likely members of the NAP region based on criteria involving infrared excess, Li\,I 6708\,\AA\ absorption, X-ray emission, parallax, and proper motions. The spectral types of individual spectra are derived by fitting them with templates that are either empirical spectra of pre-main sequence stars, or model atmospheres.  The templates are artificially veiled, and a best-fit combination of spectral type and veiling parameter is derived for each star. We use the spectral types with archival photometry to derive $V$-band extinction and stellar luminosity. From the H-R diagram, the median age of the young stars is about 1~Myr, with a luminosity dispersion of $\sim$0.3--0.4~dex. We investigate the photometric variability of the spectroscopic member sample using ZTF data, and conclude that photometric variability, while present, does not significantly contribute to the luminosity dispersion.  While larger than the formal errors, the luminosity dispersion is smaller than if veiling were not taken into account in our spectral typing process. The measured ages of stellar kinematic groups, combined with inferred ages for embedded stellar populations revealed by Spitzer, suggests a sequential history of star formation in the NAP region. 


\end{abstract}



\section{Introduction} \label{sec:introduction}

{\newrev Galactic O-type stars are rare, with only $\sim$600 cataloged \citep{goy1973,garmany1982,2017hsa9.conf..509M}. The closest is $\sigma$ Sco (09.5) at $<200$ pc. Within about 1 kpc of the Sun there are just a few tens of O-type stars. This small sample includes the famous $\theta^1$~Ori~C (O7) and  $\theta^2$~Ori~A (O9.5) responsible for ionizing the Orion Nebula \citep{2017arXiv170306191O}, as well as several other late-O type stars (such as $\iota$ Ori, $\sigma$ Ori, $\zeta$ Ori, the embedded IRS~2b in NGC~2024, and  $\lambda$ Ori) in the Orion molecular cloud complex at $\sim400$ pc,  and S~Mon (O7) associated with NGC 2264 at $\sim800$ pc.  Each of these O stars is young enough to still illuminate, and to some extent ionize, the surrounding gas and dust out of which it was recently formed.}

The North America and Pelican nebulae (NGC 7000 and IC 5070) are parts of 
an ionized region known as W80 \citep{1980A&A....88..267M,2008hsf1.book...36R}. 
This \HII\ region is ionized {\newrev 
mainly \footnote{There is a second ionizing source, the O6 type star HD 199579 which is offset from the center of the H\,II region and moving further towards the periphery; see \cite{2020ApJ...899..128K} for further discussion.}
} 
by 2MASS~J20555125+4352246 \citep{2005A&A...430..541C}, {\newrev also called the Bajamar star}, which has spectral type O3.5-O5 and sits behind the Lynds Dark Nebula LDN~935 \citep{1962ApJS....7....1L}. The dark region that separates the North America and the Pelican  nebulae is the foreground molecular cloud associated with W80, and has a total mass of about 2-5$\times$10$^{4}$~\Msun\ (\citealt{1980ApJ...239..121B}; \citealt{2014AJ....147...46Z}; \citealt{2020ApJ...899..128K}). A distance of 795 pc based on Gaia DR2 astrometry \citep{2016A&A...595A...1G,2018A&A...616A...1G} is also {\newrev calculated} by  \cite{2020ApJ...899..128K}, placing the region somewhat further away than the previous estimates \citep[see][for a review]{2008hsf1.book...36R}.

Numerous surveys have targeted the young stellar population associated with the region encompassing the North America and Pelican nebulae, and LDN~935. Beginning in the 1950s (see e.g. \citealt{1955ApJ...121..611M}, \citealt{1958ApJ...128..259H}, \citealt{1973A&AS....9..183W}), early studies revealed associations of emission-line T Tauri stars near the nebulae. 

\begin{figure*}
\begin{center}
\includegraphics[angle=0,width=2\columnwidth]{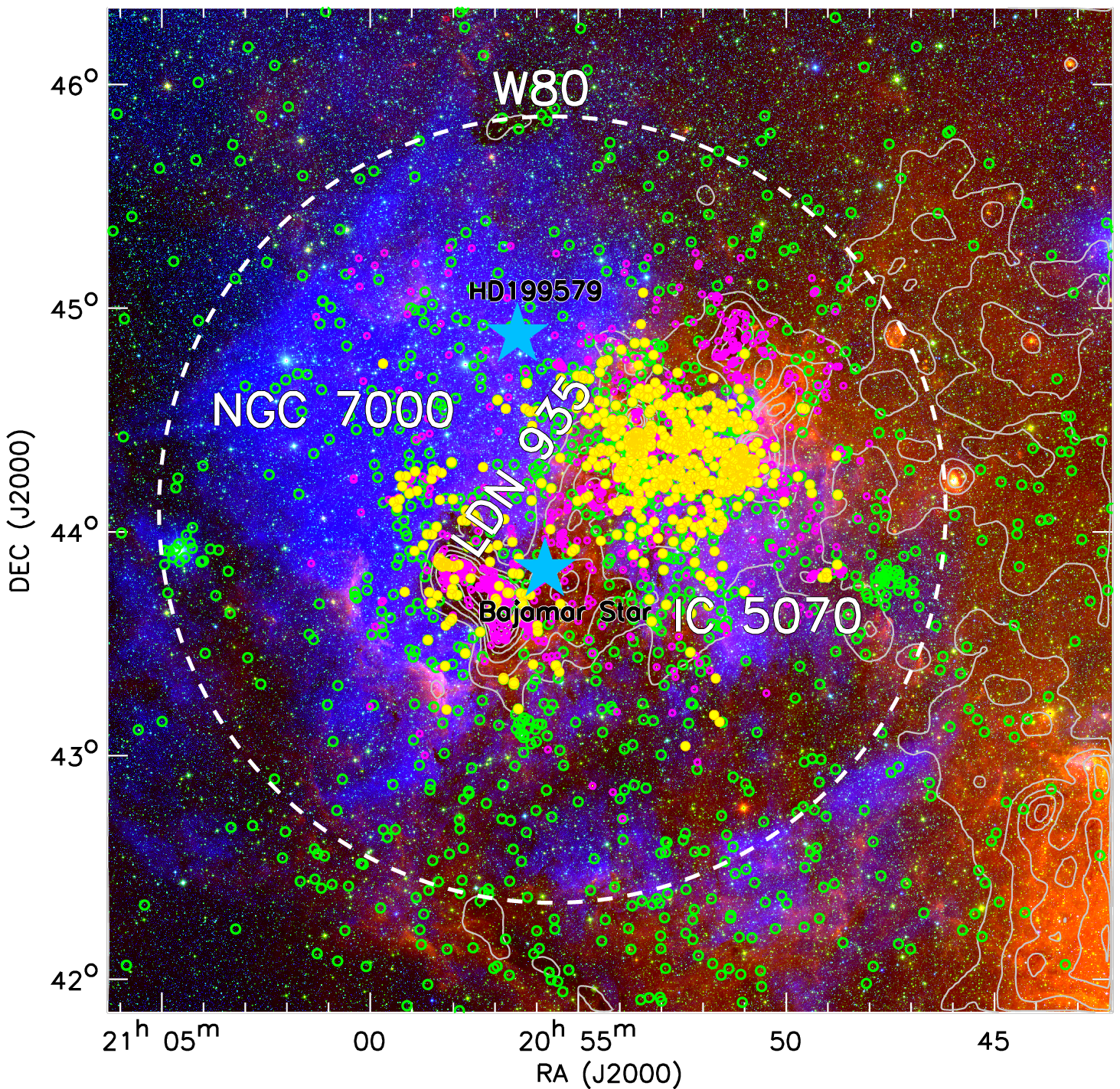}
\caption{Three-color image of  North America Nebula  created with the DSS1 R-band (blue),  WISE-W1 band(green), and WISE-W3 band (red). The gray contours are the Planck 857\,\mum\ dust emission. {\newrev White dashed circle indicates the approximate boundary of the H\,II region \citep{1983A&A...124..116W}.} All the YSO candidates in the field are shown as open circles, magenta for those with infrared excess \citep{2011ApJS..193...25R}, and green for those identified with \textit{Gaia} \citep{2020ApJ...899..128K}. Our spectroscopically confirmed YSOs are shown as yellow filled circles. The blue filled star symbols mark the \HII\ region's {\newrev main} ionizing source, the Bajamar Star, identified by \cite{2005A&A...430..541C}, and another member O star HD~199579 in the NAP.}
\label{Fig:YSO_NAN}
\end{center}
\end{figure*}

A significant contribution in understanding the star formation in LDN~935 come from the imaging surveys with the Spitzer Space Telescope \citep{2009ApJ...697..787G,2011ApJS..193...25R}. The Spitzer imaging surveys discovered more than 2000 young stellar object (YSO) candidates with infrared excess emission, and demonstrated the existence of more densely clustered as well as more loosely distributed YSOs \citep{2009ApJ...697..787G,2011ApJS..193...25R}.  A majority of the sources {\newrev with infrared excess} are within the boundaries of LDN~935, as opposed to in the nebular regions, evidencing an active star formation process in this region.  The more {\newrev area-}limited X-ray survey of LDN~935 with the XMM-Newton and Chandra Space telescopes  by \cite{2017A&A...602A.115D} confirmed the youth properties of many of the YSO candidates identified with Spitzer, and further revealed hundreds of new YSO candidates without infrared excesses. 
Figure~\ref{Fig:YSO_NAN} shows the distribution of YSO candidates in this region. {\newrev Many of these YSO candidates have been monitored photometrically and show light curve variations characteristic of YSOs  \citep{2013ApJ...768...93F,2019A&A...627A.135B}. }

The North America and Pelican nebulae region (hereafter, NAP) stands
in contrast to other nearby star-forming regions having large, well-studied stellar populations, {\newrev including those associated with the O stars in the Orion complex and the NGC 2264 cluster}.   Here, only about sixty of the YSO candidates have been studied with spectroscopy.   The collection of spectral types is reported in \cite{2011ApJS..193...25R} and references therein. In this work, we provide a first extensive -- though not yet comprehensive -- spectroscopic study of YSOs in the NAP. 

\section{New Observations}
We obtained {\newrev low}-resolution ($R\approx$ 1300-2000) optical spectra of more than 3400 prospective members of the NAP region over the time period 1998 to 2017. Selection criteria varied over the years, and included: red optical colors, photometric H$\alpha$-R excess, infrared excess, optical photometric variability, and X-ray emission.   In assembling targets we were guided by a legacy KPNO/0.9m CCD survey in BVI, H$\alpha$, and R that is partially described in \cite{2011ApJS..193...25R}, IRAC photometry presented in \cite{2009ApJ...697..787G}, MIPS photometry presented in \cite{2011ApJS..193...25R}, and  H$\alpha$ and X-ray data presented in \cite{2017A&A...602A.115D}.

A variety of multi-object spectrographs were used, with the observations catalogued in Table~\ref{tab:obs_log} and the instruments, {\newrev data acquisition,} and data reduction described in the sub-sections below.

\begin{figure*}
\begin{center}
\includegraphics[width=2\columnwidth]{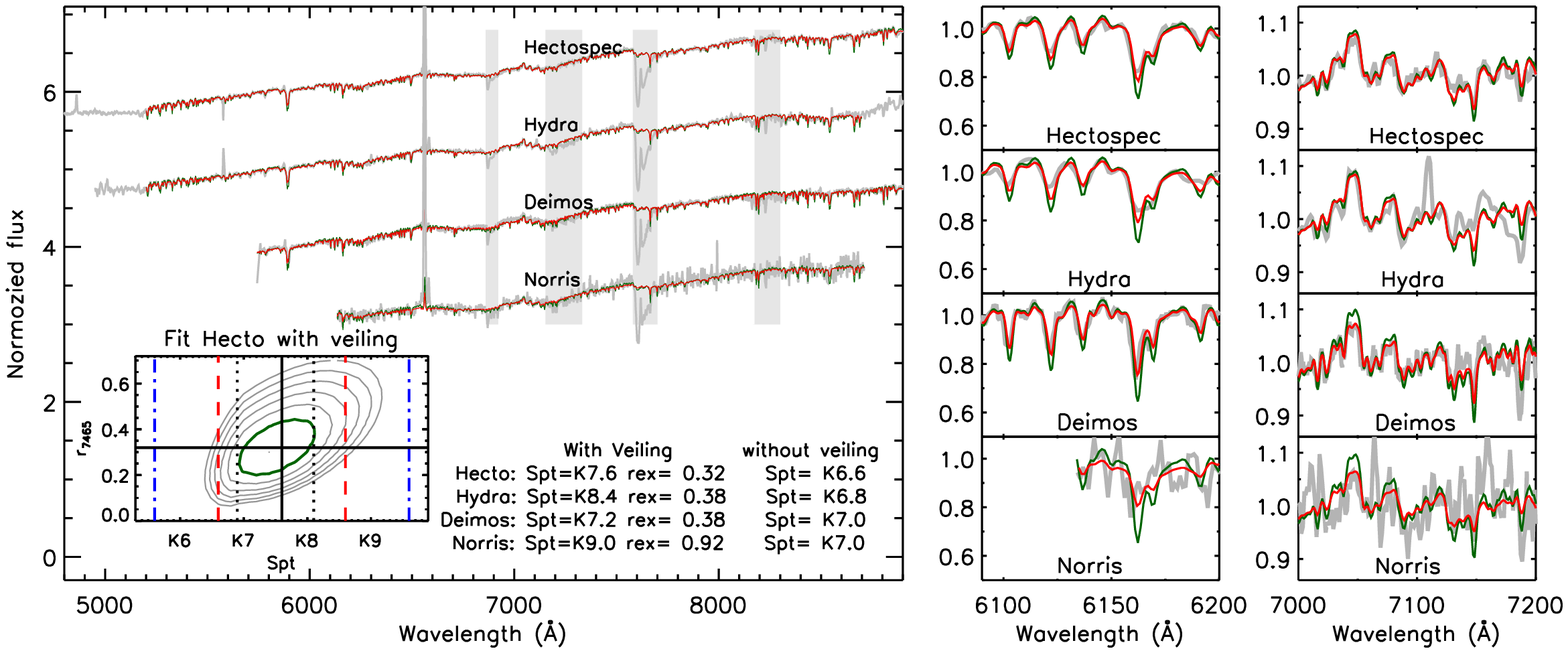}
\caption{Illustration of our spectral template fitting method for source  ID~103, which was observed with all four spectrographs used in this study.  Left: Best-fit X-shooter empirical spectral templates (red including veiling in the template model and green without veiling) over-plotted on the observed spectra (gray).  Vertically oriented gray bars indicate regions masked due to potential contamination from emission lines or telluric features. Inset contours show the distribution of the reduced $\chi^2$ derived from fitting the Hectospec spectrum with X-shooter templates with different combinations of SpT and $r_{7465}$. The green contour is for the minimum reduced $\chi^2$+0.05. The solid lines show the SpT$^{\rm best}$ (vertical solid line) and $r_{7465}^{\rm best}$ (horizontal solid line)  with minimum reduced $\chi^2$. The two vertical dotted lines show the spectral type range with $\chi^{2}_{r}$ within $\chi^{2}_{r, min}$+0.05. 
Vertical red dashed and blue dash-dotted lines are used to qualify our spectral classification (see \S~\ref{sect:spectral_classification}). Right: Zoom-in comparison of the target spectra and the best-fit template with (red) and without (green) veiling within 6090--6200\,\AA\ and 7000--7200\,\AA.  The veiled model (red) is a better fit to the observed spectrum (gray).}\label{Fig:Example-spt1}
\end{center}
\end{figure*}

   \begin{figure*}
\begin{center}
\includegraphics[width=2\columnwidth]{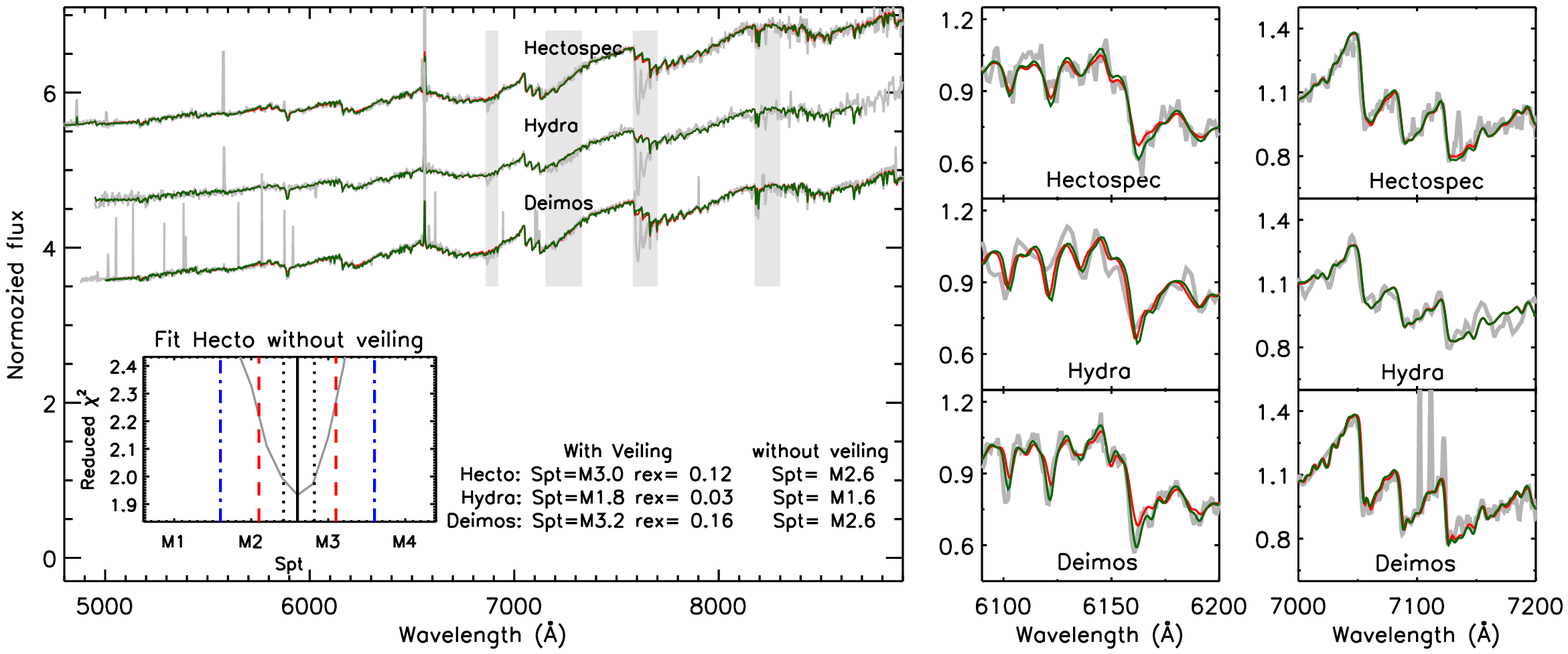}
\caption{Same as Figure~\ref{Fig:Example-spt1} but for another source ID~232. In this case the non-veiled model (green) is the preferred fit to the observed spectrum (gray).
In the left panel, inset shows the distribution of reduced $\chi^2$ derived from fitting the Hectospec spectrum with X-shooter templates having different SpT. Vertical lines represent the same quantities as in Figure~\ref{Fig:Example-spt1}. 
}\label{Fig:Example-spt2}
\end{center}
\end{figure*}

\subsection{Palomar / Norris Spectra}


We used the now-decommissioned Norris multi-object spectrograph \citep{1993PASP..105.1308H} on the 5m Palomar telescope to obtain spectra of 1294 sources. Norris was a fiber-fed spectrograph with 150 working fibers, each having a $1.^{\prime\prime}5$ aperture.
The 20\arcmin\ field of view was mapped onto a 2048x2048 CCD.

Data were taken in a total of 17 different fiber configurations by observers L. Hillenbrand, T. Small, and J. Carpenter.
A 600 l/mm grating blazed at 5000 \AA\ was used to achieve $R\approx 2000$. 
On the first observing date the spectra cover 5120--7750 \AA\ while the remainder of the observations cover 6100--8750 \AA.

In each configuration, between 30--126 fibers were assigned to a stellar position and between 5--59 fibers were assigned to sky positions.
In addition, the set of observations for each configuration included several (typically 3) long exposures on-target, and (typically 1-2) interspersed sky exposures offset
$\pm6^{\prime\prime}-10^{\prime\prime}$ from the on-targets position. 
The sky exposure permits correction on an individual star basis for nebulosity, which is spatially variable.
Calibration was achieved using dome flats and internal spectral reference lamps.

Because no filter was used during the Norris observations, some second-order light of
bright blue stars in the field was scattered into other fibers, contaminating
several spectra longward of 8100 \AA.

\begin{figure*}
\begin{center}
\includegraphics[width=0.95\columnwidth]{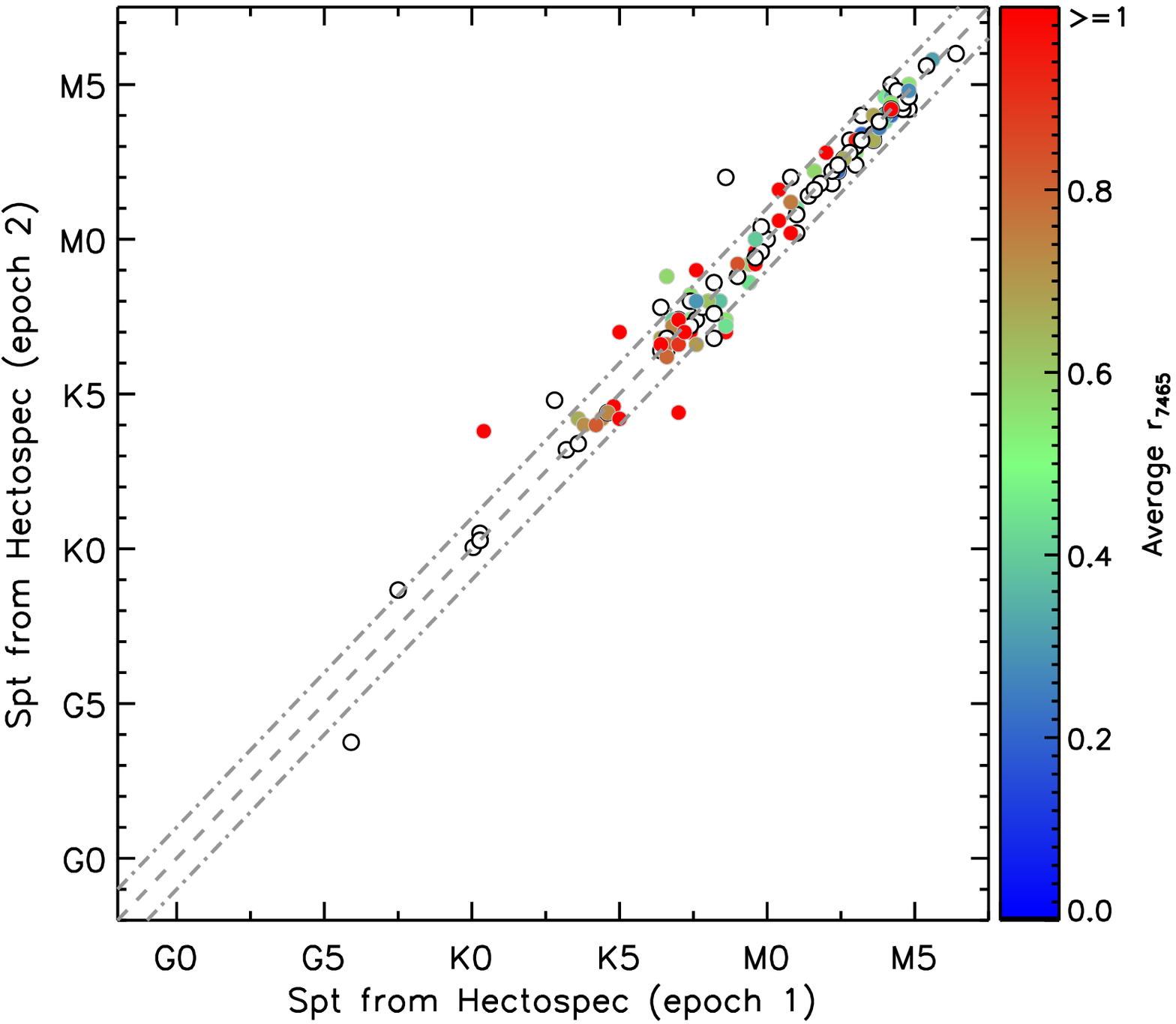}
\includegraphics[width=0.95\columnwidth]{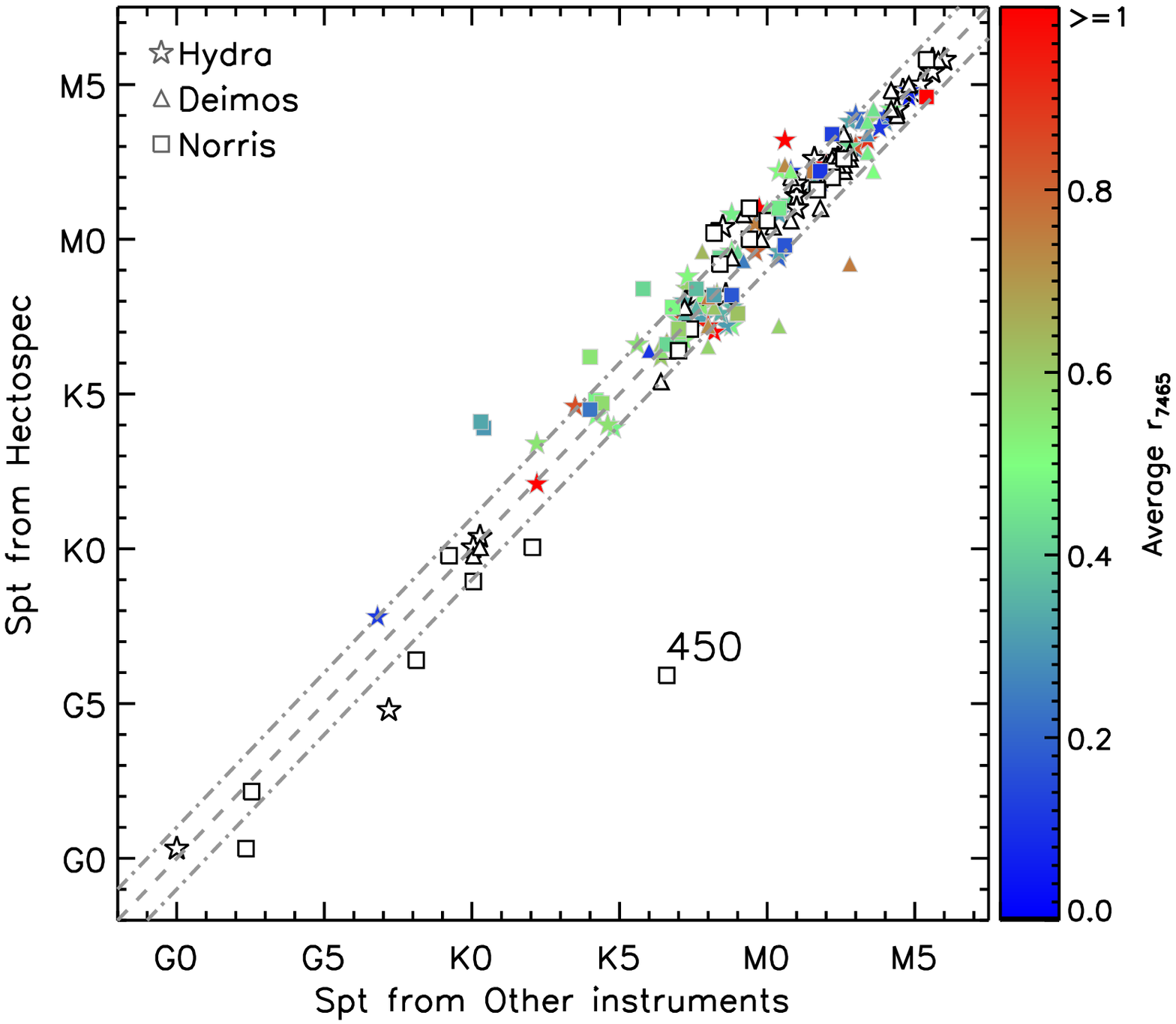}
\caption{Left: Comparison of spectral types derived from the Hectospec spectra for the same stars, observed at different epochs. Right: Comparison of spectral types derived from Hectospec with those derived from other instruments for the same stars.  In both panels, dashed lines indicate the one-to-one relationship while dash-dot lines are offset by $\pm 2$ spectral subclasses from the one-to-one lines. In each panel,  the symbols are color-coded according to their average $r_{7465}$ (open symbols represent the sources with $r_{7465}$=0.)}\label{Fig:Twospt}
\end{center}
\end{figure*}

\subsection{WIYN / HYDRA Spectra}


We used the HYDRA multi-object spectrograph \cite{1995SPIE.2476...56B} on the 3.5m KPNO/WIYN telescope to obtain 645 spectra.  HYDRA consists of a 1024x2048 CCD fed with 99 fibers, each with a $2^{\prime\prime}$ aperture. 
The 316@7 setting of the bench spectrograph with a GG-495 filter
resulted in spectra from 5000-10000 \AA\ at $R\sim1500$.

A total of 6 fiber configurations were acquired by KPNO queue observers D. Harmer and D Wilmarth.  The observing strategy was similar to that for the Norris observations, with several on-target exposures and a shorter offset sky exposure.

\subsection{Keck /DEIMOS Spectra}


A total of 258 spectra were obtained  using the multi-slit mode of the Keck/DEIMOS \citep{2003SPIE.4841.1657F} spectrograph, in 22 different slit mask configurations.  A 600 l/mm grating delivered $R\approx 1350$ spectra covering 5000-9000 \AA.
The observers were K. Findeisen and L. Hillenbrand.

\subsection{MMT / Hectospec Spectra}


With the MMT and its Hectospec \citep{2005PASP..117.1411F} fiber spectrograph, we obtained spectra in 14 different fiber configurations.  There were three separate programs running, with 517 (PI: Fang), 524 (PI: Hillenbrand), and  
 705 (PI: Fang) stars observed, respectively, and significantly overlapping selection criteria of candidate members.  
 The observer was often M. Caulkins. 

A 270 l/mm grating was used, resulting in spectra covering 3700-9150 \AA\ at $R\approx 1300$. As with the observations taken using the other fiber instruments, for the MMT sequences three on-source exposures were taken along with one offset sky exposure.

\section{Spectroscopic Data Reduction}
The Norris and HYDRA observations were reduced by G. Herczeg
using custom routines written in {\it IDL}.  
The bias in each image was corrected for using the overscan
region on the CCD.   We determined the trace of each fiber on the detector
using the dome flat.  The absolute position of each fiber on the
detector was determined independently for each individual exposure.
We corrected for cosmic rays by identifying deviations from the
illumination profile across each fiber on the detector.  We corrected for
scattered light in the detector by fitting a spline to pixels between
fiber positions. 
We then resampled the counts in each fiber onto
a sub-pixel scale  across the dispersion axis to ensure that the
extraction window remained constant for each separate on-target, sky,
and flat exposure within a given configuration.  
We measured and
subtracted background counts for each fiber based on the counts
between fiber positions.  We then extracted the spectrum in each fiber
using a 5 pixel windows in the HYDRA spectra and 6 pixels in the
Norris spectra.  The observations from Aug. 1998 were poorly focused
and required 10 pixel extraction windows.
The 2nd-order wavelength solution (a FeAr lamp for the Norris
observations and a CuAr lamp for the HYDRA observations) and the flatfield correction were
calculated independently for each fiber.

We obtain a spectrum for a fiber by summing the counts extracted for
that fiber from each on-target integration.  We correct for sky and
nebular emission by subtracting counts extracted for the same fiber in the sky
exposure, scaled to the difference in observing times.  In several
configurations the sky emission was scaled by an additional 10--20\% to account for changes in the sky transmission.  We calculate a
master sky spectrum for each configuration by combining the spectra
obtained from fibers assigned to the sky, accounting for
fiber-to-fiber sensitivity differences.  We correct for changes
in the sky emission after comparing the master sky spectrum obtained in
the on-target integration and that from the sky integrations. 
In many cases a fiber in our sky integrations randomly landed on a
star or on a region with a much different nebular emission than was detected in
the on-target integration.  For the configurations with two sky
integrations, one or a combination of the two sky fibers was used to
subtract sky and nebular emission.  In some configurations, 
the master sky spectrum was used for sky subtraction.


The DEIMOS spectra were reduced by K. Findeisen using a modified version of the DEEP2 pipeline \citep{2012ascl.soft03003C,2013ApJS..208....5N}. Images were first bias-corrected and flat-fielded using dome flats.  After spectral extraction, the one-dimensional spectra were wavelength-calibrated but not flux-calibrated. We then corrected for sky and for nebular emission by subtracting a fit to the adjacent off-source spectrum within each slit. The final spectra span approximately $\approx4400-9500$ \AA, though the range of any particular star shifts by up to $\pm 500$ \AA\ depending on the slit location within the mask.  Cosmic rays were left uncorrected by the pipeline, and were cleaned by hand from the final spectra.


The Hectospec spectra with PI Hillenbrand were processed by S. Tokarz using the Hecto Pipeline \citep{2007ASPC..376..249M}.  Following bias correction, flat-fielding, spectral extraction, and wavelength calibration, the sky exposures were scaled and subtracted from the averaged on-source exposures. The Hectospec spectra with PI Fang were reduced using IRAF routines following the standard procedures, as described in \cite{2013ApJS..207....5F}.

\begin{figure}
\begin{center}
\includegraphics[width=\columnwidth]{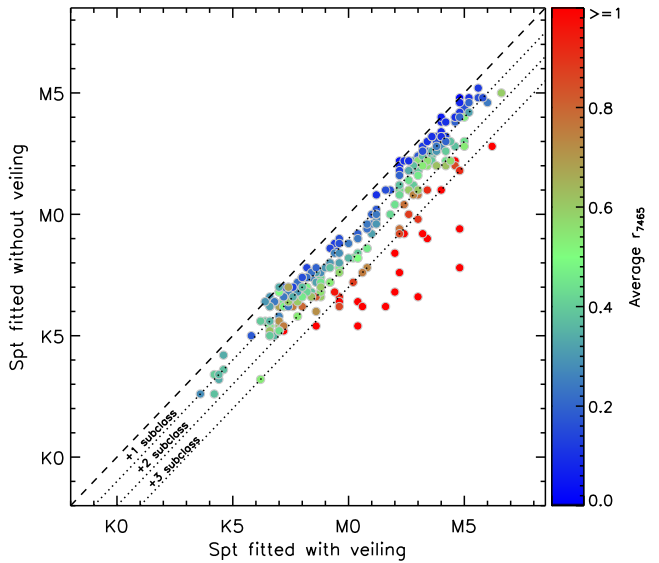}
\caption{Comparison of spectral types derived for the same stars by fitting without and with accounting for spectral veiling. Dashed line indicates the one-to-one relationship while dotted lines are offset by 1, 2, and 3 spectral subclasses from the one-to-one lines. The symbols are color-coded according to their average $r_{7465}$}\label{Fig:Twospt_veiling_noveiling}
\end{center}
\end{figure}

\begin{figure*}
\begin{center}
\includegraphics[width=1.8\columnwidth]{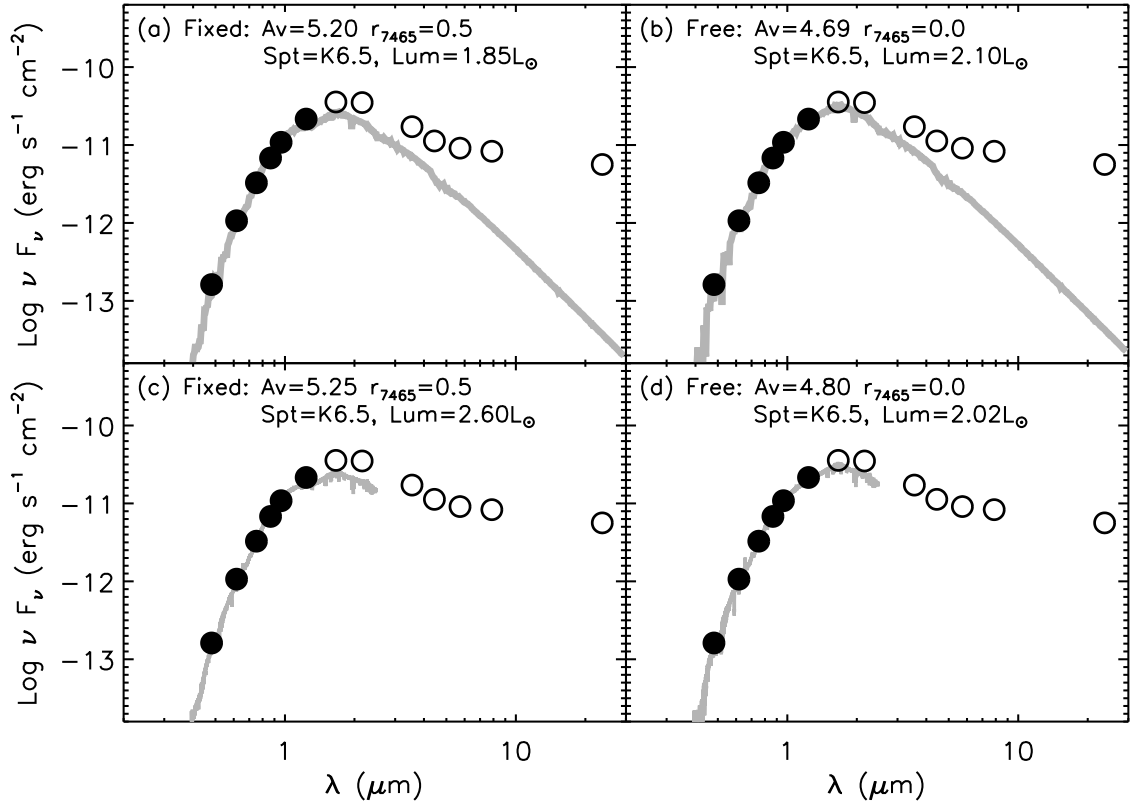}
\caption{Example of fitting broad-band photometry for Source ID~12 using model atmospheres (top panels) and X-shooter empirical templates (bottom panels). Filled circles represent the photometry used for the SED fitting, while open circles are not used to constrain the fits. 
}\label{Fig:Foursed}
\end{center}
\end{figure*}

\section{Assembly of Catalog Photometry and Variability Data}

In order to construct the spectral energy distribution of each source, we used optical photometry in 
the $g$, $r$, $i$, $z$, and $y$ bands from  Pan-STARRS \citep{2016arXiv161205560C}
and $G$, $G_{BP}$, and $G_{RP}$ bands from Gaia DR2 \citep{2016A&A...595A...1G,2018A&A...616A...1G},
near-infrared photometry in the $J$, $H$, and $K_S$ bands from the Two-Micron All Sky Survey \citep[2MASS, ][]{2006AJ....131.1163S}, and near- and mid-infrared photometry in Spitzer [3.6], [4.5], [5.8], [8.0], and [24].  The Spitzer data came from two different sources: for the YSO candidates identified with infrared excess emission we used the photometry of \cite{2009ApJ...697..787G} and \cite{2011ApJS..193...25R}, while for other sources (mostly foreground/background field stars) we {\newrev used} the aperture photometry in the 3\farcs8 from the Spitzer Enhanced Imaging Products \footnote{We noticed that some of sources have unreliable SEIP photometry and discuss this in Appendix~\ref{Appen:IRAC}). For these sources, we use point-spread-function fitting photometry newly derived as described in \cite{2009A&A...504..461F}. This mixing of photometry does not affect the results for most of members of the NAP {\newrev whose photometry are collected from \cite{2011ApJS..193...25R}}, but it does clean up the SEDs of field main-sequence stars which have large scatter using the aperture photometry from SEIP.}
\citep[SEIP;][]{2012AAS...21942806T}. 


We also inspect the variability of the young stars in the NAP using photometric data from the Zwicky Transient Facility (ZTF), which is a time-domain survey mainly in $g$ and $r$ bands starting in 2017 at Palomar Observatory \citep{2019PASP..131a8003M}. In this work, we 
{\newrev search ZTF DR2 with the data taken betwen 2018 April and 2019 June for the NAP region.}
Only $r$ band photometry is used, as our targets typically are somewhat extincted sources.

\section{Data Analysis}

\subsection{Spectral Classification}\label{sect:spectral_classification}

Our spectral data are collected from multiple instruments. We derive the instrument response function for each observation set using one early-type star observed with the corresponding instrument and spectrograph setup. A list of the stars used for the individual instruments is given in Table~\ref{tab:table_RF}. The detailed procedure is as follows.  

First, we obtain the BOSZ Kurucz model atmosphere \citep{2012AJ....144..120M} appropriate for each source. Next, we fit the broad-band photometry from {\it Gaia} and 2MASS for each source using the aforementioned model with two free parameters: extinction and stellar angular radius, as in  \cite{2009A&A...504..461F} and \cite{2013ApJS..207....5F}. Then, we shift and rotationally broaden the best-fit model atmosphere and degrade it to the  spectral resolution of the individual instruments. The ratio between the observed and model spectrum is fitted with separate polynomial functions for different wavelength ranges.  From the fitting, we derive the instrument response function of the individual instruments. 

The response function is then employed to correct the observed spectra to the same relative flux scale as a function of wavelength.  We stress that this procedure does not achieve truly flux-calibrated spectra, but instead removes the instrumental signatures in the spectra. The corrected spectra are then normalized by their flux at 7465~\AA.

We perform spectral classification of the normalized spectra by fitting them with two sets of spectral templates: one collected from VLT/X-shooter for pre-main sequence stars  with spectral types ranging from G6 to M9.5 {\newrev (\citealt{2013A&A...551A.107M}; \citealt{2017A&A...605A..86M}; \citealt{2018A&A...609A..70R}; see also in Fang et al. 2020)}, and the other from a combination of model atmospheres with $T_{\rm eff}$=2,775--35,000~K and surface gravity log$~g$=3.5--5 and $vsin~i$=0--150~\kms, with models for $T_{\rm eff}$=2,775--12,000~K  from  \cite{2013A&A...553A...6H} and for $T_{\rm eff}$ above 12,000~K from \cite{2012AJ....144..120M}. For the X-shooter templates, we interpolate the templates into two sets for use as described below, one with a grid spacing of 1.0 spectral subclass, and the other with a grid spacing of 0.2 spectral subclass.  For the model atmospheres, we interpolate the grid in steps of 25~K for $T_{\rm eff}$=2,775--7,000~K, 50~K for $T_{\rm eff}$=7,000--12,000~K, 500~K for $T_{\rm eff}$=12,000--20,000~K, and 1000~K for $T_{\rm eff}$=20,000--35,000~K, and in steps of 0.25 for log$~g$.   Before fitting to our data, the spectral templates are degraded to the spectral resolution of the observed spectra (typically 1500--2000) {\newrev which have been} corrected by the instrument response function. 

To mimic the filling effect on the photospheric absorption lines due to excess emission from the accretion shocks,  an excess flux is added to the spectral template, parameterized as $r_{\rm ex,~7465}=\frac{F_{\rm excess,~7465}}{F_{\rm phot,~7465}}$, where $r_{7465}$ is the veiling at 7465\,\AA, $F_{\rm excess,~7465}$  is the excess flux at \,7465\,\AA, and $F_{\rm phot,~7465}$ is the underlying photospheric emission at  7465\,\AA.  The shape of accretion continuum spectrum  is approximated as a constant, following e.g. \cite{2014ApJ...786...97H}.   Finally, the veiled spectral templates are reddened, parameterized by $A_V$ and using the extinction law from \citet{1989ApJ...345..245C}, adopting a total to selective extinction value $R_{\rm V}=$3.1 (typical of interstellar medium dust). 
For comparison to the data, each processed template is then re-normalized by the flux at 7465~\AA. 

\begin{figure}
\begin{center}
\includegraphics[angle=0,width=0.98\columnwidth]{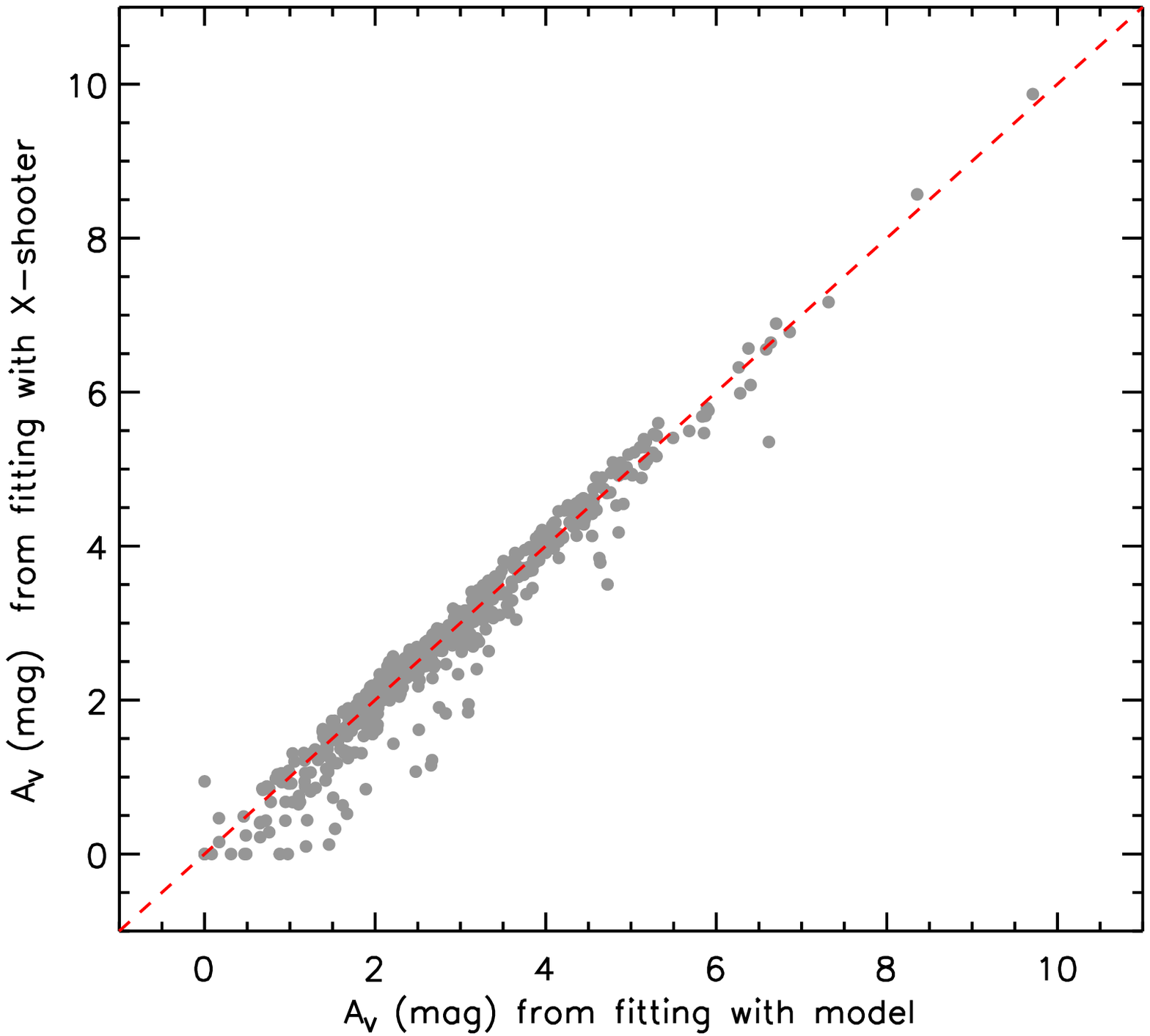}
\includegraphics[angle=0,width=0.98\columnwidth]{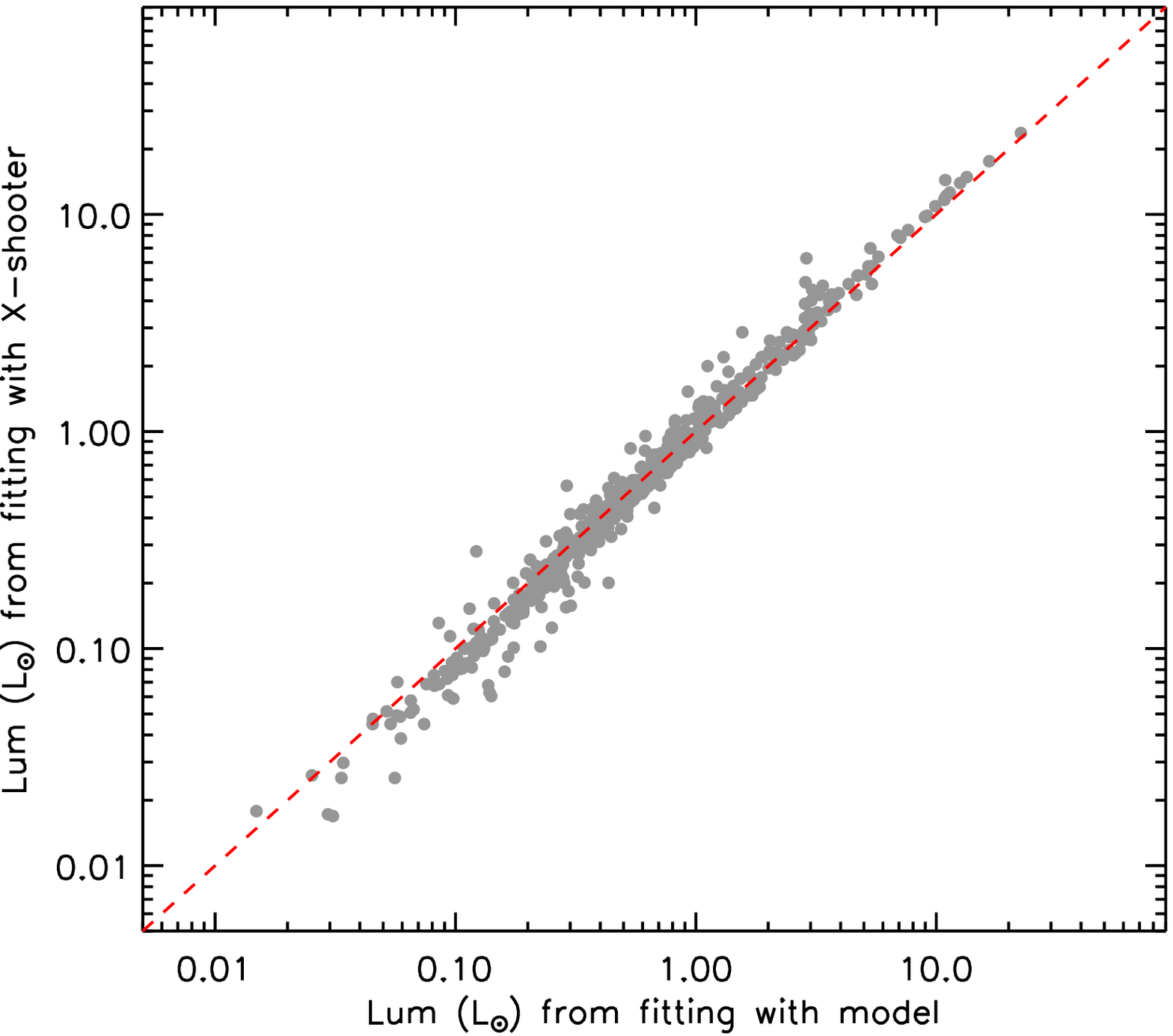}
\caption{Comparison of extinction in $V$ band (top) and stellar luminosity (bottom) from spectral fitting using model atmosphere and X-shooter templates. In each panel, the red dashed line indicates a 1:1 relation }~\label{Fig:com_Lum_ext}
\end{center}
\end{figure}

These steps are described in further detail below, where we make use of both the empirical spectral templates and the model atmospheres. We perform the fitting of each observed spectrum considering two cases: one without veiling and the other including veiling. For the fitting, we have four free parameters: spectral type (SpT), relative radial velocity ($\Delta RV$) between the template and the target spectrum, $r_{7465}$, and visual extinction ($A_{\rm V}$). We divide our procedure into two steps:
  
  \begin{itemize}
      \item {\bf Step 1}: We start the fitting using the templates with the coarse grid (1 subclass). First, we set $\Delta RV$=0\,\kms\ and normalize the template spectra by the flux at 7465\,\AA, add  $r_{\rm ex,~7465}$  to the normalized templates, and redden the veiled templates  with a visual extinction $A_{\rm V}$. Then, we divide the reddened templates by the target spectrum, and fit the ratio by an order 3 polynomial function so as to correct for the possible difference in shape between the templates and the target spectrum. The polynomial function is then applied to the templates.
      For individual templates,  $r_{7465}$ is fixed to be zero for non-veiling fitting and varies from 0 to 5.7 with a grid step of 0.1, and $A_{\rm V}$ varies from 0 to 10 mag stepped by 0.5~mag. The best-fit template is obtained by minimizing the reduced $\chi^{2}$ ($\chi^{2}_{r}$). After finding the best-fit template, we then vary $\Delta RV$  from $-$80\,\kms\ to  $+$80\,\kms\ with a grid step of 2\,\kms\, and shift the best-fit template by the $\Delta RV$ to find the best $\Delta RV$ value by minimizing the $\chi^{2}_{r}$.  During the fitting, the emission lines and telluric features have been excluded. We then use the best $\Delta RV$ as an input, and repeat Step 1. Finally, we obtain the best-fit values for the 4 parameters SpT$^{\rm best}$, $A_{\rm V}^{\rm best}$, $\Delta RV^{\rm best}$, and $r_{7465}^{\rm best}$ in the case of the veiling fitting, or 3 parameters (without $r_{7465}^{\rm best}$) in the case of the non-veiling fitting. 
     
     \item {\bf Step 2}: In this step, the procedure is the same as in Step 1, but we use the template grid with the finer spacing of 0.2 subclass, and we vary spectral type from SpT$^{\rm best}$-2 to  SpT$^{\rm best}$+2, $A_{\rm V}$ from $A_{\rm V}^{\rm best}$-2 to $A_{\rm V}^{\rm best}$+2 with a grid spacing of 0.2, $\Delta RV$ from  $\Delta RV^{\rm best}$-20 to $\Delta RV^{\rm best}$+20 with a grid spacing of 1\,\kms, and  $r_{7465}$ from  $r_{7465}^{\rm best}$-0.4 to $r_{7465}^{\rm best}$+0.4 with a grid spacing of 0.04, along with a value of 0 for the non-veiling fitting. We find the best combination of $r_{7465}$ and SpT when fitting with veiling, and the best SpT when fitting without veiling, by minimizing the $\chi^{2}_{r}$ (the minimum $\chi^{2}_{r}$ defined as $\chi^{2}_{r, min}$). The $A_{\rm V}$ resulting from the fitting is not used given that we have done some normalization during the fitting. The $A_{\rm V}$ is derived instead from the broad-band photometry (see \S\ref{Sect:Star_properties}).
     
     \item{\bf Step 3}: 
    We now use the model atmospheres in place of the empirical spectral templates, and  we repeat Step 1 to re-fit each target spectrum.  In quoting our final results, we use the best-fit model atmospheres only for spectra earlier than type K. This is motivated by the fact that for these earlier spectral types, the empirical templates are not well established, and the available grid is sparse.

  \end{itemize}

Figure~\ref{Fig:Example-spt1} shows one example of our spectral fitting for Source ID~103 compared to a spectral model based on the empirical X-shooter templates. This object has been observed with all four instruments. Best-fit templates are illustrated for each observation in both the non-veiling and including veiling cases. In order to determine which templates can fit the observations, we check the fits within two spectral ranges: 6090--6200\,\AA, where there are several metallic lines, e.g. Fe~I, Ca~I,  and 7000--7200\,\AA, where there are TiO bands. 
For this source, the fitting with veiling better matches the observations. For three spectra from Hectospec, Hydra, and Deimos with large wavelength coverage  and good data quality, the spectral types from them agree with each other within one subclass. Figure~\ref{Fig:Example-spt2} shows another example of our spectral fitting for Source ID~232. In this case the non-veiling fitting can better fit the observed spectra of the source.  

In general, we quantify the spectral classification by comparing the derived spectral type range (SpT1-SpT2) given by $\chi^{2}_{r}$ within $\chi^{2}_{r, min}$+0.05, considering thresholds defined based on the spectral types.  In our figures, we set two thresholds given by the vertical red dashed lines and vertical blue dash-dot lines.
For values of SpT$^{\rm best}$ earlier than K0, these lines correspond  to SpT$^{\rm best}\pm$2 and SpT$^{\rm best}\pm$4, respectively. 
For values of SpT$^{\rm best}$ between K0 and M0, they are SpT$^{\rm best}\pm$1 and SpT$^{\rm best}\pm$2, respectively. 
And for values of SpT$^{\rm best}$ later than M0, they are SpT$^{\rm best}\pm$0.5 and SpT$^{\rm best}\pm$1, respectively. 
We consider the spectral classification as well determined (Rank 1 in our tables) when the spectral type range SpT1--SpT2 is within narrower threshold, 
as fairly determined (Rank 2 in our tables) when SpT1--SpT2 is between the thresholds, and as poorly determined (Rank 3 in our tables) 
when SpT1--SpT2 is beyond the wider threshold. 

In our dataset, there are 130 sources with Hectospec spectra observed at two epochs.  There are also 90 sources observed with both Hectospec and Hydra, 64 with  both Hectospec and Deimos, and 47 with both Hectospec and Norris. Figure~\ref{Fig:Twospt} compares the spectral types derived for the same objects taken with Hectospec at different observational times (left), and with different instruments (right). The spectral types from different spectra for the same sources agree well with each other, especially for those sources with high veiling, demonstrating the robustness of the instrumental response corrections and the spectral normalization procedures described above. While there are no systematic offsets between these spectral types, the median difference among the Hectospec data is 0.2 subclasses, and between Hectospec and the other instruments, 0.4--0.6 subclasses. These differences are within the typically quoted spectral typing accuracy of 0.5--1 spectral subclass. 

We highlight in the right panel of Figure~\ref{Fig:Twospt} that there is one source, ID~450, which shows a large variation in the derived spectral types.  We find K6.6  from the Norris spectrum and G5.9 from the Hectospec spectrum. This object is a known FU~ori star PTF~10qpf (also named as LkH$\alpha$ 188-G4 and HBC 722) that started its outburst around 2010 \citep{2010A&A...523L...3S,2011ApJ...730...80M}. The object was observed in its pre-outburst stage on 1998 August 14 with Norris, and in its post-outburst stage on 2017 June 5. The spectral types are consistent with those in the literature  for both the pre-outburst \citep{1979ApJS...41..743C}
and the post-outburst stages \citep{2011ApJ...730...80M}.  
In this work, we adopt K6.6 as the spectral type, in order to derive the stellar properties of the FU Ori progenitor.

Figure~\ref{Fig:Twospt_veiling_noveiling} compares the spectral types for the same spectra fitting with and without veiling. As expected, the fitting without consideration of veiling tends to produce an earlier type than the fitting with veiling.  The offset depends on the strength of the veiling, with the difference typically 0--1 subclass for $r_{7465}\lesssim$0.2, 1--2 subclass for $r_{7465}\sim$0.2--0.6, and $>2$ subclasses for $r_{7465}\gtrsim$0.6. The comparisons emphasize the necessity of considering veiling effects when performing spectral classification for accreting young stars, as already shown in \cite{2014ApJ...786...97H} and Fang~et~al.~(2020). The results on the spectral classification for the members of the (see \S~\ref{sect:member}) are listed in Table~\ref{tabe_allspt}.

\subsection{Extinction and Stellar Luminosity}\label{Sect:Star_properties}

Although we have a preliminary $A_V$ estimate from the optical spectrum fitting process described above, {\newrev the optical spectra are not flux-calibrated very well.}. We thus improve our confidence in the $A_V$ values by considering the broader spectral energy distribution to fit for both the extinction and the stellar luminosity of each young star in our sample.   In general, we use the photometry in $g'$, $r'$, $i'$, $z'$ from Pan-STARRS, in $G$, $G_{BP}$, and $G_{RP}$ from Gaia, and in $J$ band from 2MASS.  For our SED fit, we match the optical and near-infrared photometry with a veiled and reddened model atmosphere having the effective temperature derived from our spectral classification, and a veiled and reddened X-shooter empirical spectral template having the same spectral type as the source.  In most cases there is agreement between the methods, with the details discussed below.

The SED fitting procedure employs three parameters: the source angular diameter $\theta$, the veiling $r_{7465}$, and the extinction $A_{\rm V}$. We first veil the model atmosphere adopting a value of $r_{7465}$ 
(either that derived from our spectral fitting above, or as a free parameter in the SED model).  
The form of the excess emission flux as a function of wavelength can then be characterized as follows:
\begin{equation}
F_{\lambda}^{ex}=\left\lbrace
\begin{array}{lll}
F_{\rm 7456}^{phot}\times r_{7465}~~~~~~~~~~~~~~\text{ if $\lambda<0.9\mu$m}\\
F_{\rm 7456}^{phot}\times r_{7465}\times\frac{1.3-\lambda}{1.3-0.9}~~\text{ if 0.9$\mu$m$\leq\lambda\leq1.3\mu$m}\\
0~~~~~~~~~~~~~~~~~~~~~~~~~~~~~~\text{ if $\lambda>1.3\mu$m}  & 
\end{array}
\right.
\end{equation}
\noindent Where the $F_{\rm 7456}^{phot}$ is the photospheric emission at 7465\,\AA, and $\lambda$ is wavelength in units of $\mu$m. 
According to \cite{2011ApJ...730...73F}, the excess emission from accretion (which peaks in the ultraviolet) can still contribute at $J$ band in some cases. Furthermore, in order to avoid any sharp jump in the flux of our model SED,  we allow the excess emission flux from veiling to decrease from a value that we have assumed is flat with wavelength in the optical, to zero over the wavelength range 0.9\,\mum\ to 1.3\,\mum.  Beyond 1.3\,\mum\ we assume there is no veiling \footnote{{\newrev This could create an artificial jump in the SED for some small number of sources with strong accretion (high $r_{7465}$).}}

\begin{figure}
\begin{center}
\includegraphics[angle=0,width=1\columnwidth]{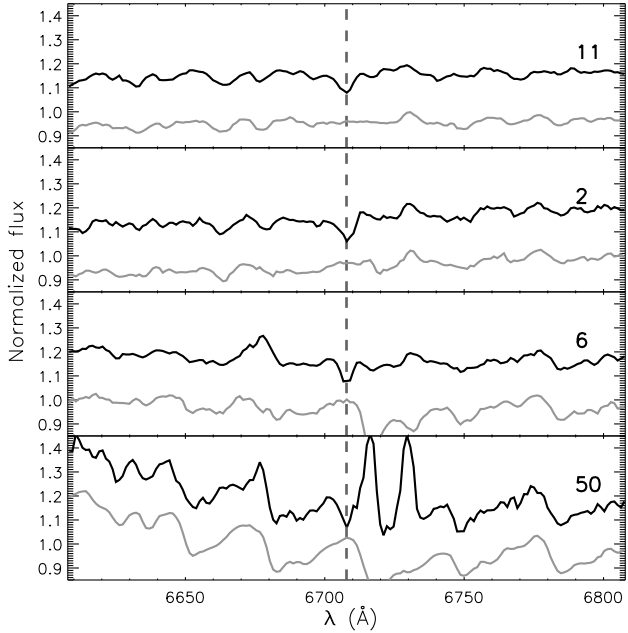}
\caption{Examples of \LiI\ detection for four of our targets, IDs~11, 2, 6, and 50 {\newrev ordered by their spectral types $\sim$K4 to M3 from upper to bottom}. In each panel, there are two spectra: one for a young star (black) and the other for a field star (gray) of similar spectral type.  For comparison purposes, the spectra are normalized and shifted.\label{Fig:Li}}
\end{center}
\end{figure}

\begin{figure}
\begin{center}
\includegraphics[angle=0,width=1\columnwidth]{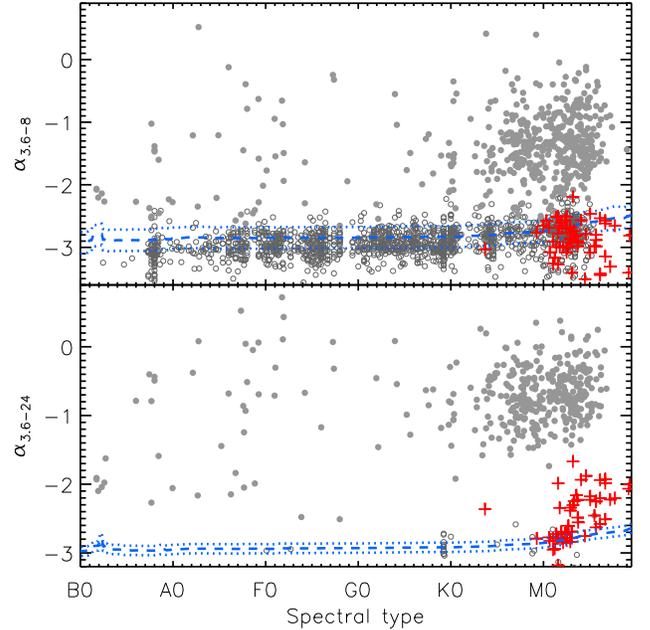}
\caption{Dereddened Spitzer infrared spectral slopes, $\alpha_{3.6-8}$ and $\alpha_{3.6-24}$ vs. spectral type for all the sources with spectroscopic observation in the field of the NAP region. In each panel, the open gray circles show the diskless stars, and the filled circles are for the sources with infrared excess emisson according to their $\alpha_{3.6-8}$ (top panel) and $\alpha_{3.6-24}$ (bottom panel). The red plus symbols mark likely Asymptotic Giant Branch (AGB) stars identified by their spectral features and their locations in the H-R diagram. The dashed line show the infrared spectral slope of the photospheric emission calculated with the BT-Settl atmospheric models, and the dotted lines are the 1$\sigma$ standard deviation, assuming a 10\% uncertainty in Spitzer photometry.}\label{Fig:alpha}
\end{center}
\end{figure}

As above, we use the extinction law of \cite{1989ApJ...345..245C} to redden the veiled model atmospheres, adopting $R_{\rm{V}}$=3.1. Synthetic photometry is calculated in the bands of the assembled SEDs by integrating the intensity of the (reddened) model atmospheres over the spectral response curve for each filter, and compared  with the observations.  The model atmospheres are the same as the ones used for the spectral classification, but considering only log$~g=4.0$. The high spectral resolution models from \cite{2013A&A...553A...6H} extend to 2.5\,\mum, and 
for the longer wavelengths  we use the BT-Settl model atmospheres \citep{2011ASPC..448...91A} with solar abundances from \citet{2009ARA&A..47..481A} at the corresponding $T_{\rm eff}$. 
This combination of models does not affect the SED fitting since our SED fitting goes only to 2MASS-$J$ band. By minimizing the $\chi^{2}$, the optimum values for the three free parameters are achieved. To investigate how the veiling can affect the final results, we consider two cases in the fitting: (1) setting  $r_{7465}$ to be the same value as the one from the spectral fitting as described above, (2) setting $r_{7465}$ to be a free parameter.  Although both cases are improper for any source showing strong brightness variability, and neither case may be proper in other circumstances, the SED fitting method can give us a sense of the uncertainty on the stellar properties when the results are compared to the results on veiling and extinction derived from the spectral fitting process.  

As above for the spectral fitting, here for the SED fitting, we use model atmospheres for spectral types G5 and earlier, and  we replace the model atmosphere with the X-shooter empirical templates for the majority of the sample with spectral types from G6 to M. In the fitting of the broad-band photometry, we have three parameters: $F_{\rm 7456}^{phot}$, $A_{\rm V}$, and $r_{7465}$.  $F_{\rm 7456}^{phot}$ is the photospheric emission at 7465\,\AA, and the other two parameters are the same as in the earlier fitting processes.  In Figure~\ref{Fig:Foursed}, we show an example illustrating the best fit results with the models and with the X-shooter empirical templates.

In Figure~\ref{Fig:com_Lum_ext}, taking the confirmed members in NAP (see \S\ref{sect:member}),  we compare our methods for deriving the extinction values at $V$ band (upper panel) and stellar luminosities (bottom panel) through SED fitting using model atmospheres versus X-shooter templates. The earliest type of the sources shown in the Figure is G6 due to the limit of spectral type range of X-shooter templates. {\newrev For most of the sources}, there are no significant shifts between $A_{V}$ from the two methods with the mean difference between them about 0.10~dex. {\newrev But when there are differences, model fitting tends to give larger $A_{\rm V}$. This is especially true at $A_{\rm V}<$2.}  The  $L_{\star}$ comparison between the two methods shows general consistency when $L_{\star}\gtrsim0.4~L_{\odot}$; however, at lower luminosities there is a systematic shift between the $L_{\star}$ values between the two methods with the X-shooter template values 0.05~dex lower.
This shift may be due to the increasing difference between stellar models and observed spectra toward the later spectral type. 

\begin{figure}
\begin{center}
\includegraphics[angle=0,width=1\columnwidth]{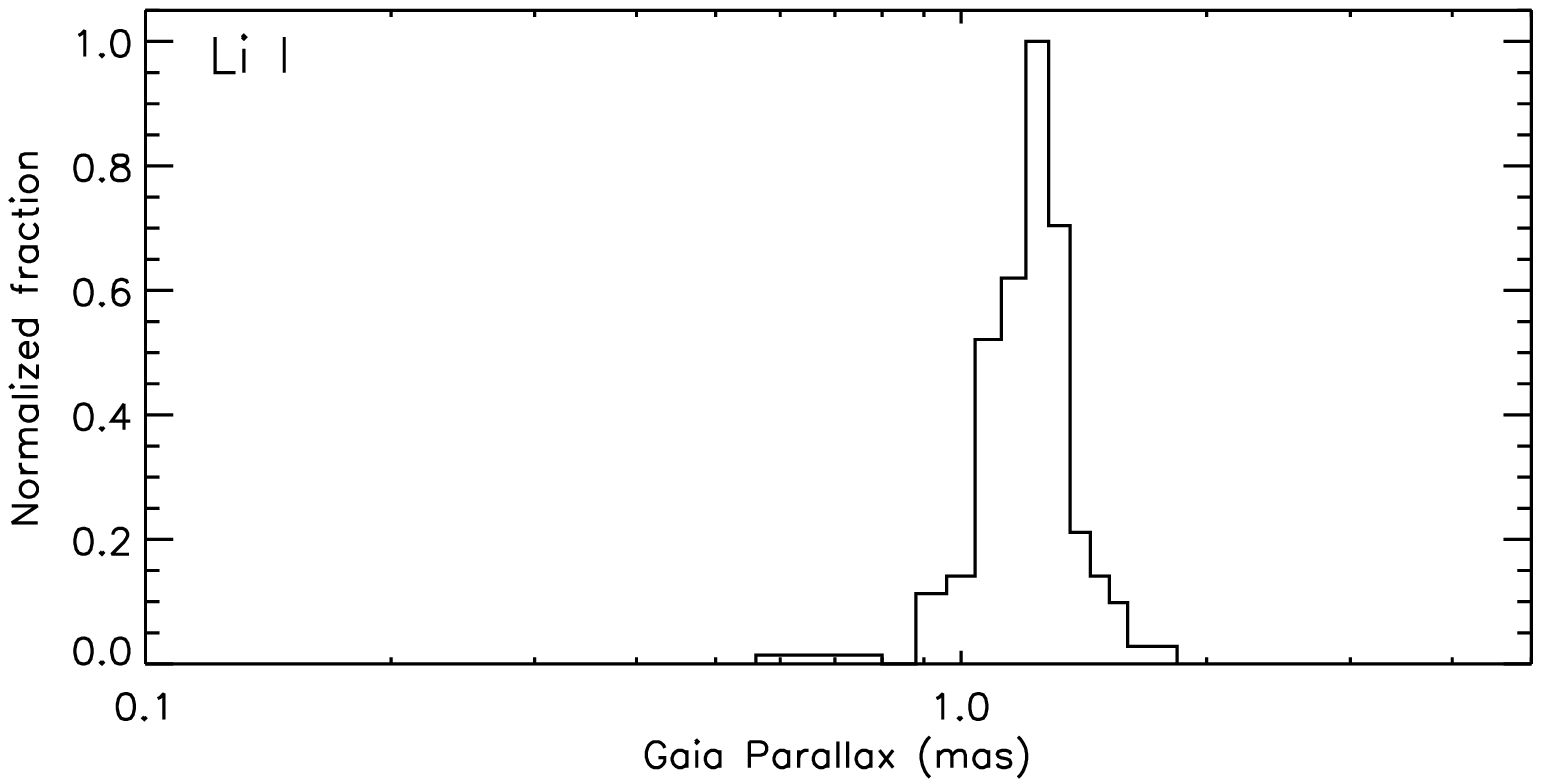}
\includegraphics[angle=0,width=1\columnwidth]{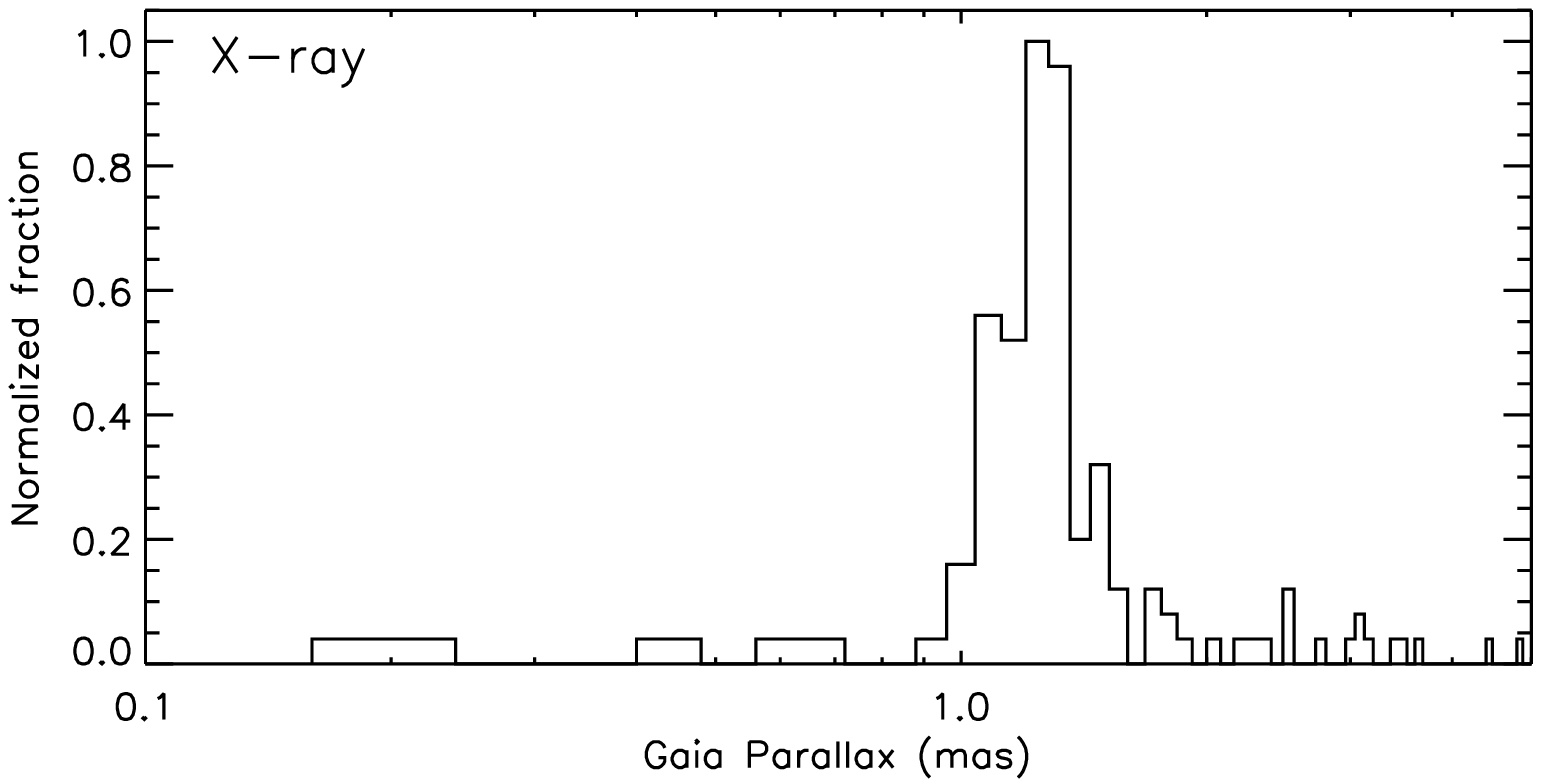}
\includegraphics[angle=0,width=1\columnwidth]{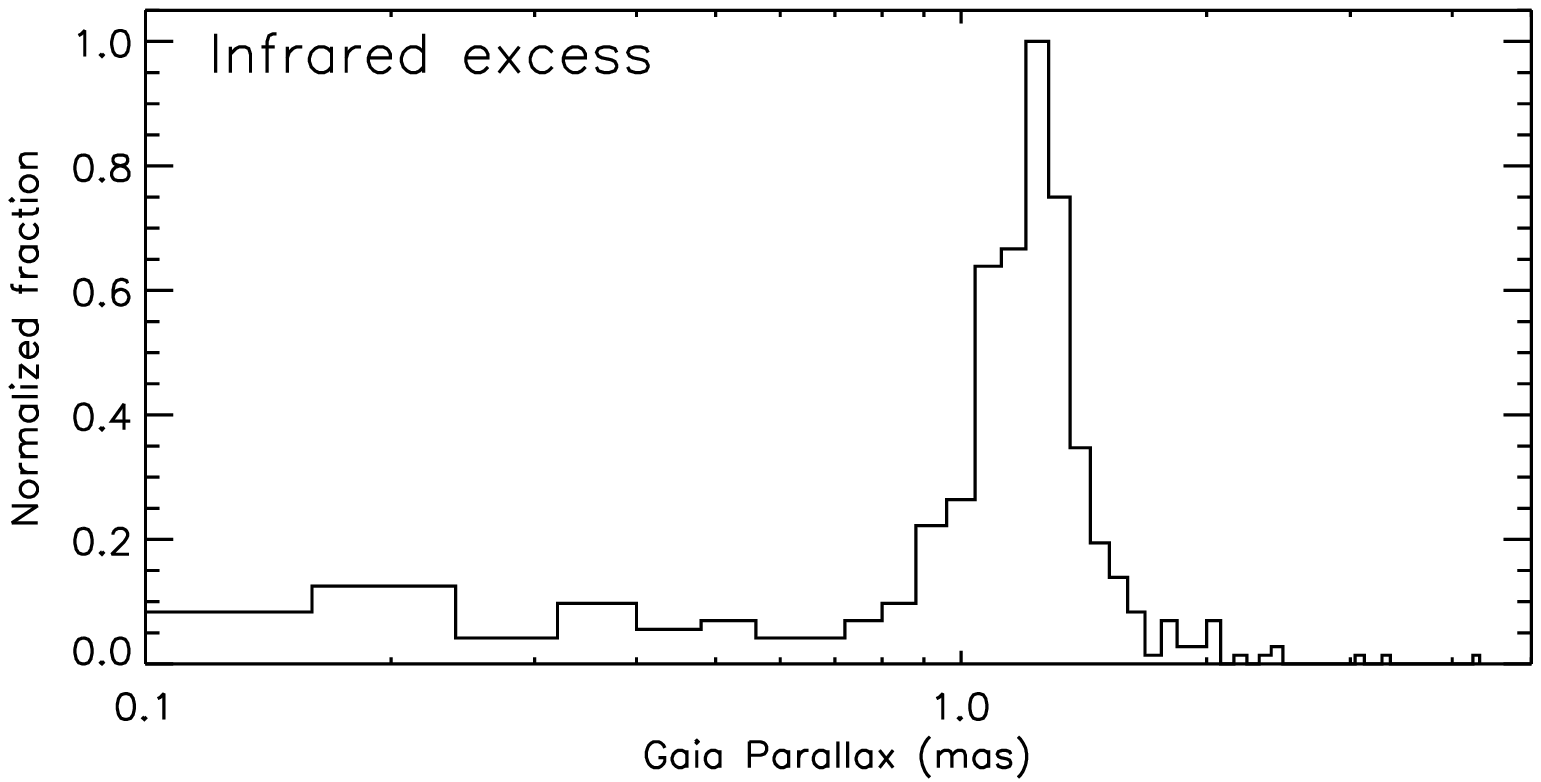}
\caption{Distribution of {\it Gaia} parallax for the sources with Li\,I absorption (upper), with X-ray emission (middle), and those with infrared excess emission (bottom).
}\label{Fig:GAIA_YSO}
\end{center}
\end{figure}

\subsection{Membership Selection}\label{sect:member}

In the previous section we assessed the foreground extinction and the basic stellar properties of temperature and luminosity for our sample. In addition, we derived the spectral veiling at red optical wavelengths, which is a proxy for the infall and accretion of material onto the star.  The evidence for ongoing accretion is one indicator of stellar youth.   Other stellar youth indicators are investigated in this section. These include: lithium absorption in the spectra, which would indicate an age younger than the lithium depletion time, X-ray emission indicative of coronal activity, which is known to decline with age, and infrared excess indicative of a dusty circumstellar disk that can feed the disk-to-star accretion that is detected as the spectral veiling.  We investigate each of these measures of stellar youth in the first three subsections below.  Then in the fourth subsection we use Gaia astrometry to confirm the membership of the lithium, infrared excess, and X-ray selected samples. {\newrev All the sources discussed in the section are among our spectroscopic sample.}

\begin{figure}
\begin{center}
\includegraphics[angle=0,width=1\columnwidth]{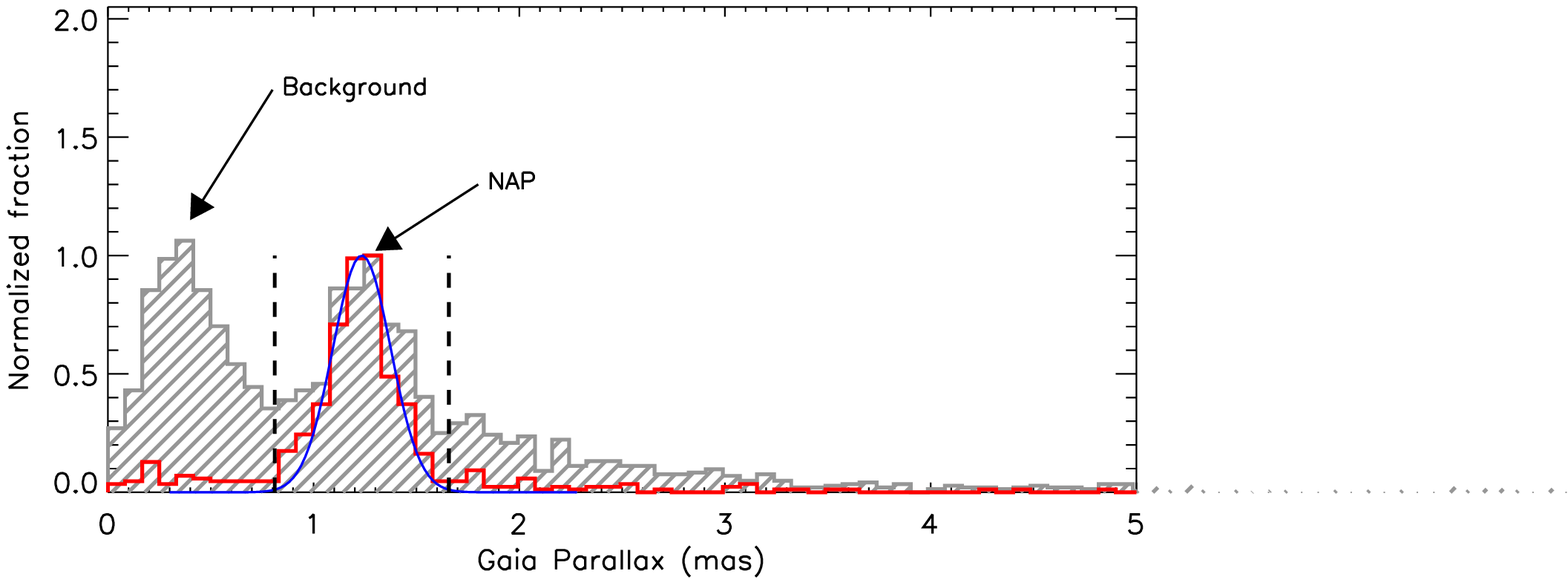}
\includegraphics[angle=0,width=1\columnwidth]{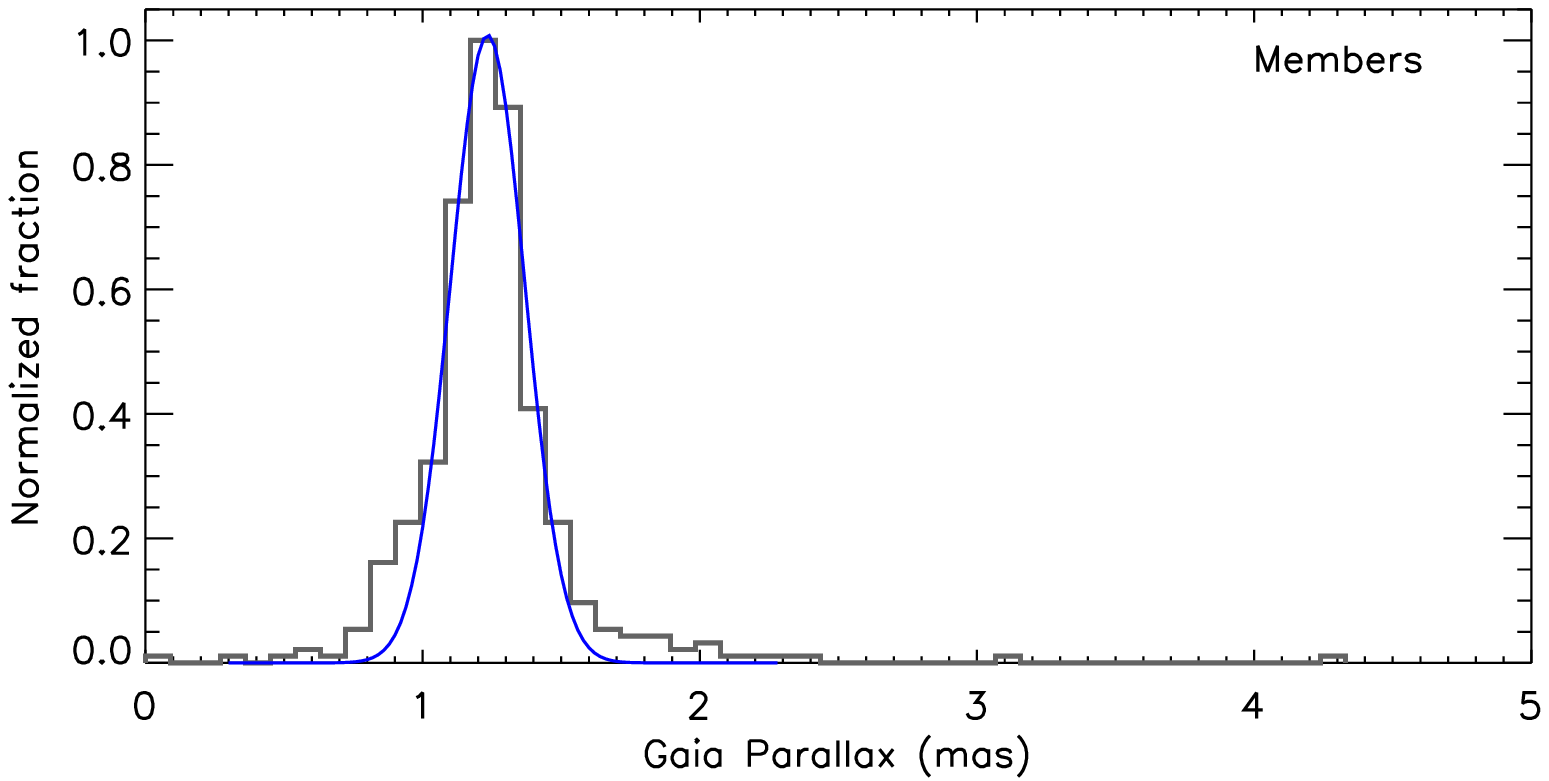}
\caption{Upper: Distribution of {\it Gaia} parallax for our spectroscopic targets (grey line filled  histogram), and the YSO candidates (red open histogram) before membership refinement shown in the upper panel. Bottom: Distribution of {\it Gaia} parallax for the identified members after membership refinement in the NAP.   
}\label{Fig:GAIA_all}
\end{center}
\end{figure}

\subsubsection{Li\,I Absorption}
We use the presence of the Li\,I absorption line at 6708\,\AA\  to characterize the youth of the stars. In Figure~\ref{Fig:Li}, we show examples of \LiI\ detection.  For four individual sources, we show both the object spectrum and the spectrum of an older field star with a similar spectral type, for comparison. We visually inspect all spectral targets and detect \LiI\ absorption from 350 sources.

\subsubsection{X-rays}
Parts of the have been  observed and studied with the X-ray space telescopes XMM-Newton and Chandra \citep{2017A&A...602A.115D}. We matched our targets with the X-ray sources using the positional uncertainties of the X-ray sources as the tolerances. In this way, we found 175 counterparts for the X-ray sources in \cite{2017A&A...602A.115D} in our spectroscopic sample.

\begin{figure}
\begin{center}
\includegraphics[angle=0,width=\columnwidth]{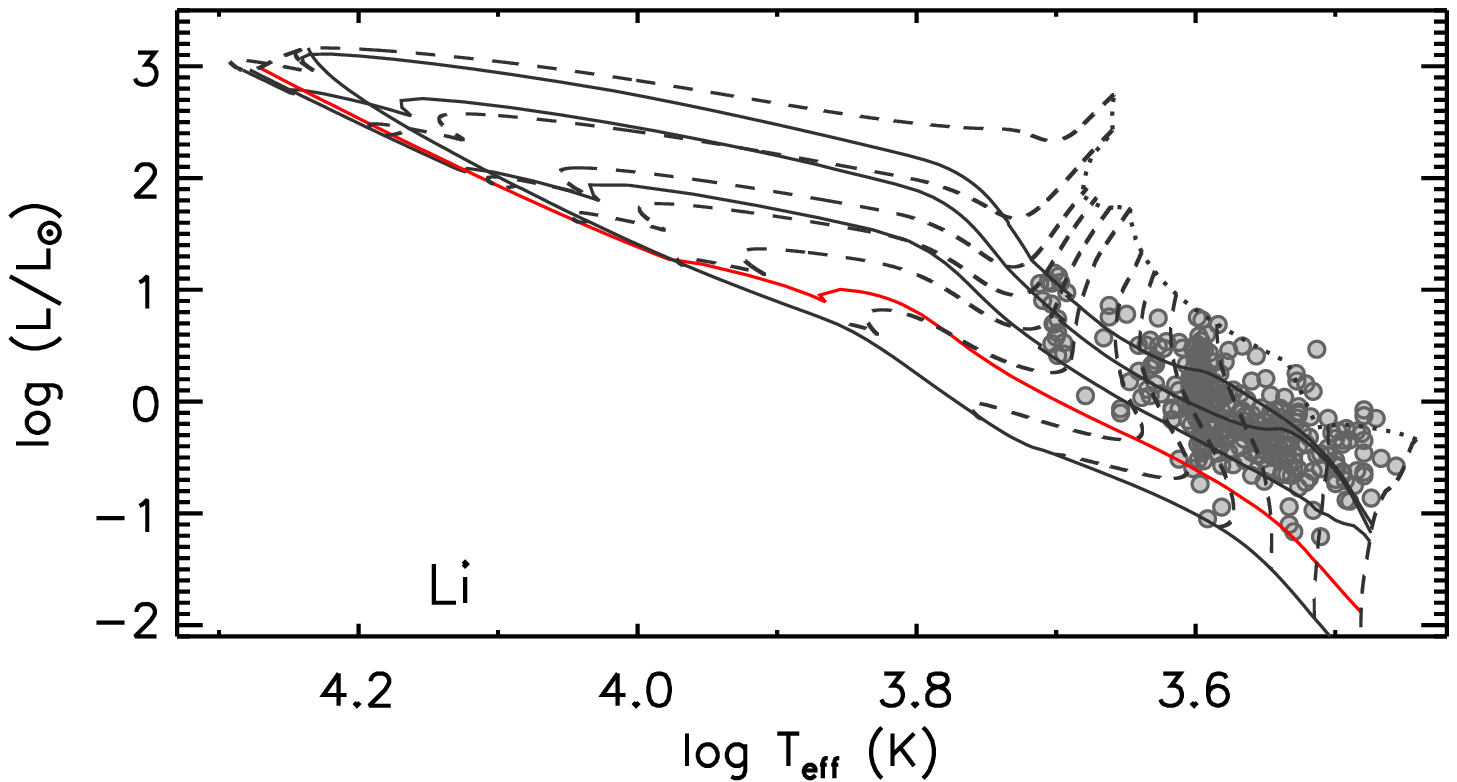}
\includegraphics[angle=0,width=\columnwidth]{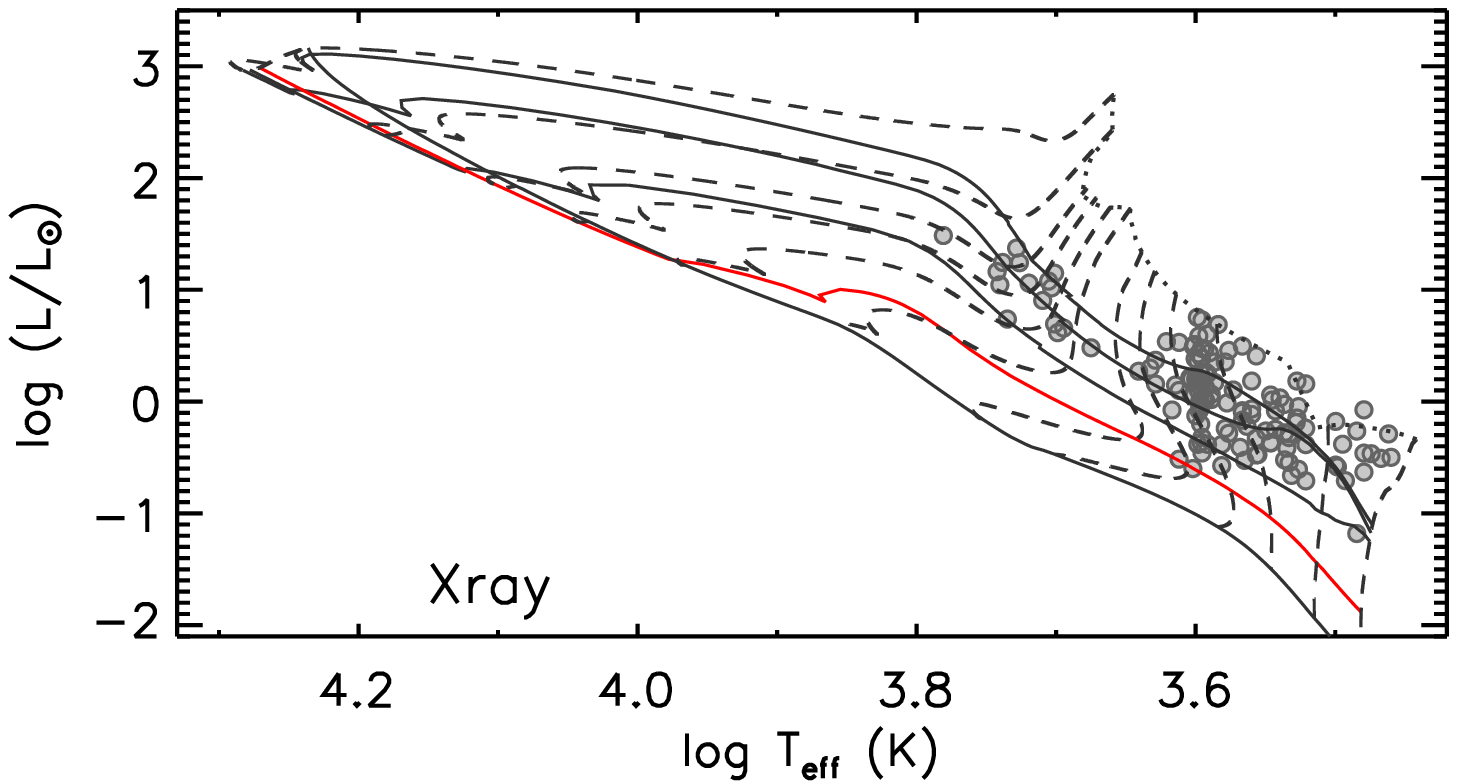}
\includegraphics[angle=0,width=\columnwidth]{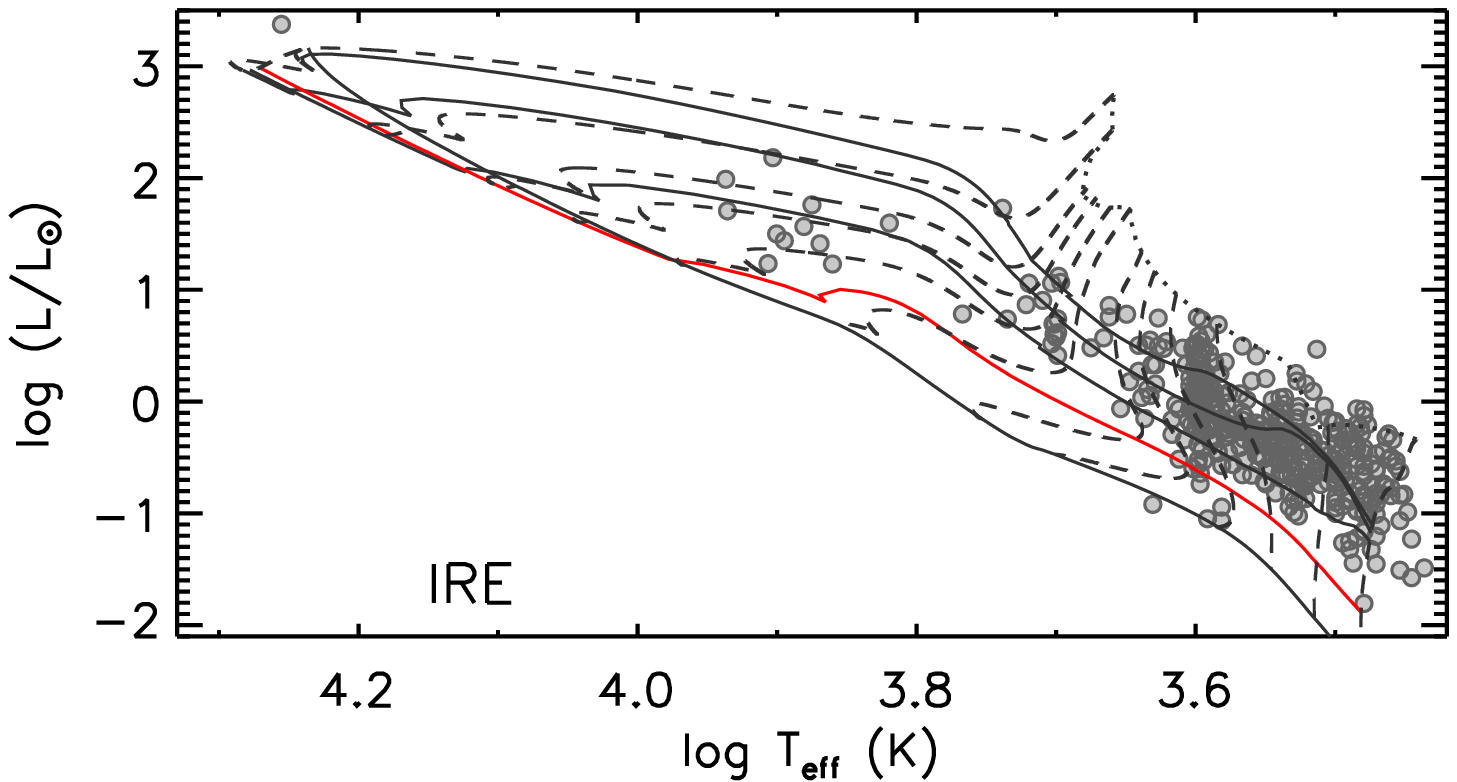}
\caption{HRD for young stars identified with Li\,I absorption (top), infrared excess (middle), and X-ray (bottom). The stellar luminosity is derived from fitting the SEDs using model atmospheres. In each panel,  the nonmagnetic evolutionary tracks (dashed lines) from \cite{2016A&A...593A..99F} are shown as a comparison. The solid lines show the isochrones at ages of 0.5, 1, 3,  5, 10, and 50~Myr from top to bottom. The dotted line {\newrev connecting the tops of the mass tracks indicates} the birth line. The dashed lines present the evolutionary tracks of young stellar objects with masses of 0.09, 0.2, 0.4, 0.6, 0.8, 1.0, 1.5, 2.0, 2.5, 3.0, 4.0, and 5.8~$M_{\odot}$, respectively. \label{Fig:HRD_YSO}}
\end{center}
\end{figure}

\begin{figure*}
\begin{center}
\includegraphics[angle=0,width=0.98\columnwidth]{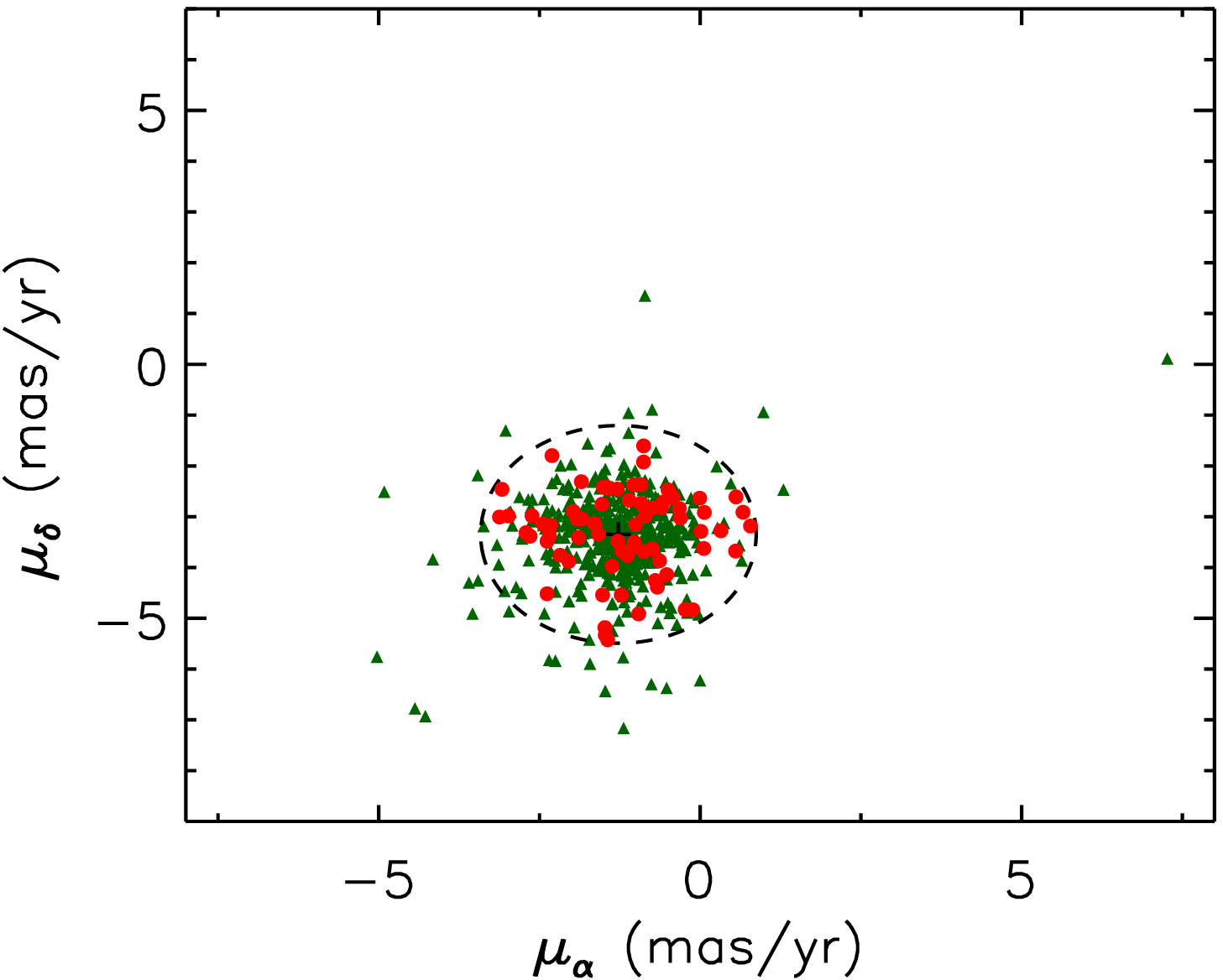}
\includegraphics[angle=0,width=0.98\columnwidth]{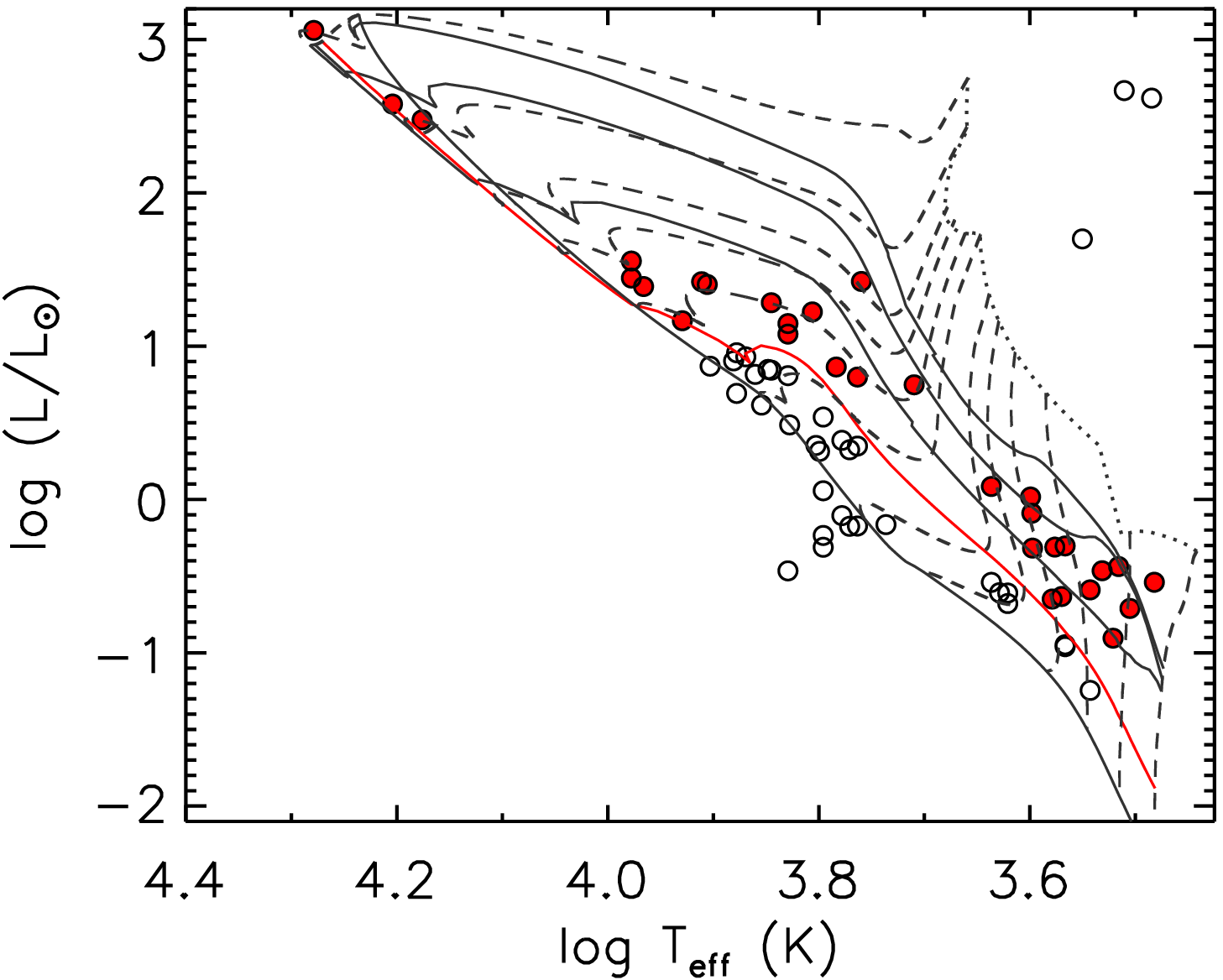}
\caption{Left: \textit{Gaia} proper motions of  members (green filled triangles) identified with Li~I absorption, X-ray emission, and/or infrared excess emission. Only the members with $RUWE\leq1.4$ are used for the plot. The mean proper motions of the members and the 3\,$\sigma$ ellipses of the standard deviation of the distribution are shown with dashed-line curve. The red filled circles are the selected member candidates based solely on their parallaxes and proper motions.  Right: H-R diagram for sources identified {\newrev as members} based {\newrev solely} on their \textit{Gaia} parallax values with stellar luminosities calculated adopting the mean distance of the (810~pc). The sources (red filled circles) apparently younger than 10~Myr are selected as memebers, and others (open circles) are considered as field stars with distances close to that of the NAP. Evolutionary tracks are the same as in Figures~\ref{Fig:HRD_YSO}.
\label{Fig:HRD_GAIA}}
\end{center}
\end{figure*}

\subsubsection{Infrared Excess}

For each source with an optical spectrum in our sample, we determine whether there is evidence for infrared excess emission by comparing the appropriate the BT-Settl model atmosphere with two versions of the infrared spectral slope, $\alpha_{3.6-8}$ and $\alpha_{3.6-24}$, calculated with the Spitzer photometric data and corresponding to the spectral range of [3.6] to [8.0] and [3.6] to [24], respectively, as done in \cite{2017AJ....153..188F}. 

Figure~\ref{Fig:alpha} shows the two infrared spectral slopes  versus the spectral types, for our sources. In the figure, the spectral slopes of the reference BT-Settl atmospheric models \citep{2011ASPC..448...91A} are calculated allowing for a 10\% uncertainty in the Spitzer photometry. Sources with infrared spectral slopes steeper than the slopes of the atmospheric models are considered as showing no infrared excess emission. For those with shallower infrared spectral slopes than the atmospheric models, we visually examine their spectral energy distributions. Sources that show infrared excess at more than 3$\sigma$ confidence level are considered to show infrared excess emission. 

There are 518 sources which show infrared excess emission. An additional 6 sources show strong infrared excess, but their optical spectra are highly veiled and thus they cannot be classified.  As they are likely true infrared excess sources, we include them for further investigation. Thus, in total, we identify 524 sources with infrared excess emission among our spectroscopic sample.

\subsubsection{Gaia Astrometry}

Combining the sources {\newrev identified} in the previous subsections having infrared excess emission, Li\,I absorption, or X-ray emission results in 655 unique YSO candidates. {\newrev In this section}, we further assess the likelihood of their membership in the NAP using the information from {\it Gaia} DR2 \citep{2016A&A...595A...1G,2018A&A...616A...1G}. {\it Gaia} astrometry can also reveal more members of the NAP which are not identified using the above three criteria.  It is expected that infrared excess and X-ray emission samples may be incomplete, and that the \LiI\ indicator is not always apparent at our relatively low spectral resolution.  

To control for only high quality {\it Gaia} parallaxes, we computed the Renormalised Unit Weight Error ($RUWE$) following the method described in the {\it Gaia} DR2 manual. When $RUWE>$1.4 it may indicate that the source is a multiple system or otherwise problematic in terms of the astrometric measurement. In Figure~\ref{Fig:GAIA_YSO}, we show the distribution of {\it Gaia} parallaxes with $RUWE\leq$1.4 for our YSO candidates selected using above three criteria. The parallax distribution in each case is strongly peaked at $\sim$1.24\,mas. In the figure, we can note that the YSO candidates with infrared excess emission are contaminated mainly by the background sources with small parallaxes, but the X-ray sample is polluted more by the foreground sources; YSO candidates with identified Li\,I absorption represent the cleanest parallax distribution and thus the best sample of likely members of the NAP. 

{\newrev In Figure~\ref{Fig:GAIA_all} we show the distribution of {\it Gaia} parallaxes with $RUWE\leq$1.4 for all of our spectroscopic sample. Unlike the distributions shown in Figure~\ref{Fig:GAIA_YSO}, the parallax distribution for the full sample shows a double peak. This is due to the background field stars at small parallaxes and then the clustered NAP members.} We combine the YSO candidates identified with the above three criteria, and obtain the distribution of {\it Gaia} parallaxes with $RUWE\leq$1.4 highlighted in red in the upper panel of Figure~\ref{Fig:GAIA_all}. We fit this distribution with a Gaussian function, limiting the fitting range to between 1.0 and 1.7\,mas so as to reduce the probable contamination. The fitting gives three parameters: $a_{0}$=1.000 is for the normalization, $P=1.234$\,mas is the centroid {\newrev and a standard deviation in the parallax $\sigma=0.141$ mas}.  Hereafter, we adopt 810~pc as the distance of the, 
consistent with the 795~pc value from \cite{2020ApJ...899..128K}.
For the YSO candidates with {\it Gaia} parallaxes within $P-3\times\sigma$ and $P+3\times\sigma$, we accept them as members of the. For those outside limits of parallaxes, we further inspect their memberships as follows. 
\begin{figure}
\begin{center}
\includegraphics[angle=0,width=1\columnwidth]{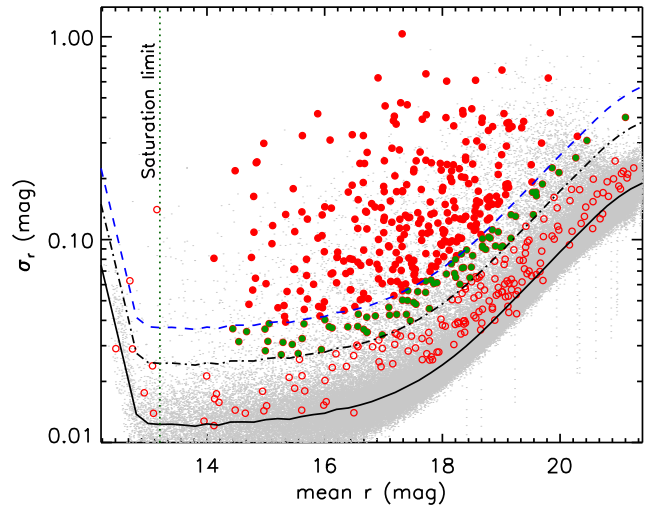}
\caption{{\newrev Photometric variability statistic} $\sigma_{\rm r}$ vs {\newrev mean magnitude} $\overline{r}$ for all stars in the NAP field with distance less than 2~kpc.  {\newrev The underlying distribution is shown as small gray-color dots.} The black solid line marks $\overline{\sigma}_{\rm r}$  within individual magnitude bins, while the black dash-dotted line indicates 2$\times\overline{\sigma}_{\rm r}$  and the blue dashed line 3$\times\overline{\sigma}_{\rm r}$.  Our identified YSOs are shown as red circles: red-filled circles are for strongly variable stars,  
green-filled circles for moderately variable stars, and open circles for stars with low variability.
}
\label{Fig:Var}
\end{center}
\end{figure}

\begin{figure}
\begin{center}
\includegraphics[angle=0,width=0.98\columnwidth]{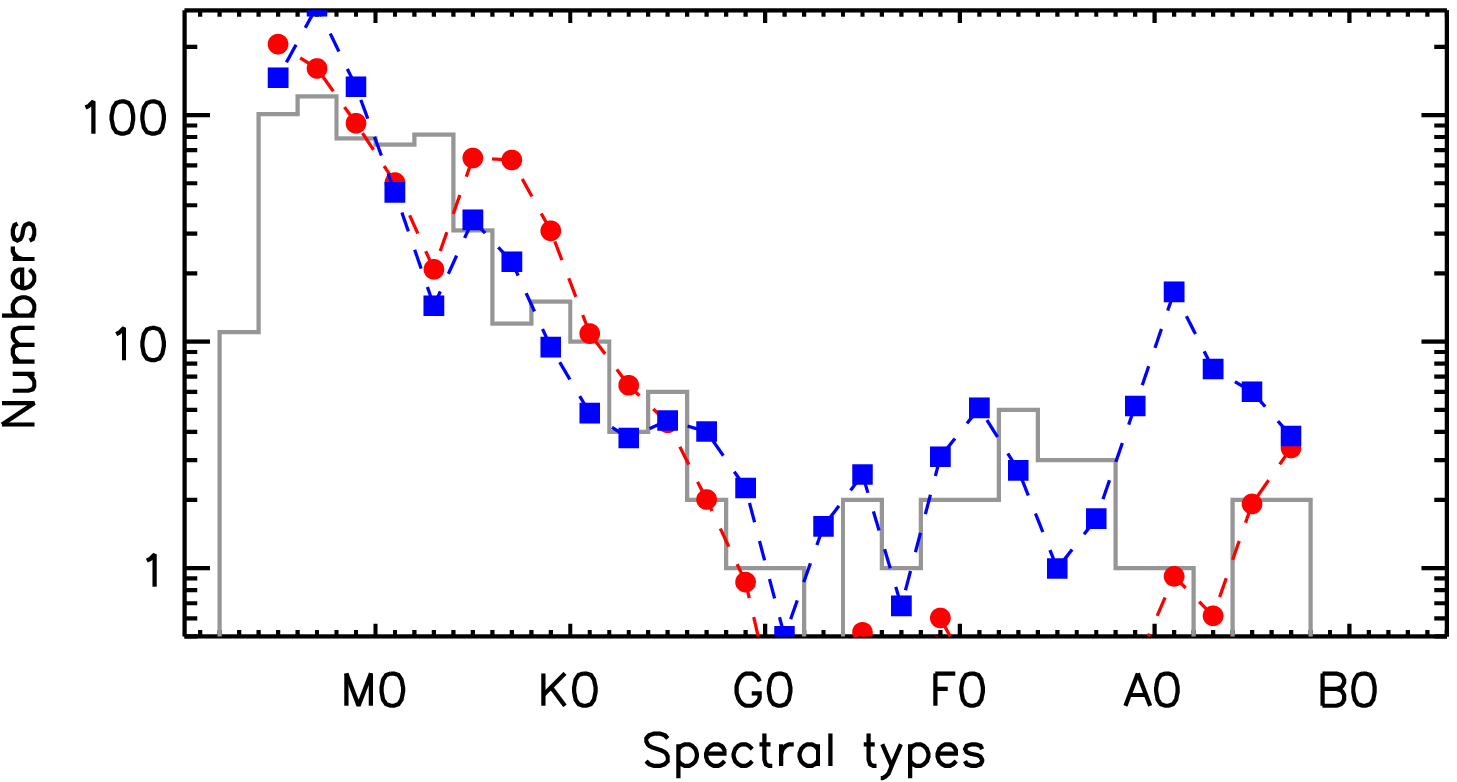}
\includegraphics[angle=0,width=0.98\columnwidth]{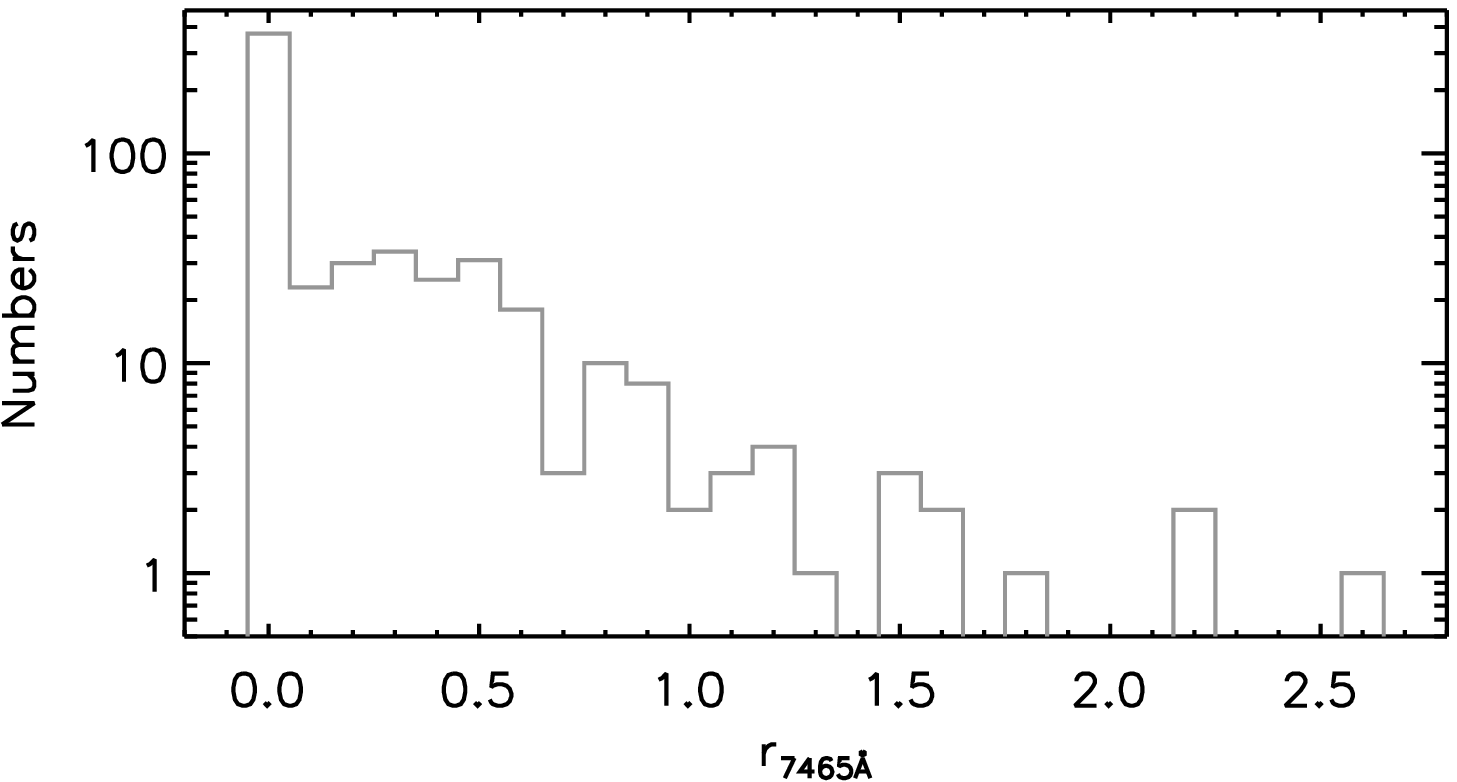}
\includegraphics[angle=0,width=0.98\columnwidth]{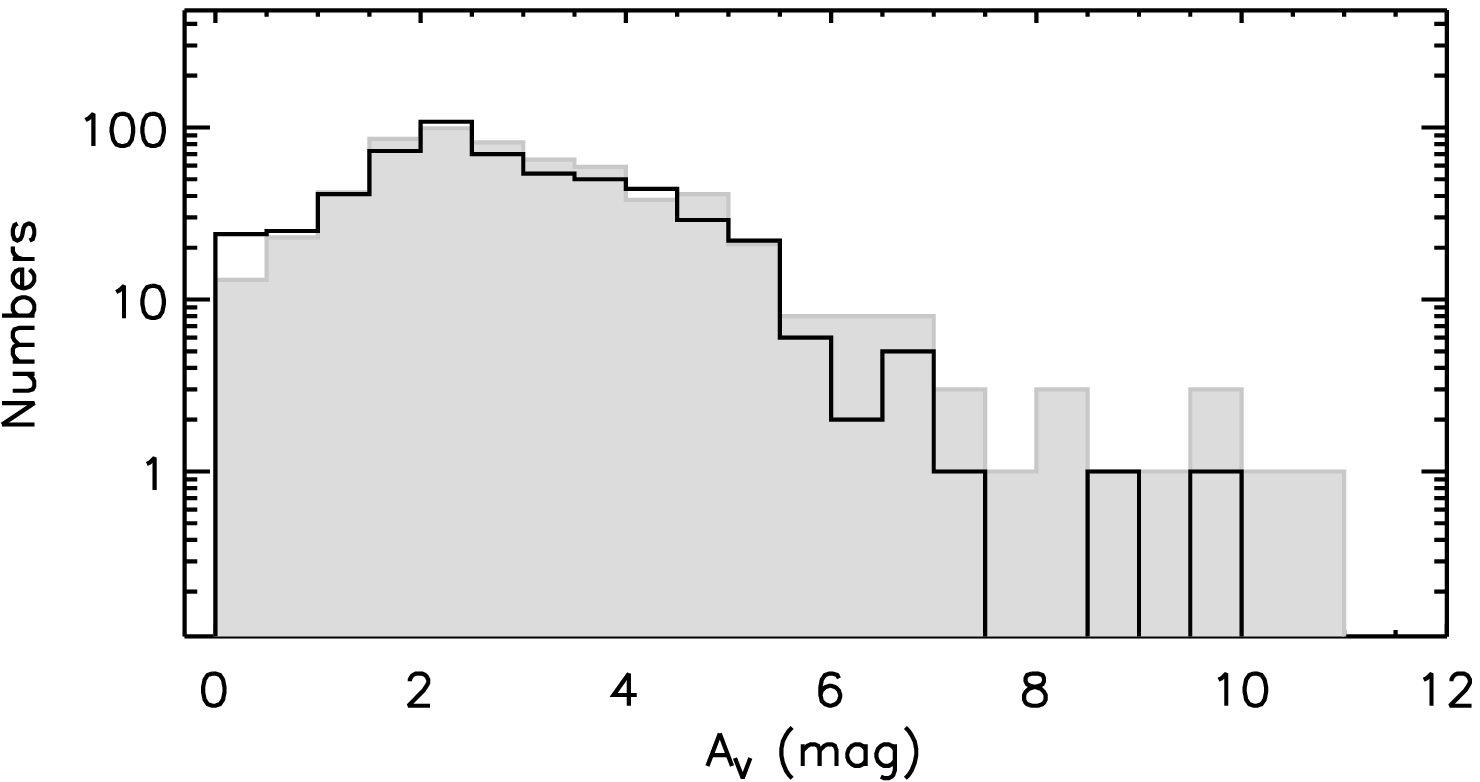}
\caption{{\newrev Results of our spectral and SED fitting processes for members of the NAP.} Upper: Distribution of spectral types.  The red dashed line connecting filled circles shows the predicted distribution of spectral type for a 1\,Myr model cluster with an IMF like that of the Trapezium cluster while the blue dashed line connecting filled boxes is for a  10\,Myr model cluster. Middle: Distribution of veiling at 7465\,\AA\ . Bottom: Distribution of $V$-band extinction, derived from fitting SEDs using model atmospheres (gray filled histogram) or   X-shooter templates (black open histogram). \label{Fig:Spt_veiling}}
\end{center}
\end{figure}

{\noindent\it Li\,I absorption:} 
Among the 350 YSO candidates identified with  Li\,I absorption, 284 sources are within the established parallax limits. Of the other 66 sources, 26 have no estimate of parallax or a negative parallax is reported. We thus inspect their membership likelihood by considering their locations in the H-R diagram assuming they have a parallax of 1.234\,mas and using the nonmagnetic evolutionary tracks from \cite{2016A&A...593A..99F} as a comparison (hereafter we only use these evolutionary tracks). {\newrev We include all of them as the members of the based on their locations in H-R diagram.}
For the other 40 sources, 33 have $RUWE>1.4$, and thus have unreliable astrometric measurement. We therefore continue to include them as member candidates. For the 7 sources with $RUWE\leq1.4$, 4 agree with the mean parallax of the (1.234\,mas) within $3\times\sigma_{plx}$, where $\sigma_{plx}$ is the individual uncertainty on the parallax for each source. These sources are thus also considered as possible member candidates of the NAP. For the remaining 3 sources, an inspection of their SEDs suggest they show infrared excess emission, which is consistent with their being young stars, but also allows the possibility that they could be lithium-rich, dusty giants. In Figure~\ref{Fig:HRD_YSO}~(top), we show the H-R diagram for all YSO candidates identified with Li\,I absorption adopting a distance of 810~\,pc. For these three sources, their locations in the H-R diagram are consistent with other young stars. Thus, we list them as potential members of the NAP with a note indicating they could be excluded as members by our formal parallax criterion. In Figure~\ref{Fig:HRD_YSO}(top), we also note two sources are below the 10~Myr isochrone.  While unusual, it could be the case that these sources harbor an edge-on disk, and thus have their luminosities underestimated.   In summary, we identify 350 likely members of the NAP with Li\,I absorption.

\begin{figure}
\begin{center}
\includegraphics[angle=0,width=0.98\columnwidth]{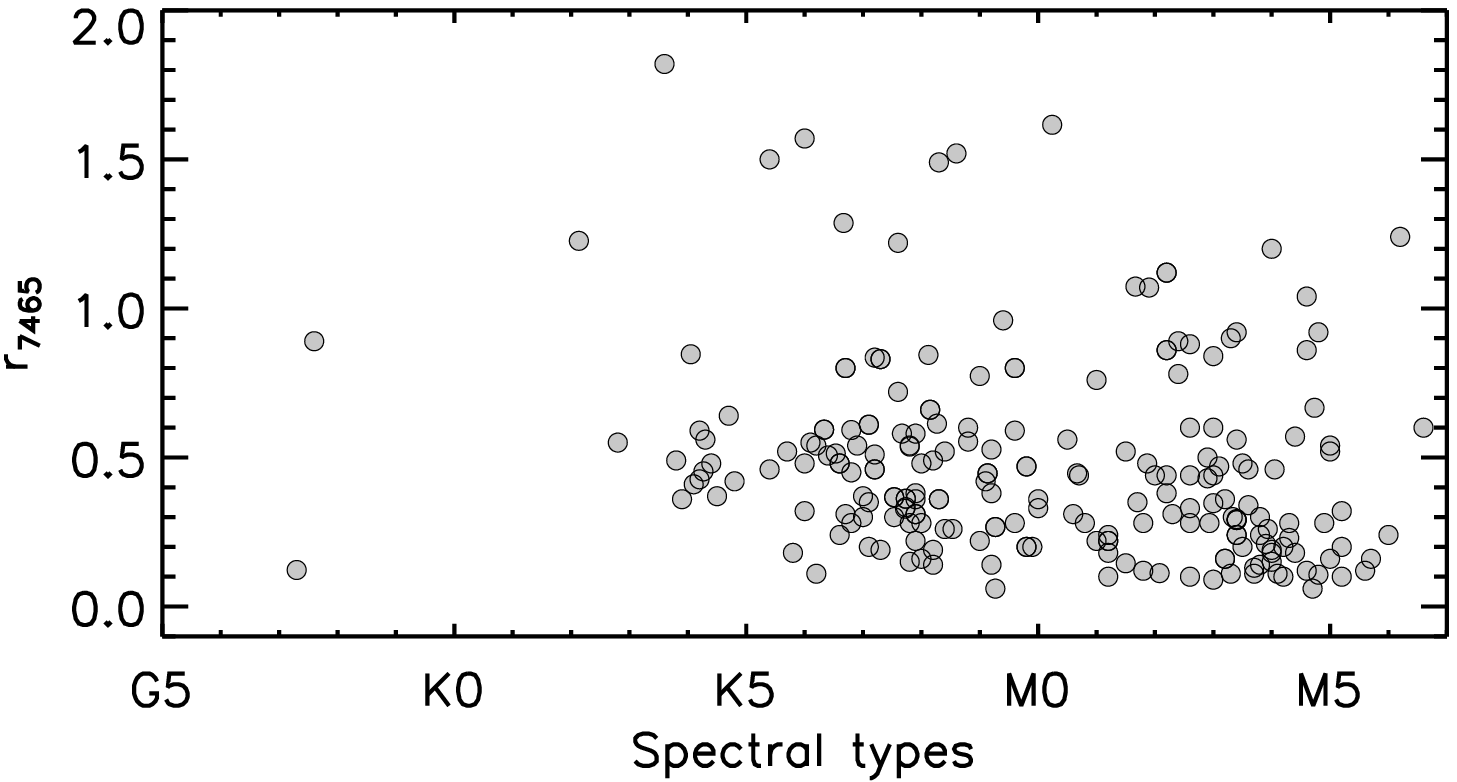}
\includegraphics[angle=0,width=0.98\columnwidth]{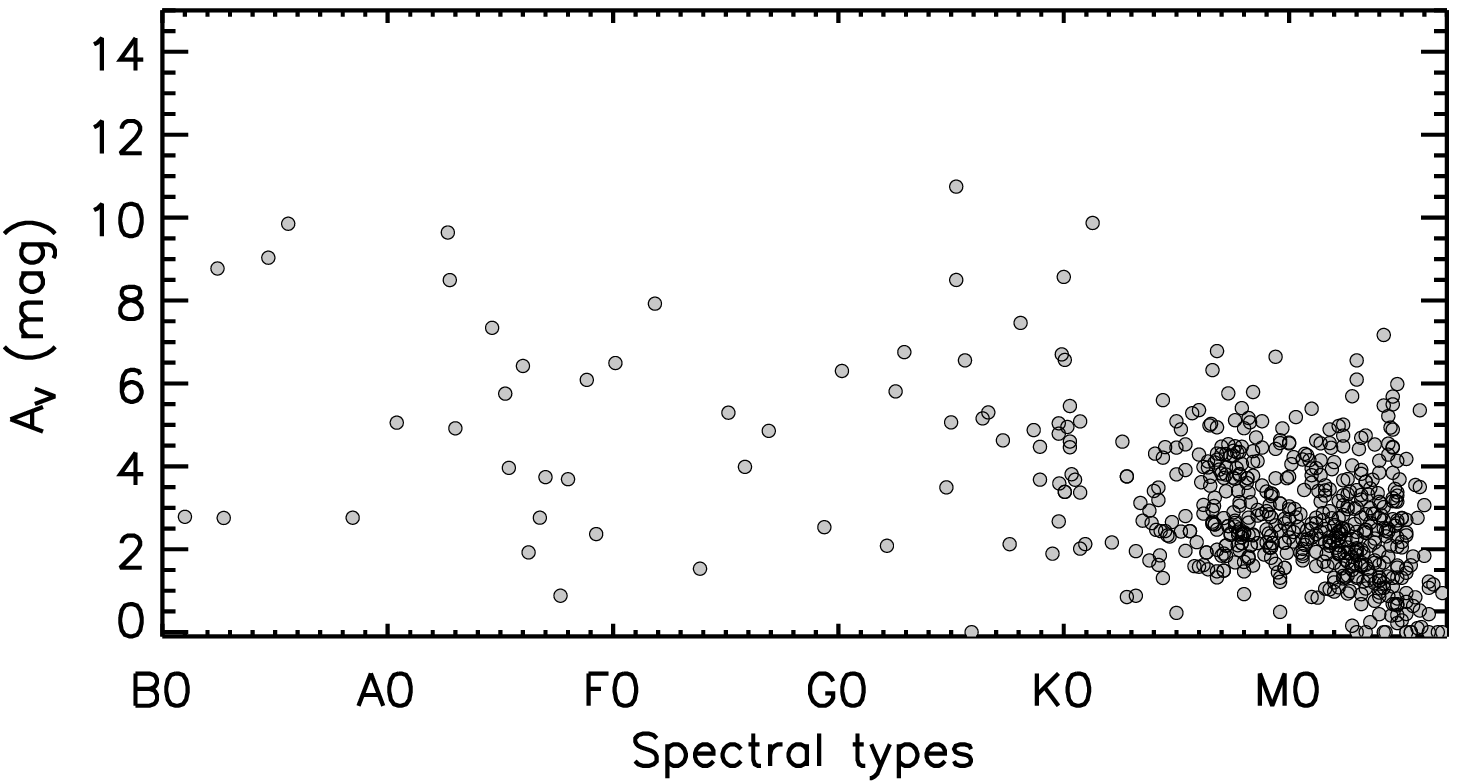}
\caption{Veiling (top) and visual extinction (bottom) as a function of spectral type for the members of the NAP.  Top panel includes only the sub-sample with measurable veiling in the spectra while bottom panel includes all members of the NAP studied in this work.\label{Fig:veiling_Av}}
\end{center}
\end{figure}

{\noindent\it X-ray:} In our spectroscopic sample, we found 175 counterparts for the X-ray sources. Among them, 106 also present Li\,I absorption and have already been included above as members of the NAP. For the remaining 69 sources, 36 are within the parallax limits we have established for the NAP, while two have no estimate of parallax or a negative parallax, four have $RUWE>1.4$, and one agrees with the mean parallax of the NAP (1.234\,mas) within $3\times\sigma_{plx}$.  We include all the aforementioned 43 sources as member candidates for further inspection of their SEDs and their locations in the H-R diagram. Among the 43, 11 sources do not show infrared excess emission and have H-R diagram locations consistent with those of main sequence field stars.  One star, known as LkH$\alpha$~170 or V751~Cyg, which shows infrared excess and has been misclassified as a Herbig~Ae/Be star, is an apparent VY Scl star \citep{1999A&A...343..183G}. A detailed description of this source is presented in Appendix~\ref{Appen:vyscl}.  The remaining 31 sources should be good members of the NAP, 15 of which show infrared excess emission. Among the 69 sources mentioned above, 43 have been accounted for, leaving 26 remaining sources that have $RUWE\leq1.4$ but are outside of the established limits on parallax,  Of these, 24 should be foreground/background field stars based on their locations in H-R diagram. For the other three, all of them should be young stars based on their locations in H-R diagram; one has spectral type A8 and the other has spectral type M3, with extinction $A_{\rm V}$=3.2 mag which is consistent with the other members. Thus, we include the M-type star as a member of the NAP.  The A-type star has non-measurable extinction and is thus considered a foreground star. In conclusion, we confirm 32 members of the NAP based on X-ray data. The middle panel of Figure~\ref{Fig:HRD_YSO} shows the H-R diagram for the members identified with X-ray emission.

\begin{figure}
\begin{center}
\includegraphics[angle=0,width=0.98\columnwidth]{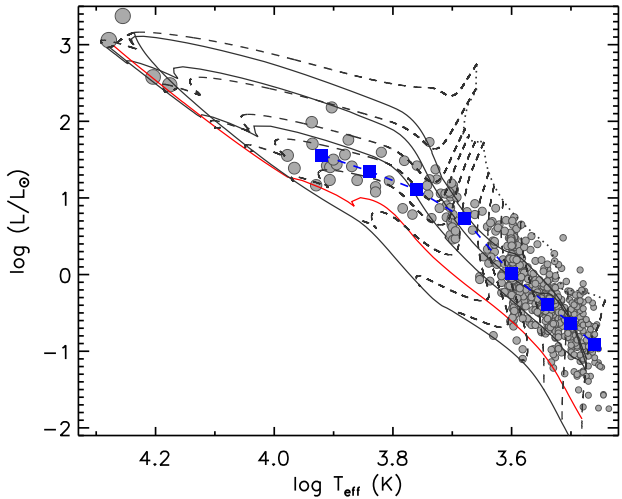}
\caption{H-R diagram for the members (gray filled circles) of the NAP. The sizes of the circles are scaled with the $T_{\rm eff}$ of the sources. The evolutionary tracks are the same as in Figure~\ref{Fig:HRD_YSO}.
\label{Fig:HRD_final}}
\end{center}
\end{figure}

\begin{figure*}
\begin{center}
\includegraphics[angle=0,width=2\columnwidth]{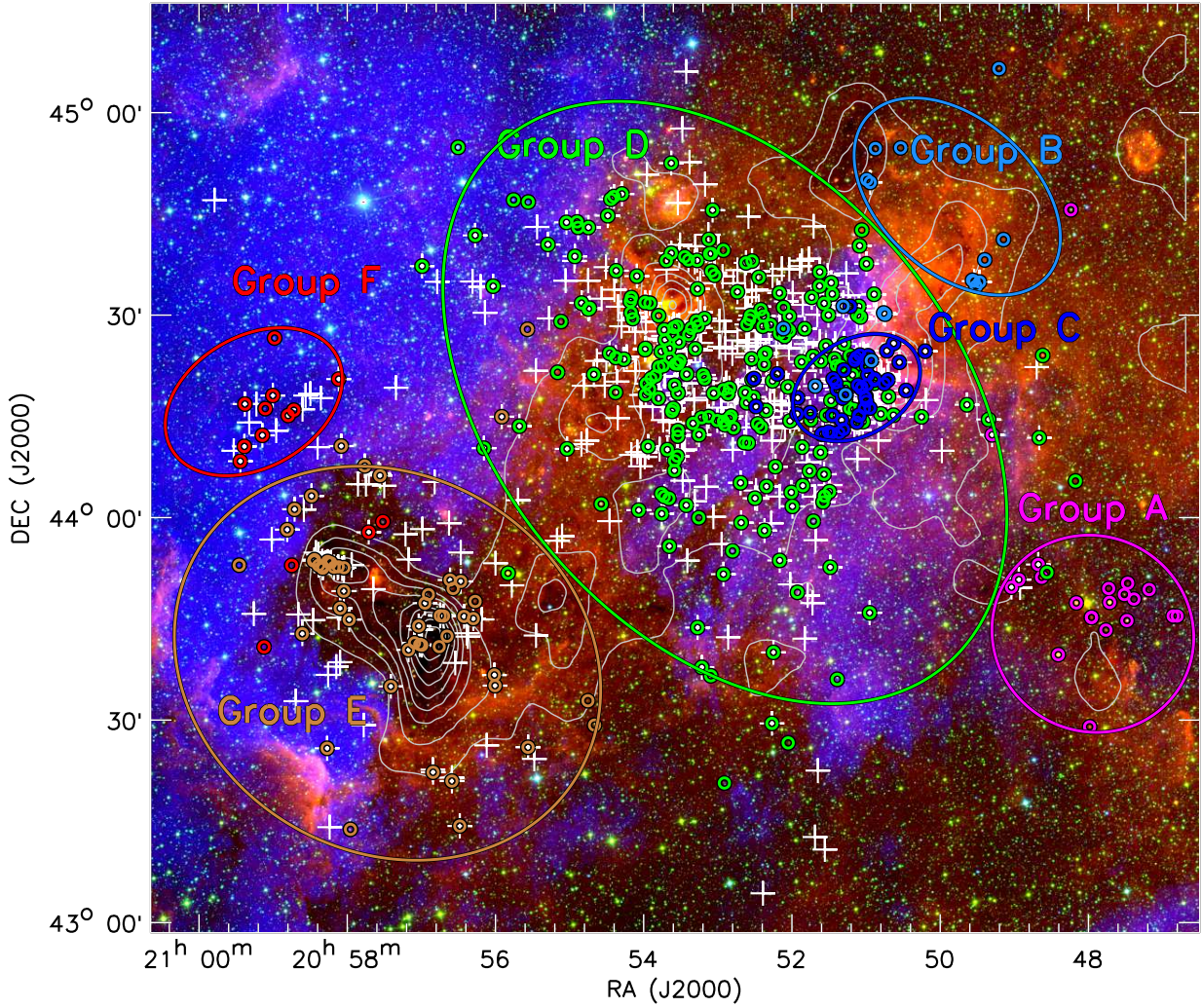}
\caption{NAP members (white pluses) identified in this work overplotted on  three-color image of  North America Nebula created in the same way as in Figure~\ref{Fig:YSO_NAN}. The color-coded open circles are the members of individual kinematic groups identified by  \cite{2020ApJ...899..128K}. The color-coded large ellipses show the boundaries of kinematic groups depicted in this work to group the YSOs without reliable astrometric measurements.}
\label{Fig:YSO_group}
\end{center}
\end{figure*}

{\noindent\it Infrared excess:} Among 524 YSO candidates with infrared excess emission, 289 sources have been identified as members of the NAP above, based on Li\,I absorption and/or X-ray emission.  Among the remaining 235 sources, 120 are within the parallax limits we have established for the NAP, 35 have no estimate of parallax or negative parallaxes, 18 sources have $RUWE>1.4$, and 23 agree with the mean parallax of the NAP (1.234\,mas) within $3\times\sigma_{plx}$. Among the 196 total sources, 6 sources have strong infrared excess and highly veiled spectra.  For these, we visually inspect their locations in the H-R diagram and identify 164 sources as likely members of the NAP . For the 6 sources with highly veiled spectra, we cannot place them in the H-R diagram since their spectra cannot be classified. We simply include them as  probable members of the NAP.  For the other 39 sources, 2 of them might be the members of the NAP based on their locations in the H-R diagram. In total, we identify 172 members of the NAP based on infrared excess emission. We also find that 40 nonmembers with spectral types earlier than G show infrared excess emission; they are presented in Appendix~\ref{Appen:IRE_nonmember}. 
The bottom panel of Figure~\ref{Fig:HRD_YSO} shows the H-R diagram for the members identified with infrared excess emission.

{\noindent \textit{New \textit{Gaia} members}: }Besides the 554 members identified with the above three criteria,  we also consider as candidate members of the NAP sources { \newrev with spectroscopic observations} that are within $P-3\times\sigma$ and $P+3\times\sigma$ of the nominal parallax, and that have proper motions consistent within 3$\sigma$ of the mean proper motion (see the left panel of Figure~\ref{Fig:HRD_GAIA}). In this way, we obtain 62 additional candidates among our spectroscopic sample. We further consider their membership likelihood using the H-R diagram, and select only those apparently younger than 10~Myr as members. In the right panel of Figure~\ref{Fig:HRD_GAIA}, we show these 26 newly appreciated members, as well the other 36 field stars with distances close to that of the NAP.

In summary, by considering the three samples selected on the basis of YSO criteria in the last section, plus additional sources with kinematics consistent with those of stars meeting the YSO criteria, we have  identified a total of 580 members of the NAP from our spectroscopic sample. Among them, 461 show infrared excess emission, and are likely disk-bearing young stars, while the other 119 sources do not show infrared excess emission (based on the available Spitzer data). The criteria on which the individual members are identified are given in Table~\ref{tabe_YSO}.   The distribution of the parallax for these sources (restricted to $RUWE\leq1.4$) is shown in the bottom panel of Figure~\ref{Fig:GAIA_all}. A Gaussian fit to the distribution {\newrev of the parallax} also gives a mean distance of $\sim$810\,pc. 
The spatial distribution of all the spectroscopically identified members of the NAP was shown in Figure~\ref{Fig:YSO_NAN}; while illustrated there, 
we note that our spectroscopic survey does not cover the immediate vicinity of either of the two massive O-type stars in the region: the Bajamar star and HD~199579.

 \begin{figure*}
\begin{center}
\includegraphics[angle=0,width=\columnwidth]{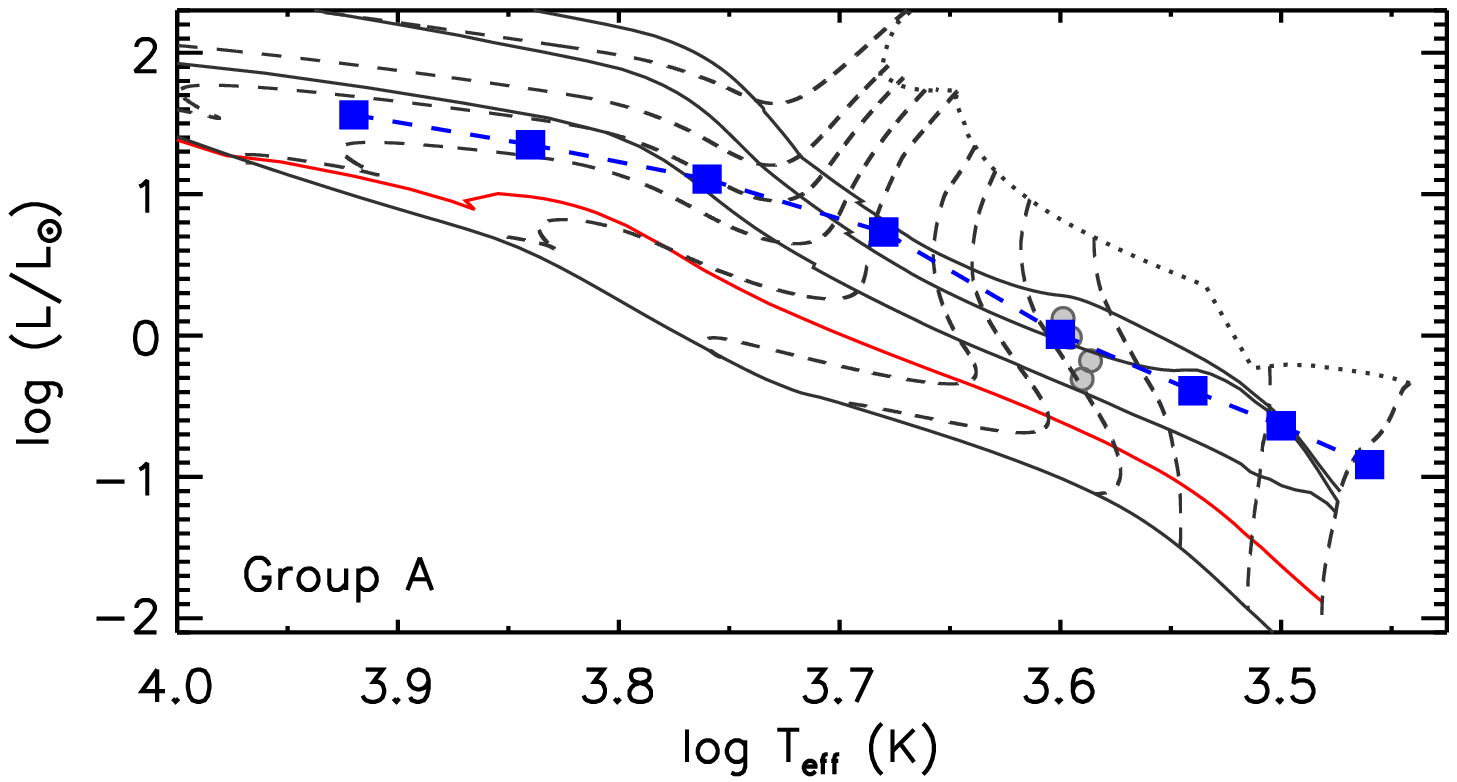}
\includegraphics[angle=0,width=\columnwidth]{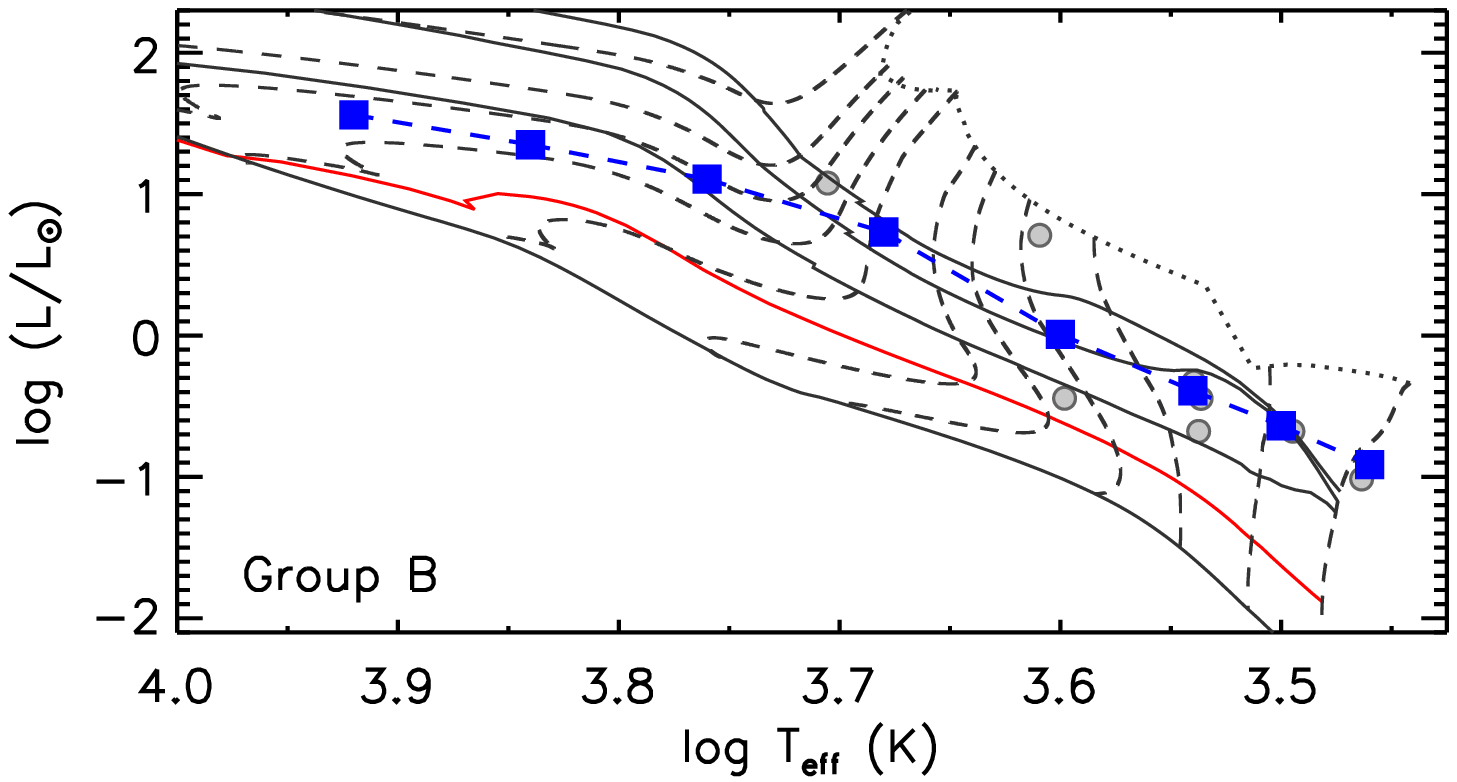}
\includegraphics[angle=0,width=\columnwidth]{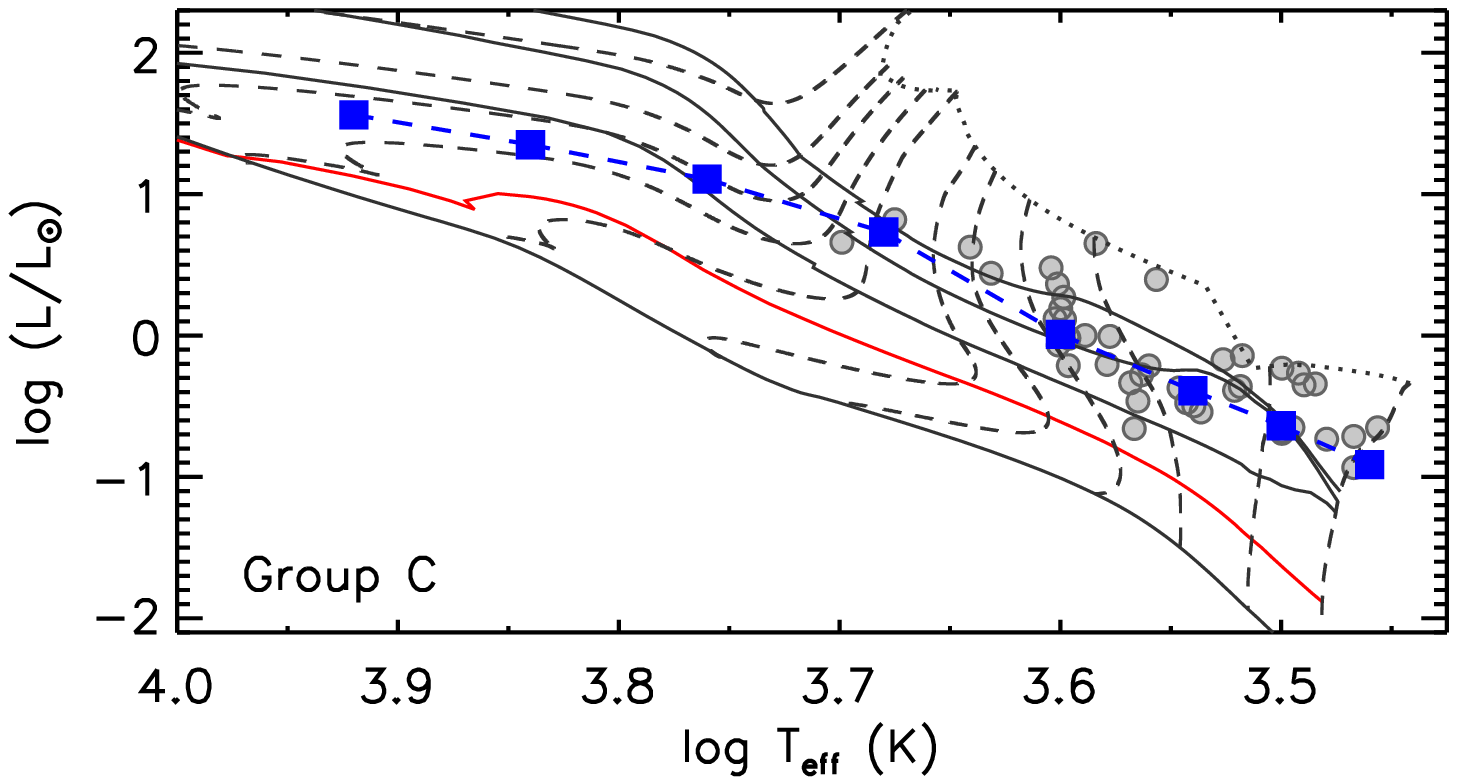}
\includegraphics[angle=0,width=\columnwidth]{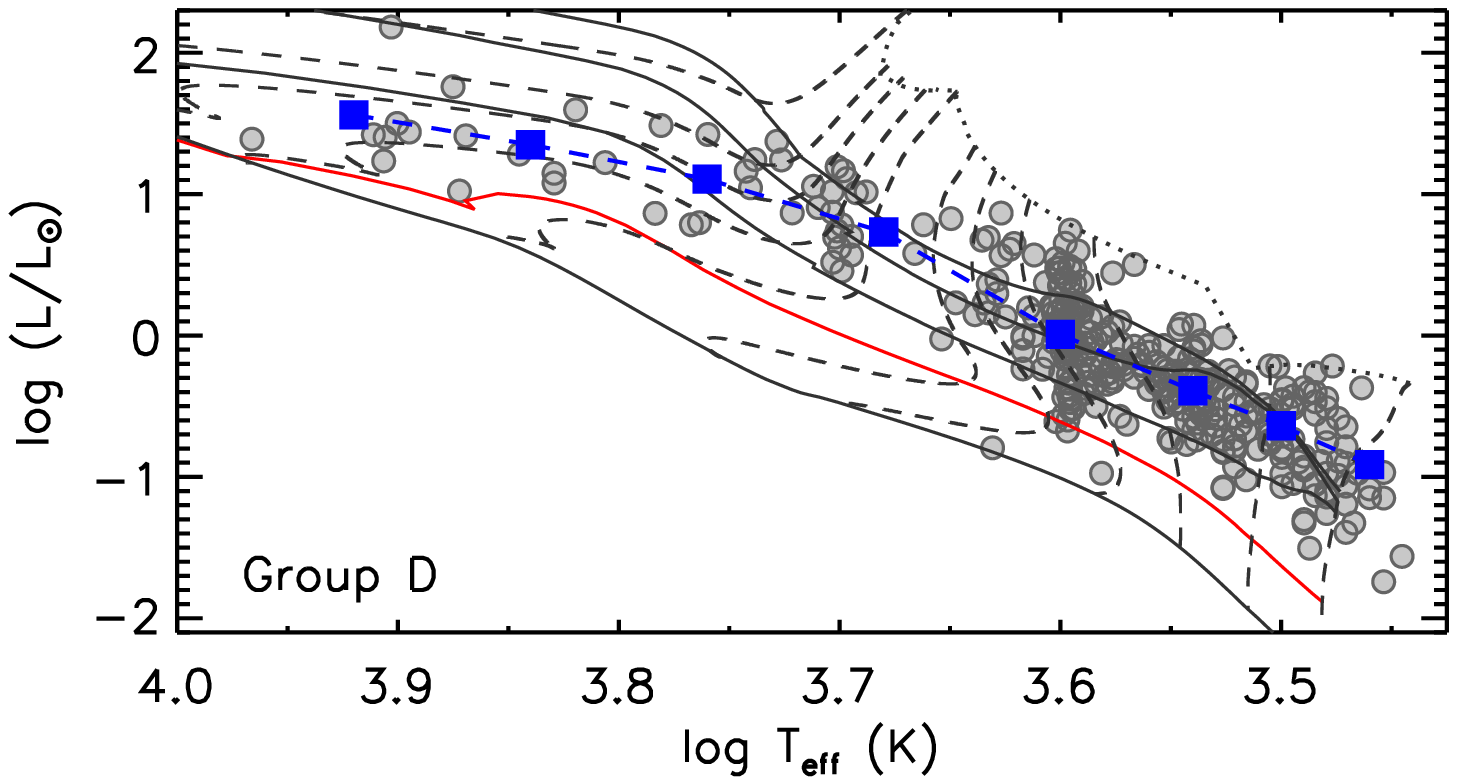}
\includegraphics[angle=0,width=\columnwidth]{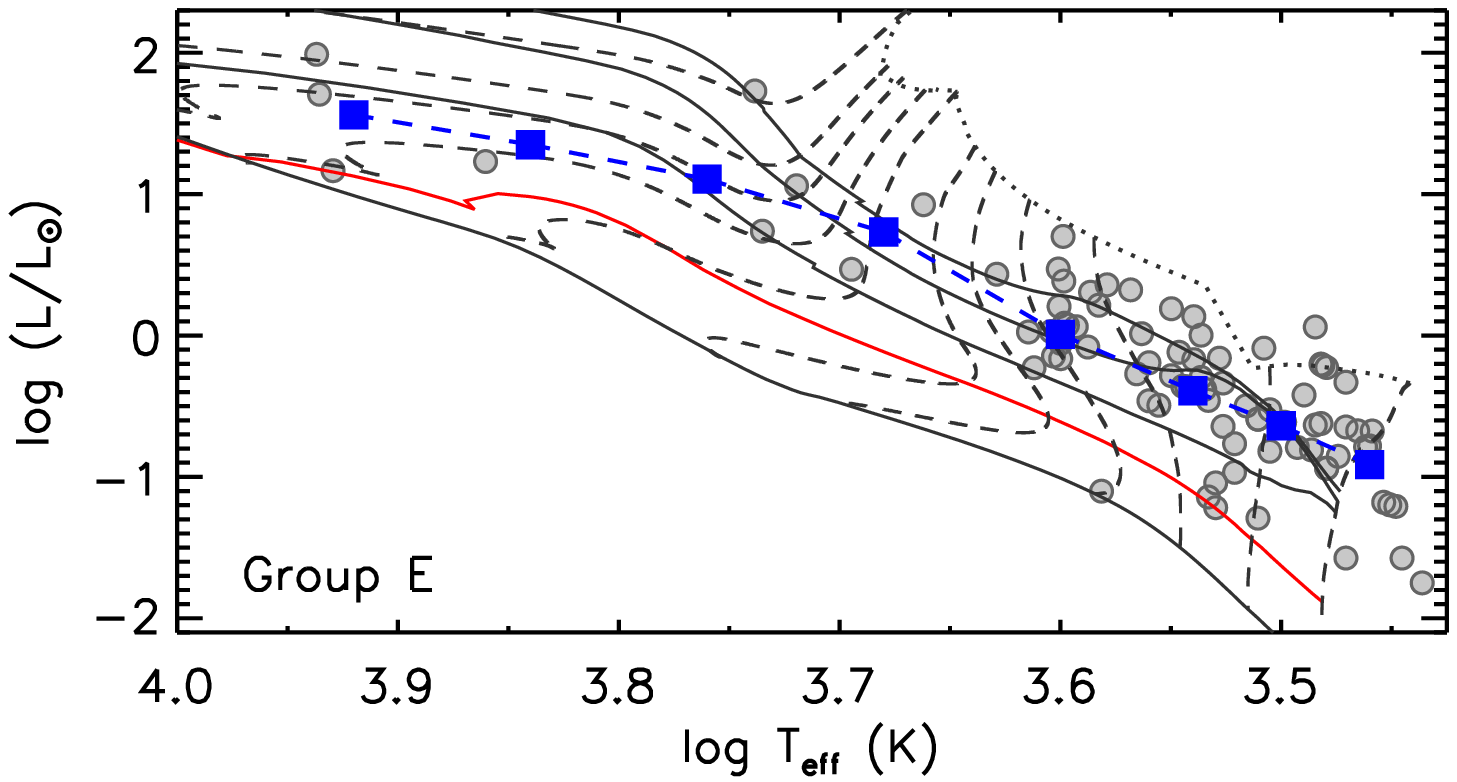}
\includegraphics[angle=0,width=\columnwidth]{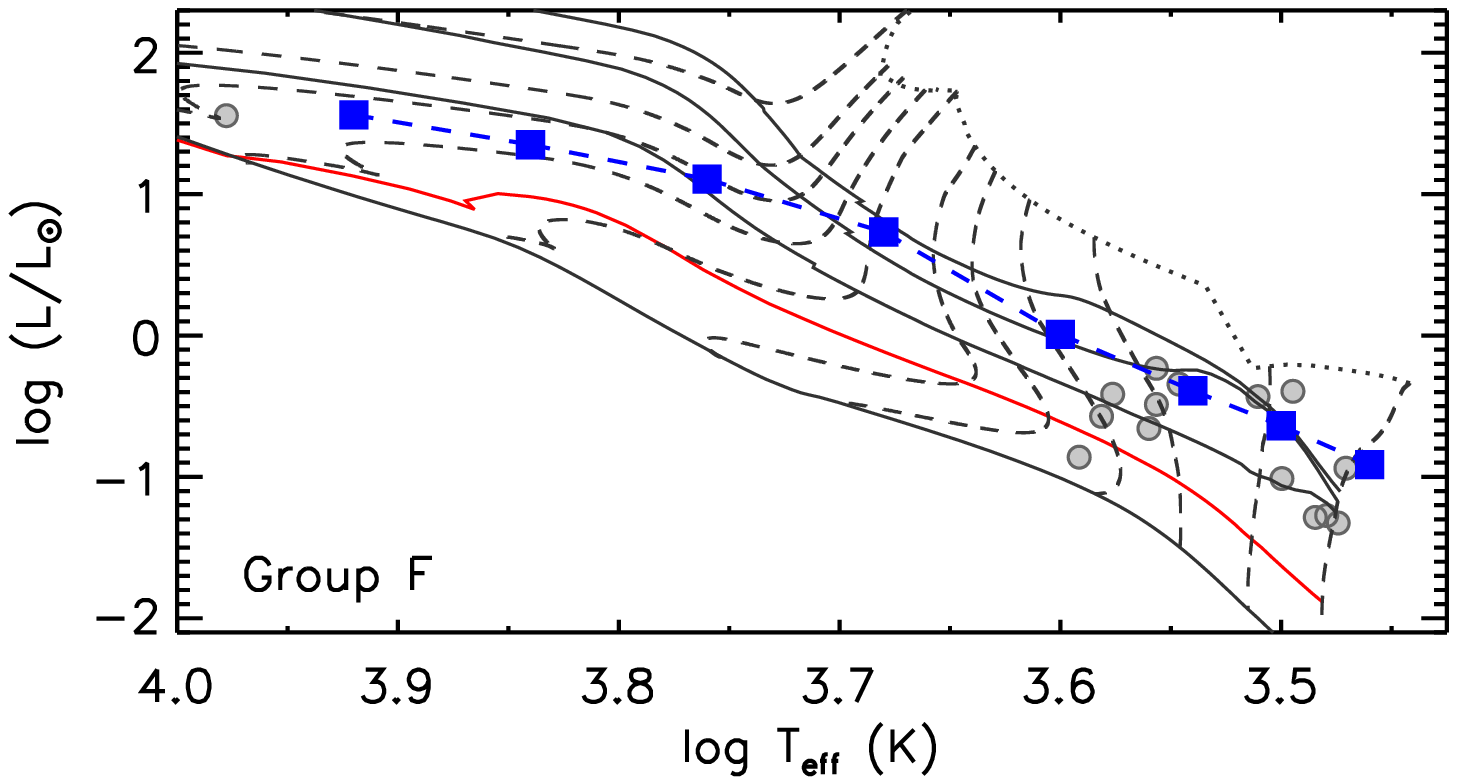}
\caption{H-R Diagram for YSOs in six kinematic groups identified by \cite{2020ApJ...899..128K}. In each panel, the blue filled boxed linked by dashed lines are the median locations of members of the NAP in H-R diagram. The evolutionary tracks are the same as in Figure~\ref{Fig:HRD_YSO}.}
~\label{Fig:HRD_group}
\end{center}
\end{figure*}

\subsection{Optical Photometric Variability}\label{Sect:var}

While we are not using variability as a selection criterion for cluster membership, discussed above in \S5.3, we do assess the implications of photometric variability on the HR diagram below, in \S6.3. 
We identify variable stars among our spectroscopic sample in the following way. 

First, we extract  the lightcurves for all stars with Gaia distances $<2$~kpc in the NAP field from the ZTF survey. For each star, we consider its photometry only between the 95th and 5th percentile points in the $r$-band light curve, in order to exclude the possibility of contamination from bad pixels, cosmic rays, spurious measurements, etc. For those stars with least 30  measurements, we then  calculate the standard deviations ($\sigma_{\rm r}$) and mean ($\overline{r}$) of the selected photometric measurements. 

In Figure~\ref{Fig:Var}, we show $\sigma_{\rm r}$ as a function of $\overline{r}$ for those stars. Then, we divide the
$\overline{r}$  into different magnitude bins with a bin size of 0.2~mag, and calculate the  3$\sigma$-clipping mean ($\overline{\sigma}_{\rm r}$) of the $\sigma_{\rm r}$ for all the stars in each magnitude bin. In Figure~\ref{Fig:Var}(a), we also show the relation between $\overline{\sigma}_{\rm r}$ and $\overline{r}$. It is expected that $\overline{\sigma}_{\rm r}$  
increases towards fainter $\overline{r}$, which is indeed seen at 
$\overline{r}\gtrsim$13.2~mag. However, the relation inverts when $\overline{r}\lesssim$13.2, which may be due to the saturation (the 
saturation limits lie within $\sim$12.5 to 13.2 mag in $r$ band). Thus, in this work, we only investigate the variability of stars with $\overline{r}$ fainter than 13.2~mag. 
Next, we interpolate the $\overline{\sigma}_{\rm r}$--$\overline{r}$ relation for different $\overline{r}$, and define those with $\sigma_{\rm r}>3\overline{\sigma}_{\rm r}$ as strongly variable stars, the ones with $3\overline{\sigma}_{\rm r}\geq\sigma_{\rm r}>2\overline{\sigma}_{\rm r}$ as moderately variable stars, and those with  $\sigma_{\rm r}\leq2\overline{\sigma}_{\rm r}$ as weakly variable stars. 

Among the 503 members of the NAP, as defined in \S\ref{sect:member}, having an estimate of their variability from ZTF, 53.5\% of them show high variability ($>3\sigma$) and an additional 15.7\% are moderate variables ($2-3\sigma$). The fraction of highly variable members is 60.4\%  for those YSOs showing infrared excess emisson, and only 28.8\% for those without infrared excess emission. For the stars in the general field of the NAP as shown in Figure~\ref{Fig:Var}, but not necessarily members of the young cluster, only 1.0\% are highly variable, and 1.3\% are moderately variable. The classification of variability for individual members is listed in Table~\ref{tabe_YSO}

\section{Results on The Stellar Population of the North America / Pelican Nebula}

\begin{figure}
\begin{center}
\includegraphics[angle=0,width=\columnwidth]{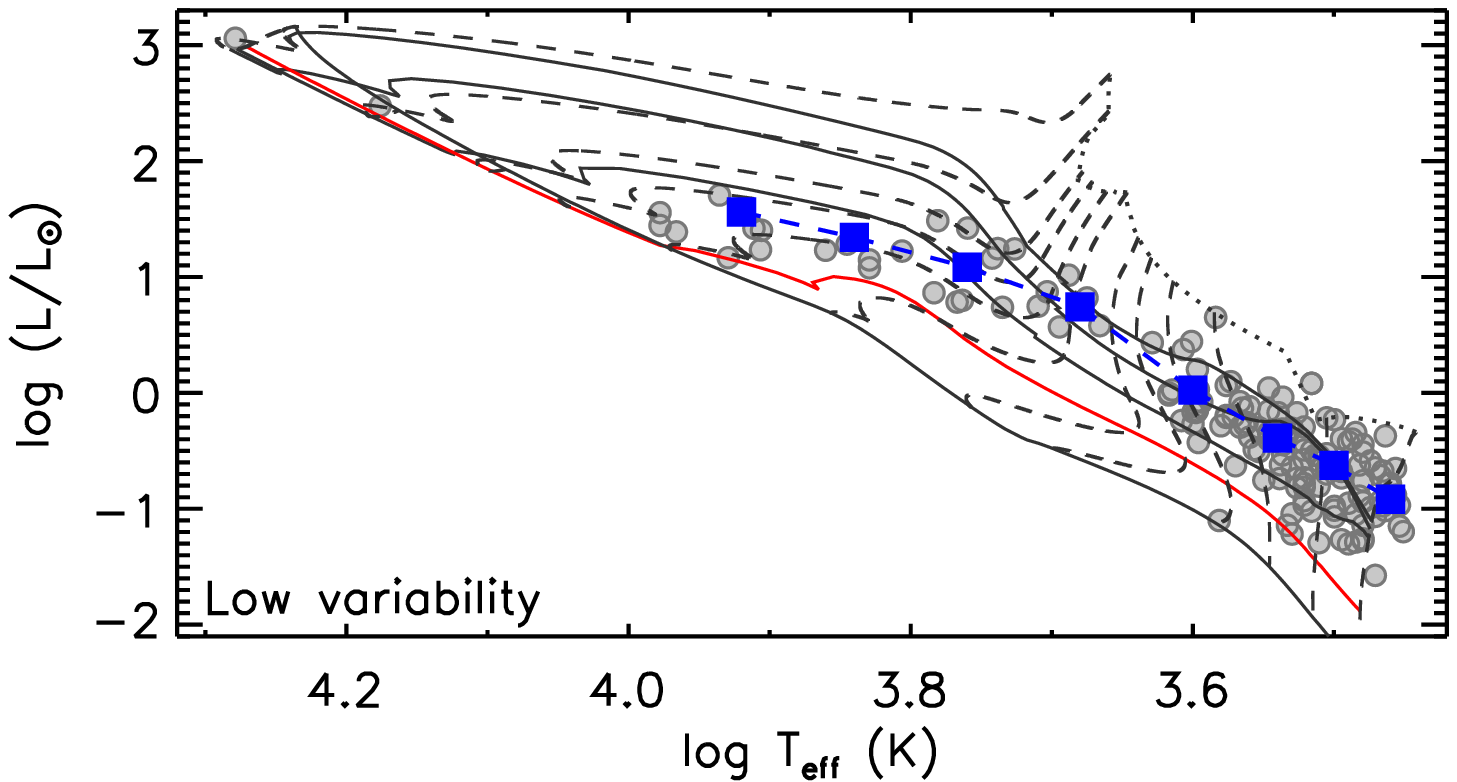}
\includegraphics[angle=0,width=\columnwidth]{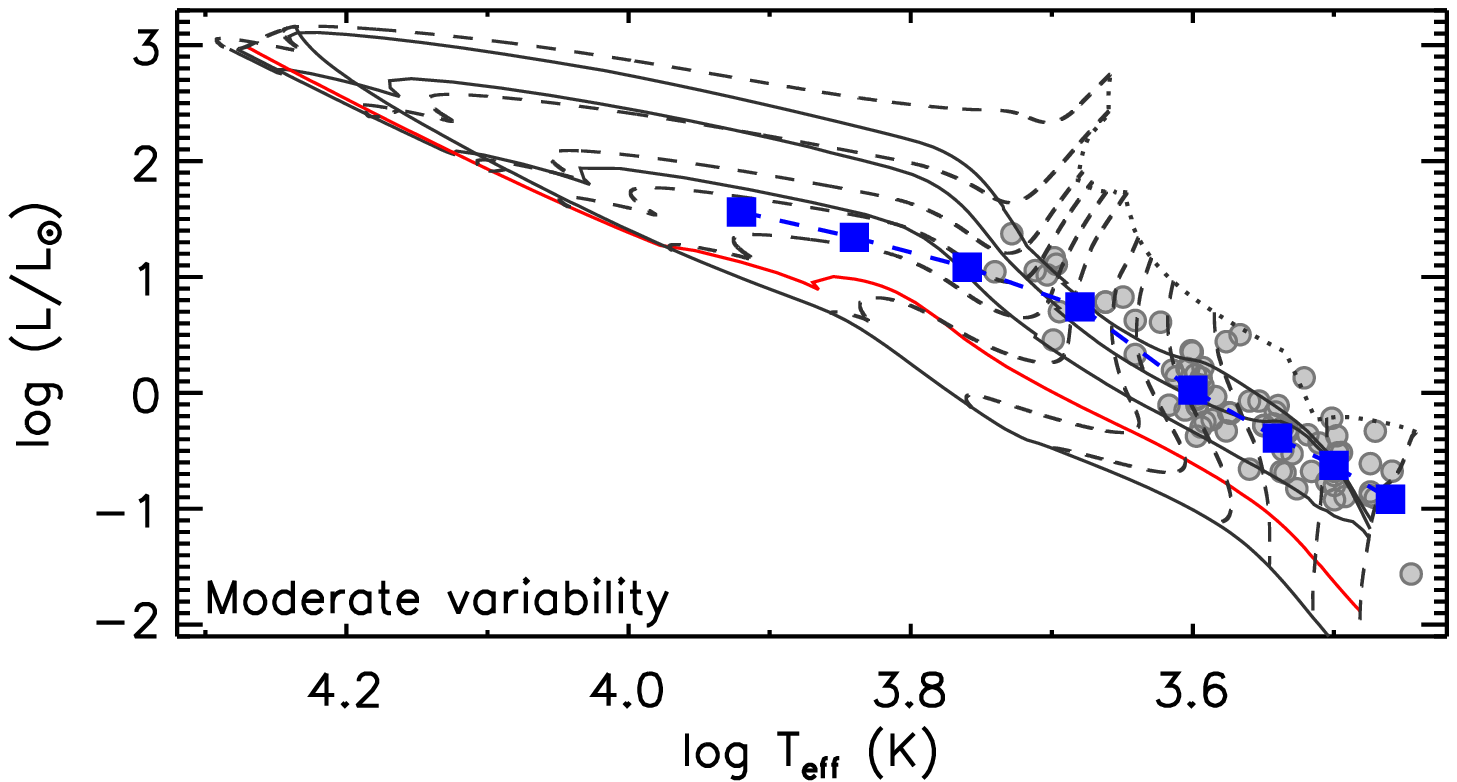}
\includegraphics[angle=0,width=\columnwidth]{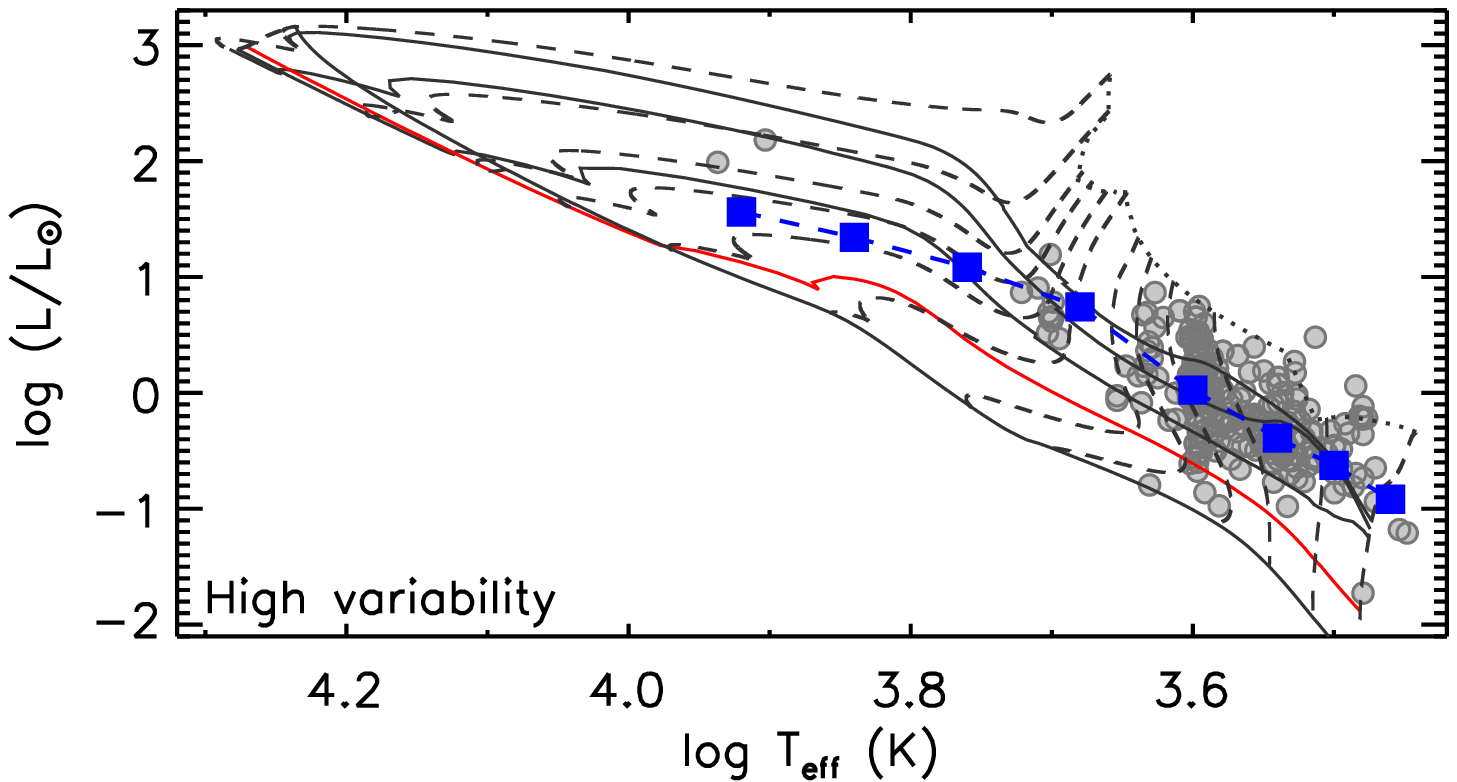}
\caption{H-R diagram for young stars with photometric variability data and showing variability at low (top panel), moderate (middle panel), or high (bottom panel) levels. The evolutionary tracks are the same as in Figure~\ref{Fig:HRD_YSO}.
\label{Fig:HRD_var}}
\end{center}
\end{figure}

\begin{figure}
\begin{center}
\includegraphics[angle=0,width=\columnwidth]{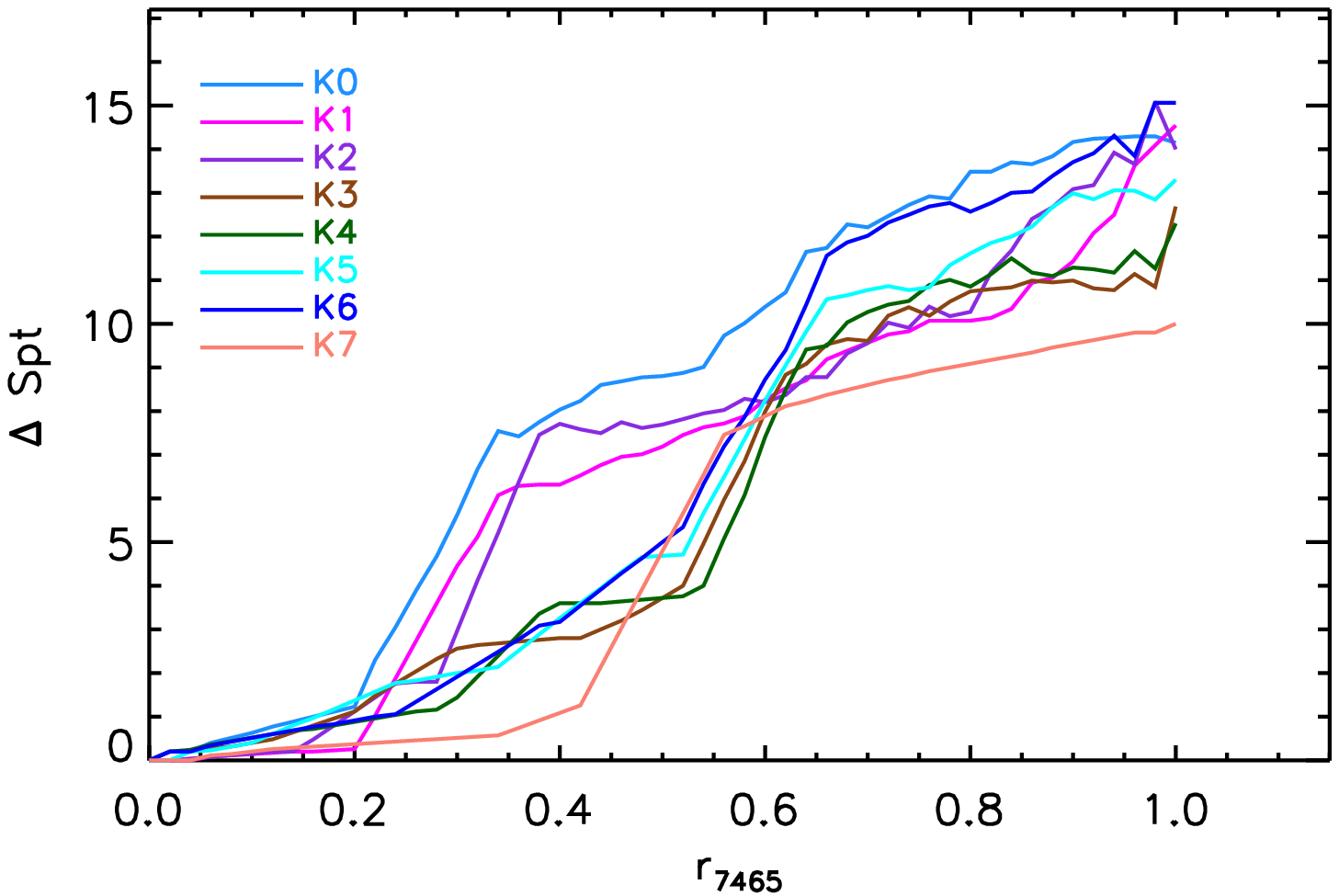}
\includegraphics[angle=0,width=\columnwidth]{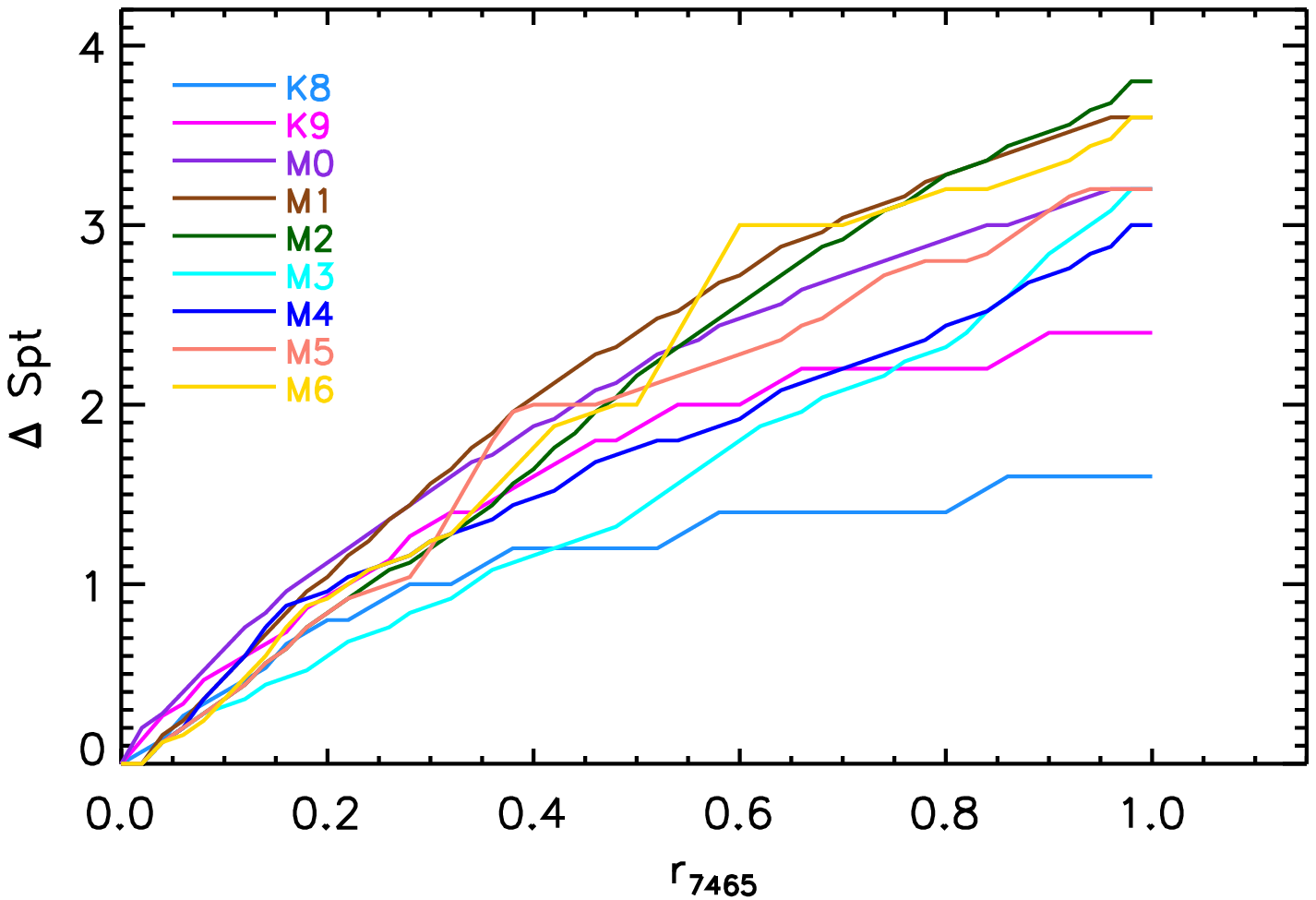}
\caption{The difference between the input spectral types and the spectral types derived from the veiled spectra without the consideration of veiling as a function of veiling strength $r_{7465}$. Each colored line is for a different input spectral type.
\label{Fig:deltaSpt}}
\end{center}
\end{figure}

\begin{figure}
\begin{center}
\includegraphics[angle=0,width=\columnwidth]{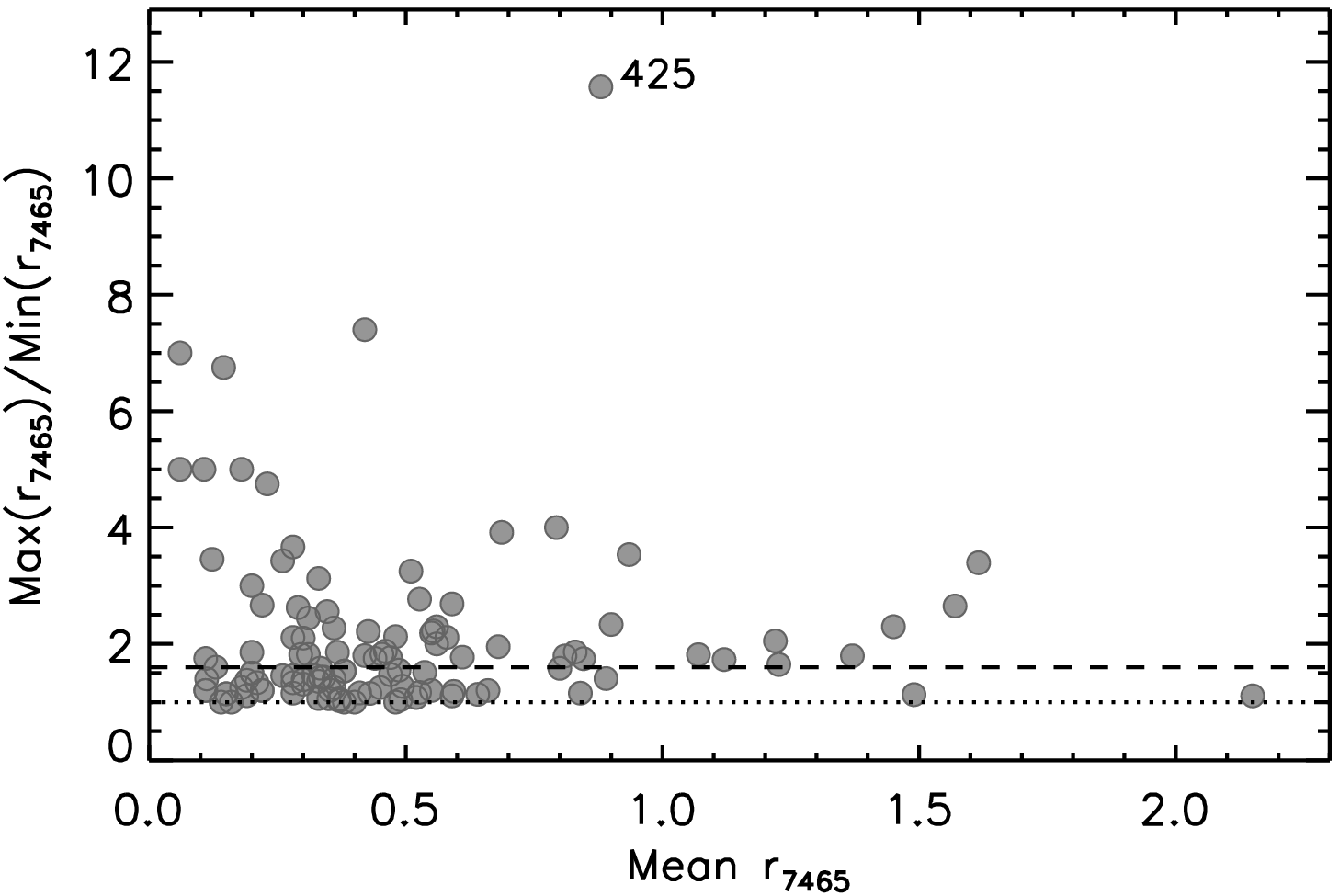}
\caption{Ratio of the maximum and the minimum of $r_{7465}$ vs. the mean $r_{7465}$ for accreting young stars in our sample with multiple-epoch spectroscopic observations. The dotted line shows the equal ratio, and the dashed line shows the median of the distribution, which is  1.6. An outlier in the distribution, with both large mean veiling and large variability in the veiling is Source 425 (
2MASS~J20573729+4406439); see Figure~\ref{Fig:Example-veiling_var} for detail. \label{Fig:acc_var}}
\end{center}
\end{figure}

\subsection{Properties of the Final Catalog of Stellar Parameters}

Among our large spectroscopic sample, we have identified 580 members of the NAP, and we obtain the spectral types of {\newrev 574} of these sources.
Figure~\ref{Fig:Spt_veiling} shows the distribution of spectral types, veiling values, and extinction values derived for these sources using the methods described in \S\ref{sect:spectral_classification} and \S\ref{Sect:Star_properties}.  

In the top panel of Figure~\ref{Fig:Spt_veiling}, the spectral types of the members are compared to a calculated distribution of spectral types for model clusters having ages of 1 and 10\,Myr and mass function as derived in the Trapezium cluster \citep{2002ApJ...573..366M}. Assuming that the NAP and Trapezium clusters have a similar mass function, we can {\newrev search for biases in our spectroscopic survey}.  As shown in Figure~\ref{Fig:Spt_veiling}, our sample includes more sources with spectral type between F0 and A0 than {\newrev in a 1\,Myr model cluster}, suggesting that our sample is substantially incomplete for the middle-M and later spectral types. 
A 10\,Myr model cluster  predicts a distribution of stars with spectral type between F0 and A0 that is more similar to the observed NAP, but also having many more stars that are earlier than A0, than observed in the NAP.

The middle panel of Figure~\ref{Fig:Spt_veiling} shows the distribution of $r_{7465}$ derived from our spectral classification. Among the 484 disk-bearing YSOs, 46.5\% (225/484) of them show measurable veiling on top of their absorption spectra. The median $r_{7465}$ is around 0.4 with a tail to higher values.  The bottom panel of Figure~\ref{Fig:Spt_veiling} shows the distribution of  visual extinction.  The peak value is $A_{\rm V}=2.6$  mag with a tail extending to about 10 mag.

The top panel of Figure~\ref{Fig:veiling_Av} shows that there is no correlation between  the derived veiling values, $r_{7465}$, and the spectral types, SpT. The bottom panel of  Figure~\ref{Fig:veiling_Av} shows $A_{\rm V}$ vs. spectral types. While there is, reassuringly, no correlation between $A_{\rm V}$ and spectral types in our sample, there are 12 M-type sources with $A_{\rm V}\lesssim0.5$~mag,  which is much smaller than those of other members. 
The extinction of these 12 sources could be underestimated due to the discrepancy between their SEDs and the templates used for SED fitting at these late spectral types.

For placing stars in the H-R diagram, we have adopted $A_{V}$ and $L_{\star}$ from SED fitting using model atmospheres for YSOs with spectral types earlier than K0, and using X-shooter templates for those YSOs with later spectral types.  In Table~\ref{tabe_YSO}, we list the parameters as derived from each method, and the value we adopt.

Finally, in Figure~\ref{Fig:HRD_final}, we show the H-R diagram for the entire set of NAP members identified among our spectroscopic sample, compared with the nonmagnetic evolutionary tracks from \cite{2016A&A...593A..99F}.
Their B2 to M7 distribution of spectral types corresponds to stellar masses ranging from  $>5.8~M_\odot$ to $<0.09 M_\odot.$  The luminosity range of the main locus spans ages from about 0.5-5 Myr.  

We divide the NAP members into different log~$T_{\rm eff}$ bins, and estimate the median luminosity and the luminosity dispersion. The result is listed in Table~\ref{tab:medHRD} and also shown in Figure~\ref{Fig:HRD_final}. The comparisons with model evolutionary tracks suggests a median age of $\sim$1\,Myr for the NAP members.
 Following \citet{2008ASPC..384..200H},  we use the median luminosity as a function of temperature, as plotted in the Figure, to characterize the overall stellar population.
 Below, we compare this median H-R diagram for the ensemble population to different sub-populations of stars, including kinematic sub-clusters of the NAP complex, and certain categories of stars such as variable objects.

\subsection{The H-R Diagram for Different Kinematic Groups}\label{Sect:Group_HRD_Age}

\cite{2020ApJ...899..128K} investigate the kinematics of young stars in the NAP region using the data from \textit{Gaia} DR2. They confirm 395 previously identified young stars with reliable astrometric  measurements as cluster members, and divide them into 6 groups which we show in Figure~\ref{Fig:YSO_group}. Among the 395 sources, we have obtained spectral types for 302 sources, and included 299 as members of the NAP based on the analysis described in \S5.3. For the remaining 3 sources, they {\newrev do not show infrared excess emission and} have  isochrone ages older than 10~Myr and are thus excluded as members of the NAP. The majority of the 299 YSOs are in Group C, D, E, and F. 

Besides the YSOs which are already grouped, we have 316 members in our sample without assigned groups from \cite{2020ApJ...899..128K}. We tentatively assign these stars to groups based on their spatial locations. In Figure~\ref{Fig:YSO_group} we depict the boundaries of Group C, D, E, and F using four ellipses. Group E and F are spatially separated, and the ungrouped YSOs near them can be easily divided into individual groups based on their locations. Groups C and D overlap and it is hard to assign a group without reliable kinematic data. To avoid potential contamination, we only use the kinematic members identified by \cite{2020ApJ...899..128K} in the case of Group C, and  include only the ungrouped YSOs within the boundary of Group D but outside of the boundary of Group C as members of Group D. In this way, we have 43, 383, 89 and 15 YSOs in Group~C, D, E, and F, respectively. 

In Figure~\ref{Fig:HRD_group} we show the H-R diagram for the four kinematic groups. We judge relative ages of individual groups by comparing their H-R diagrams with  the median locations of all the members of the NAP (Figure~\ref{Fig:HRD_final}{\newrev )}. In the H-R diagram, 62\% of the members in Group~C, 46\% in Group~D, 54\% in Group~E, and 27\% in Group~F are above the median locations of members of the NAP, indicating that Groups C and E are younger than Group~D, while Group~F is the oldest. Groups~A and B, have only 4 and 8 stars with spectroscopic data, respectively, with only 2 stars in Group~A and 3 stars in Group~B that are above the median location of members of the NAP. Given such small numbers of members in both groups, it is hard to judge the relative ages among the groups; all we can say is that they have ages comparable to those of other groups in the NAP region.

\subsection{The H-R Diagram and Stellar Variability}

In Section~\ref{Sect:var}, we have investigated the variability of the members of the NAP, and divided them into three categories: high, moderate, and low r-band photometric variability. In Figure~\ref{Fig:HRD_var}, we show the H-R diagrams for the YSOs with different amplitudes of variability.  A similar luminosity spread can be noted as present in YSOs with low to high variability. 

To quantify this impression, we first limit the $T_{\rm eff}$ between 3000 and 4000~K to calculate the luminosity spread. 
We then fit the log$L_{\star}$--log$T_{\rm eff}$ relation with a linear function, and then subtract the calculated log$L_{\star}$ at log$T_{\rm eff}$  using the fitted linear function. In this way, we  can remove the contribution in the luminosity spreads from the general slope of the log$L_{\star}$--log$T_{\rm eff}$ relation. The luminosity spreads are then calculated using the resulting residual luminosity as 0.32, 0.24, and 0.31~dex for YSOs with low, moderate, and high variability. The luminosity spreads  are similar for YSOs at low and high variability.  The lower luminosity spread for the moderately variable YSOs may be attributed to a relatively smaller number of sources in this category.

\section{Discussion}

\subsection{The Effect of Spectral Veiling on Stellar Classification}
Figure~\ref{Fig:Twospt} shows that a veiling continuum superposed on absorption spectra of young stars can affect the spectral classification process. Among the 224 sources in our sample with measurable veiling in their optical spectra, the spectral types that result from fitting the spectra with a veiling component included are systematically later than those derived without accounting for veiling.  The size of the effect is typically $\sim$1.2 subclasses later for spectral types later than K7, and $\sim$4.2 subclasses later for earlier types.  For any individual source there will be a dependency on the veiling flux itself; the numbers quoted above can be considered appropriate for a typical veiling value $r_{7465} \approx 0.4$ as illustrated in the distribution of Figure~\ref{Fig:Spt_veiling}.

How does the presence of veiling shift the location of young stars in the H-R diagram?  To investigate this quantitatively, we re-fit the SEDs of the member stars in our sample to derive their stellar luminosity adopting the spectral type from fitting the spectra without consideration of veiling.  

We found that there are {\newrev no} significant shifts in the stellar luminosity in fitting the SEDs with veiling or without veiling.  This can be understood as due to the fact that, although there is an obvious effect on the temperatures, there is also a trade-off between temperature shift and bolometric correction shift, extinction correction shift, and veiling correction shift, such that the integrated luminosity remains relatively constant. In addition, the veiling effect is predominantly considered in the optical part of the flux distribution (or Wien side, if considering a Planck function representation of the stellar atmosphere) which is still far from the peak flux in the red or near-infrared wavelength range. 
Regarding the temperatures, there is a much larger effect however. Due to the $\sim$1--4 subclass difference for the spectral types, the locations of the veiled young stars will systematically shift to hotter temperatures, typically by $\sim$130~K for stars with spectral type later than K7, and typically by $\sim$680~K for earlier type stars. {\newrev When $L_{\star}$ stays the same, as $T_{\rm eff}$ gets hotter the stellar radius also becomes smaller, so the stellar age is older when excluding veiling in the spectral classification.} Thus, for the earlier spectral types especially, it is very important to include the potential effects of veiling when performing spectral classification.

We further test how wrong the spectral type could be with increasing veiling by taking the X-shooter templates and veiling them {\newrev artificially,} adopting a constant shape for veiling continuum flux. The veiling effect is parameterized as $r_{7465}$, as used elsewhere in this work. We then follow the spectral classification procedure for these veiled spectra without consideration of veiling. As expected, the resulting spectral types tend to be earlier than the input ones. 
Figure~\ref{Fig:deltaSpt} shows the difference between the input and resulting spectral types as a function of $r_{7465}$. When excluding the veiling in spectral classification,  {\newrev for $r_{7465}$ $\gtrsim$0.2}  the errors on the spectral types are generally larger for mid-K and early-K type stars than for the M type stars.

   \begin{figure*}
\begin{center}
\includegraphics[width=2\columnwidth]{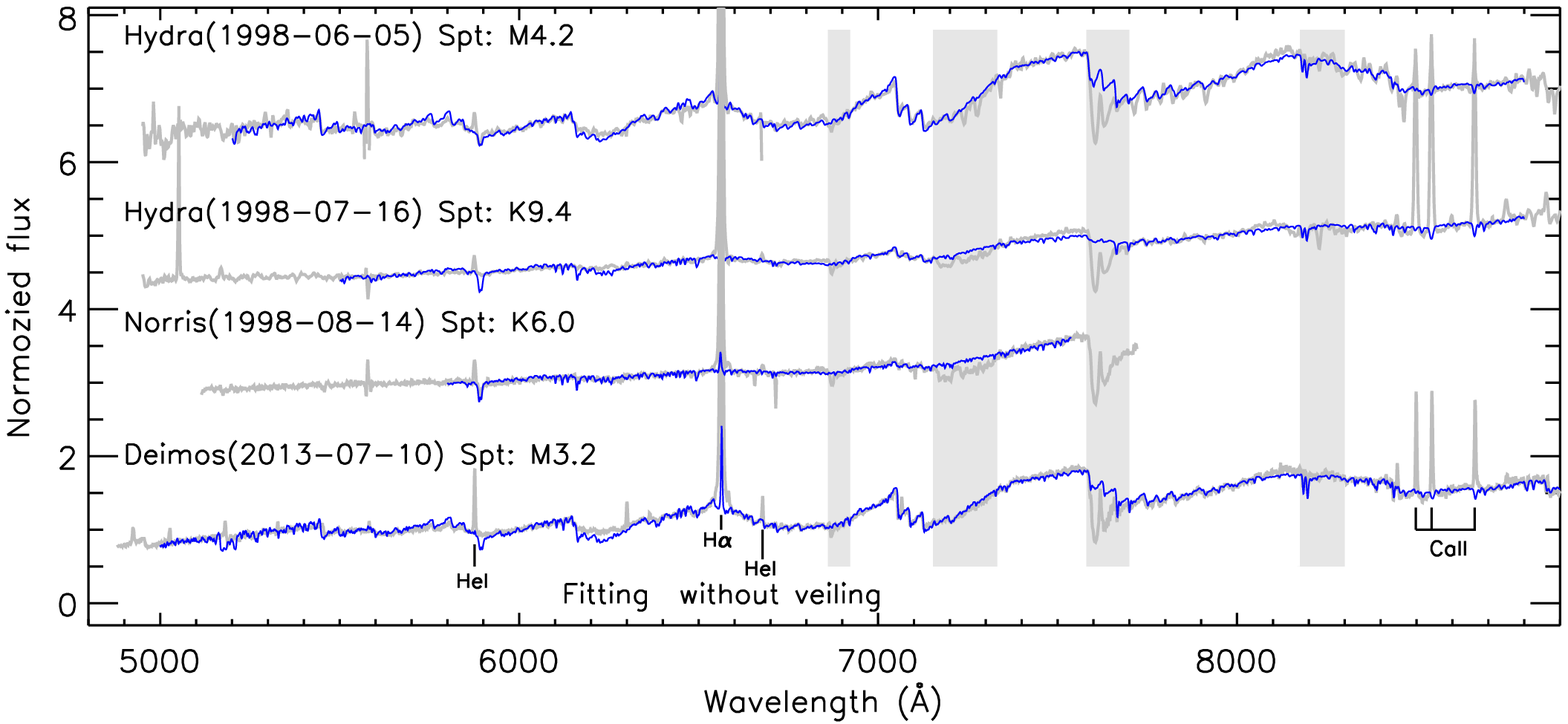}
\includegraphics[width=2\columnwidth]{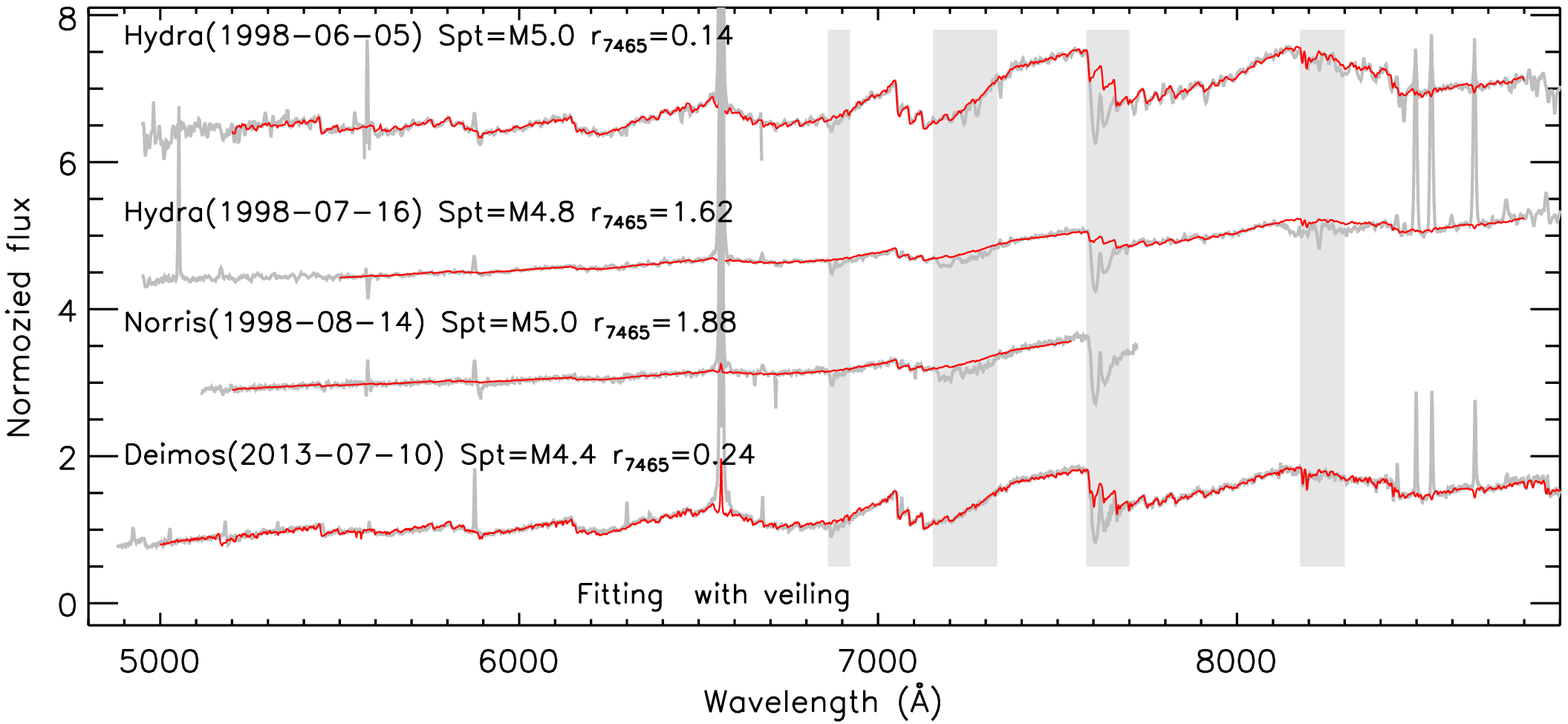}
\caption{Illustration of the spectral template fitting for Source 425 (2MASS~J20573729+4406439)  which was observed with Hydra, Norris and DEIMOS, and shows significant variations in the presented spectrum over time. 
The best-fit X-shooter empirical spectral templates (colored lines) are over-plotted on the observed spectra (gray). Vertically oriented gray bars indicate regions masked due to potential contamination from telluric features.   Fits are without veiling (top panel) and with veiling (bottom panel). 
While the non-veiled templates produce a range of spectral type solutions, the veiled templates (bottom panel) produce a consistent spectral type with a large variation on the veiling parameter.}\label{Fig:Example-veiling_var}
\end{center}
\end{figure*}

\subsection{Veiling variations}

Young stars usually show variable accretion activity, which is a likely cause of the variable veiling on the spectra. As demonstrated in Figure~\ref{Fig:deltaSpt}, if we do not consider veiling in the spectral classification process, we would obtain potentially very different and erroneous spectral types for the stars. This effect may also be the cause for the discrepancy among {\newrev reported} spectral types in the literature for some young stars with substantial variable accretion activity. 

In this work, we have identified 134 young stars which show measurable veiling in their optical spectra, and have been observed spectroscopically at multiple epochs. In Figure~\ref{Fig:acc_var}, we show {\newrev the} ratio between the maximum and minimum of $r_{7465}$ vs. mean $r_{7465}$ for these sources. The median  of the ratios between the maximum and minimum of $r_{7465}$ is about 1.6. Among these sources, there is one (ID~425 or 2MASS~J20573729+4406439) which shows a large mean $r_{7465}$ (0.88) and a high ratio (11.57) between  the maximum and minimum of $r_{7465}$. In Figure~\ref{Fig:Example-veiling_var}, we show the 4 spectra of the object observed at different dates. 
The variation in the strength of the TiO absorption bands is prominent among these spectra. The spectra also show strong accretion-related emission lines, e.g. He~I~5876~$\AA$, 6678~$\AA$, H$\alpha$, and the Ca~II ``infrared" triplet, all indicating strong accretion activity. The variations in the strength of the TiO bands, which are anti-correlated with changes in the emission line strength, are likely due to the variable excess emission induced by the accretion variability. When the veiling is excluded in doing the spectral classification, the spectral type of the object can vary from K6 to M4.2 (see the top panel of Figure~\ref{Fig:Example-veiling_var}). The spectral types from different spectra 
converge when we include the veiling in the spectral fitting (see the bottom panel of Figure~\ref{Fig:Example-veiling_var}). The source may represent an extreme case in our sample, but it is a perfect example for showing the importance of taking veiling into consideration when doing the spectral classification.

\begin{figure}
\begin{center}
\includegraphics[angle=0,width=0.98\columnwidth]{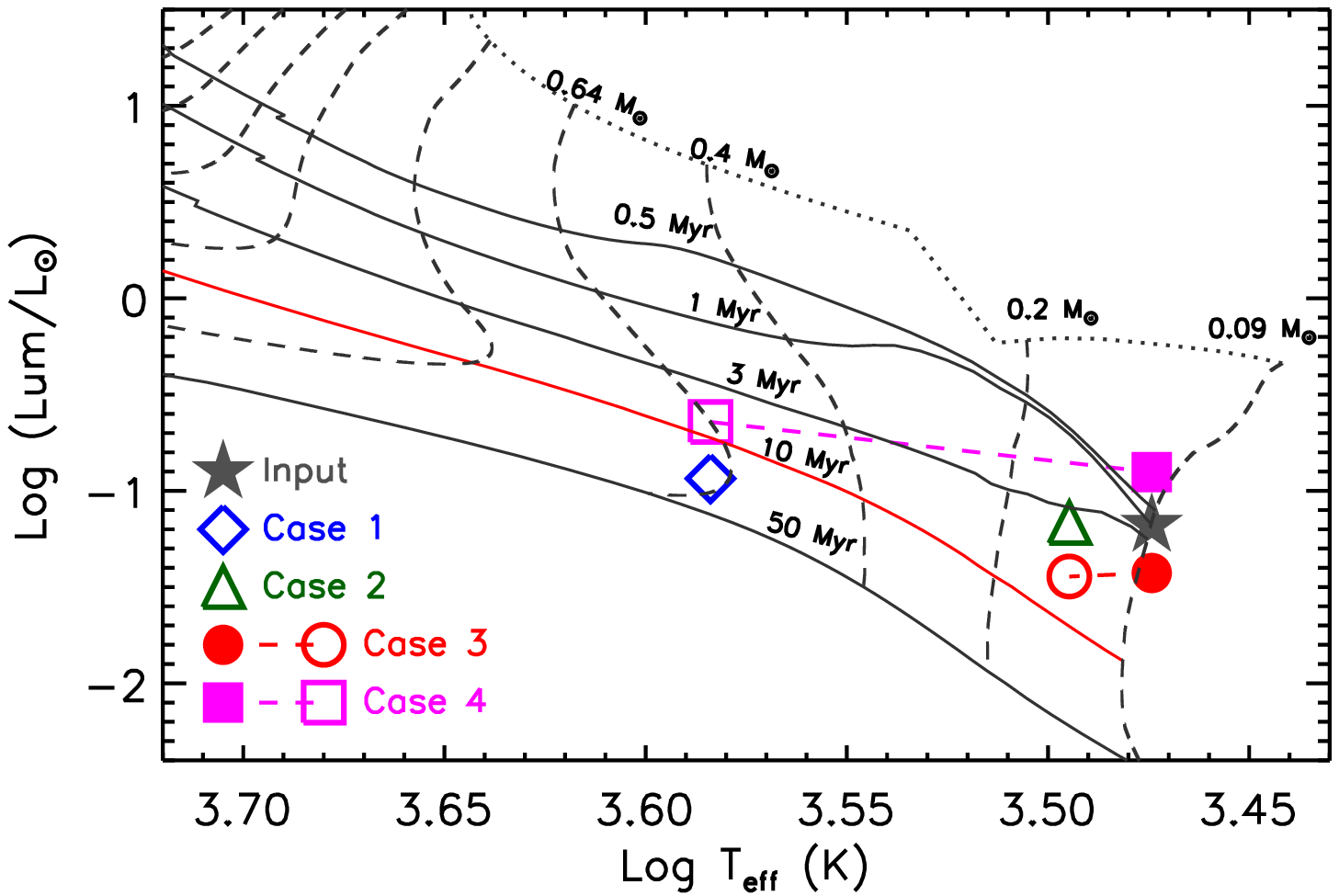}
\caption{H-R diagram for a model star with $T_{\rm eff}$ and $L_{\star}$ derived from simulated observations in 4 hypothetical cases. The gray filled star symbol marks the location of the model star with the input $T_{\rm eff}$ and $L_{\star}$. We performed four simulations, and assume that the photometry and spectra are taken simultaneously for two cases (Cases 1 and 2), and not for the other two (Cases 3 and 4). In Cases 1 and 2, the data are assumed to 
be taken with $r_{7465}=1.6$ and $r_{7465}=0.1$, respectively, and the spectral classification in the two cases is done without consideration of veiling. In Case 3, the spectra are assumed to be taken at $r_{7465}=0.1$, and the photometry at $r_{7465}=1.6$; the converse is true for Case 4. For Cases 3 and 4, the open (circle or square) symbols are for spectral classification without considering veiling, and the filled (circle or square) symbols including veiling.}\label{Fig:modelHRD} 
\end{center}
\end{figure}

For Source ID~425, assuming that the shape of veiling continuum flux is  constant, the change of $r_{7465}$ from 0.14 to 1.88 would lead to $\sim$0.8~mag brightness change in the $r$ band, which is comparable to the maximum variation ($\Delta r$=0.9\,mag)  observed in $r$ band from the ZTF survey. The ZTF survey did not cover the observation dates for our spectroscopic observations, however, which were earlier.  However, the similar variations in recent photometry and older spectroscopy indicate that the accretion variations have continued for years and possibly decades. For a source with such strong variations in the photometry, it is necessary for accurate H-R diagram placement that the photometry is taken at the same time as the spectra in order to better characterize the stellar properties. Unfortunately, this is not the case for Source ID~425.  

To quantify the impact of different, typically unknown, accretion scenarios for H-R diagram placemment,  we can simulate the observations for a case like Source ID~425, and inspect how the input parameters and the derived parameters can differ. We consider an M5 type YSO with an age of 1~Myr and $A_{\rm V}$=3~mag.
This YSO  is modelled to exhibit variable accretion activity parameterized by $r_{7465}$ that is varied from  $r_{7465}$=1.6 to $r_{7465}$=0.1. 

We model four cases.  In the first two cases, we assume that both the broad-band photometry and spectra are observed simultaneously  when $r_{7465}$=1.6 (Case~1) or $r_{7465}$=0.1 (Case~1), we derive the stellar parameters without a consideration of veiling. In Case 3, we assume the broad-band photometry is taken when $r_{7465}$=1.6,  and the spectrum is observed when $r_{7465}$=0.1. For Case 3, we derive the stellar parameters when neglecting or
considering veiling in spectral classification. Case 4 is similar to Case 3, but assumes the photometry is taken when $r_{7465}$=0.1 and the spectrum is obtained when $r_{7465}$=1.6. For individual cases, the simulated results are shown in Figure~\ref{Fig:modelHRD}. From the input $T_{\rm eff}$ and $L_{\star}$, the model star is expected to be 1~Myr old with a mass of 0.09~$M_{\odot}$. When excluding the veiling effect in spectral classification, the derived $T_{\rm eff}$ and $L_{\star}$ vary depending on $r_{7465}$, which induces large uncertainty in estimated stellar masses and ages. 
In an extreme case, the derived mass and age can be 0.64~$M_{\odot}$ and 25~Myr, which is very different from the input 0.09~$M_{\odot}$ and 1~Myr. When including veiling effects in spectral classification, the derived mass is consistent with the input one, but the age measurement can be very uncertain, from 0.3~Myr to 4~Myr. In the simulations, we have {\newrev assumed} that all broad-band photometry is observed simultaneously, but this may not be the case in reality, which could induce additional  uncertainties given that many YSOs are variable at the $>0.05-0.1$ mag level (see Section~\ref{Sect:var}).

\begin{figure*}
\begin{center}
\includegraphics[angle=0,width=2\columnwidth]{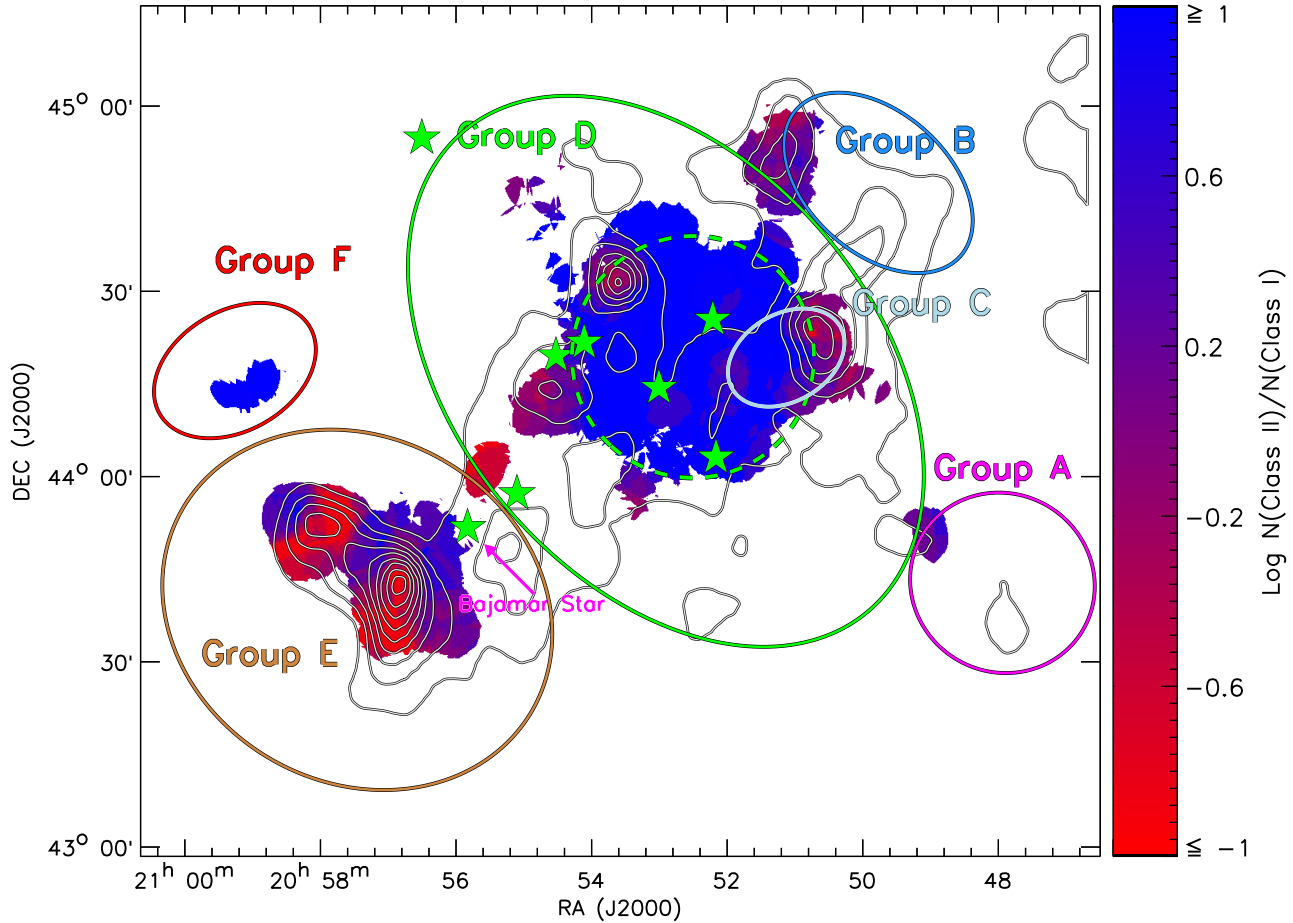}
\caption{ The number ratio between Class~II and Class~I sources in the NAP, shown in the color scale. The color bar is shown. The green filled star symbols show the O or B type stars. The contours show the Planck 857 $\mu m$ dust emission.}
\label{Fig:YSO_ratio}
\end{center}
\end{figure*}

\begin{figure*}
\begin{center}
\includegraphics[angle=0,width=2\columnwidth]{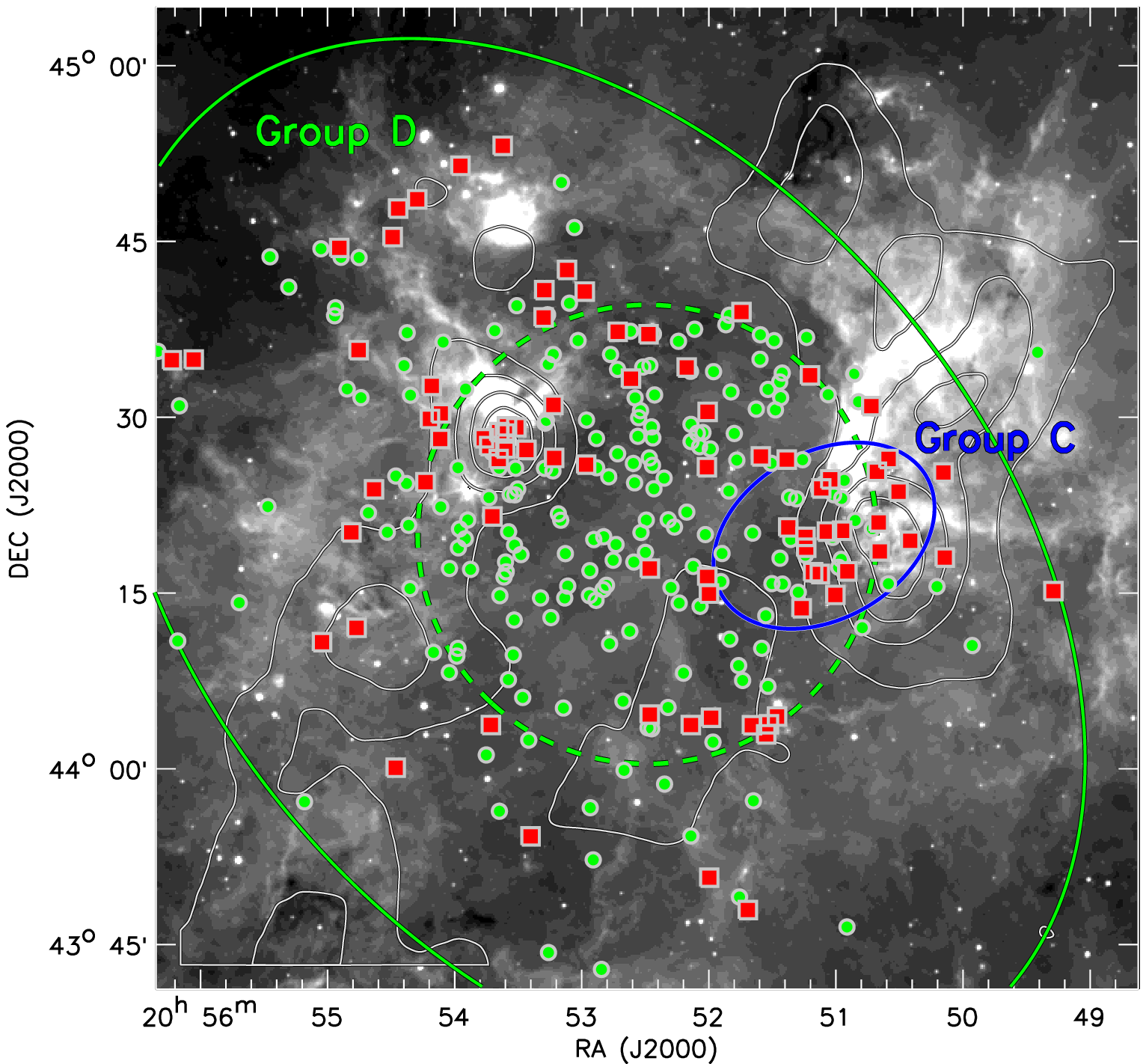}
\caption{Distribution of YSOs centered at Group D overplotted on the image in  WISE-W3 band. The red filled squares for the YSOs with isochrone ages younger than 0.5 Myr and the green filled circles are for those with isochrone ages between 1 and 10 Myr.  The contours show the Planck 857 $\mu m$ dust emission. The green dash circle is the same as in Figure~\ref{Fig:YSO_ratio}.}
\label{Fig:YSO_GroupD}
\end{center}
\end{figure*}

\subsection{Star Formation History in the NAP Complex}

In Section~\ref{Sect:Group_HRD_Age}, we have discussed the probable age difference among the kinematic groups within the NAP region, suggesting that there has been a multiple-epoch star formation process across the complex. Here, we investigate this issue from a different view using the number ratios between Class~II and Class~I type SEDs, designated henceforth as CII/CI. 

We use the YSO catalog from \cite{2011ApJS..193...25R}, and in total find 596 Class~II sources and 270 Class~I sources in the field. We grid the NAP region spatially, and calculate the CII/CI ratio at each grid point using the nearest 10 YSOs, including both Class~II and Class~I sources. In order to avoid noise from grid points where there are few or no identified YSOs, we set a threshold of 0.1~arcmin$^{-2}$ on the YSO surface density centered at each grid point. The CII/CI ratio of the grid points with the minimum YSO surface density are used in Figure~\ref{Fig:YSO_ratio} to illustrate the variation of CII/CI over the NAP field. 

Overall, the CII/CI ratio varies, and we can consider how the ratio changes among the identifiied kinematic groups. Within Group E, Class~I sources are dominant (low CII/CI), while within Group F, the number of Class~II is far more than Class~I sources (high CII/CI). Within Group~D, two distinct regions can be noted. Near the center of this large group, the region 
is dominated by Class~II sources (high CII/CI), and surrounding that area, there are several regions, including Group C, with CII/CI$\approx$1. Within Groups A and B, the surface densities of Class~II and Class~I sources are both below the threshold (0.1~arcmin$^{-2}$), but we note that their CII/CI are high at 7 and 3, respectively. Surrounding each of Group A and B, there are regions with relatively dense concentrations of Class~II and Class~I sources with CII/CI ratios$\approx$1.  This is the same structure as noted above for Group D.

The CII/CI ratio has been explored as a tracer of regional evolution of star formation \citep{2011ApJ...739...84G}. An investigation of the correlation between the gas surface density and the YSO mass surface density with different CII/CI suggests that the regions at early evolutionary stages tend to have low CII/CI and high gas surface density \citep{2020arXiv200505466P}. In the NAP, we also see the same trend that the regions with low CII/CI are associated with dense gas material traced by the Planck 857 $\mu m$ dust emission (see Figure~\ref{Fig:YSO_ratio}). 
The strong variations in CII/CI over the NAP field demonstrated in Figure~\ref{Fig:YSO_ratio} suggest that the star formation is not coeval across the cloud. 

We further investigate the region centered on Group~D using the YSOs that we have confirmed here with spectroscopy. According to their isochrone ages derived from the H-R diagram (see Figure~\ref{Fig:HRD_final}), we can divide these YSOs having $T_{\rm eff}\geq 3000~K$ into two categories: ``young'' YSOs with isochrone ages younger than 0.5~Myr, and ``old'' YSOs with isochrone ages between 1 and 10~Myr.   In Figure~\ref{Fig:YSO_GroupD}, we show the distributions of two types of YSOs. 

The majority of YSOs are distributed within the infrared bubble that is prominent in the WISE W3-band image. Intriguingly, while the ``old'' YSOs share the uniform distribution of the overall population, the ``young'' YSOs tend to be distributed near the boundary of the bubble, especially in the two dense molecular cores on the east and the west (Group C). This suggests that there has been sequential star formation within Group D -- consistent with the result presented above from the study of the CII/CI ratio. 

In combination with other results from the H-R diagram (see Section~\ref{Sect:Group_HRD_Age}), our interpretation is that star formation has progressed from the region of Group F to Group D, then propagated to the regions surrounding Group D, including Group C, with Group E currently the most active site of star formation within the NAP region.  {\newrev Groups A and B do not have enough stars for age assessment, but seem consistent with the median age. 
We note that \cite{bally2014} have discussed the most recent star formation
history in the NAP region, as traced by outflows.}

\section{summary}

We have presented the first large-scale spectroscopic study of members and prospective members of the North America and Pelican Nebula region.  The primary data set consists of $R\approx 1300-2000$ optical spectra of $\approx 3400$ stars, which we combine with multi-wavelength photometric data in order to create spectral energy distributions spanning $\sim 0.4-25 \mu$m.  We also consider the photometric variability of the spectroscopic sample using ZTF data.

We first assess the membership of each source in the spectroscopic sample by considering the evidence for: Li~I absorption in the spectra, X-ray emission from cross-referencing to published data, and infrared excess emission from the spectral slopes we derive based on published photometry or new photometric analysis that we have conducted for stars without previously published measurements.  We further assess the membership of these sub-samples using Gaia astrometry.  Considering only the likely members, we then use the spectra and the broadband spectral energy distributions to derive stellar parameters.  These include spectral type and photospheric temperature, continuum veiling, foreground extinction, and bolometric luminosity.  

Beyond presenting our catalog of $\approx 600$ confirmed members and their stellar parameters, our scientific results include the following:
\begin{itemize}
\item
The H-R diagram for our spectroscopic sample spans masses from a $\sim$6~$M_{\odot}$  B7 star to several  0.09 $M_{\odot}$ M7 stars.  We note that the region contains two more massive O-type stars that were not part of our spectroscopic survey.
\item
The median age of the sample is 1~Myr and the luminosity dispersion is $\sim$0.3--0.4~dex.  The observed photometric variability does not significantly contribute to the luminosity dispersion. 
\item
Incompleteness at the low mass end precludes us from making any conclusions about the mass distribution.
\item
Different kinematic sub-groups appear to have different ages.
The star formation history in the region began with Group F, then continued in Group D,  extended to the regions surrounding Group D, including Group C, with Group~E currently the most active site of star formation. 
\end{itemize}

We also highlight {\newrev and extensively discuss} the systematic effects of veiling on stellar spectral typing, and quantify the errors that are made when veiling is not accounted for.  One specific object with substantial veiling variability measured among our several spectra is used to demonstrate the significant impact of veiling.




\acknowledgments
Many thanks to the anonymous referee for comments that help to improve this paper. We thank Michael A. Kuhn for discussions about the NAP  region. We recognize the contributions of former Caltech SURF students Daniel DeFelippis and Danika Wellington to our efforts. This paper uses data products produced by the OIR Telescope Data
Center, supported by the Smithsonian Astrophysical Observatory. 
This work has made use of data from the European Space Agency (ESA) mission
{\it Gaia} (\url{https://www.cosmos.esa.int/gaia}), processed by the {\it Gaia}
Data Processing and Analysis Consortium (DPAC,
\url{https://www.cosmos.esa.int/web/gaia/dpac/consortium}). Funding for the DPAC
has been provided by national institutions, in particular the institutions
participating in the {\it Gaia} Multilateral Agreement.
We use ZTF data, enabled by support from the NSF under grant AST-1440341, Caltech, IPAC, the Weizmann Institute for Science, the Oskar Klein Center at Stockholm University, the University of Maryland, the University of Washington, Deutsches Elektronen-Synchrotron and Humboldt University, Los Alamos National Laboratories, the TANGO Consortium of Taiwan, the University of Wisconsin at Milwaukee, and Lawrence Berkeley national Laboratories.

\clearpage

\startlongtable
\begin{deluxetable*}{lcccrcccccrccccccccccccccccccccccc}
\renewcommand{\tabcolsep}{0.05cm}
\tablecaption{Log of Spectroscopic Observations\label{tab:obs_log}}
\tablewidth{700pt}
\tabletypesize{\tiny}
\tablehead{
                            &\colhead {Observation}& \colhead{RA}     & \colhead{DEC}    & \colhead{Exposure}   &  \colhead{Telescope and}    &Selection      &\\
\colhead{Identifier} & (UT Date)& \colhead{(J2000)}  & \colhead{(J2000)}  & \colhead{(minutes)}  & \colhead{Instrument} &Criteria\tablenotemark{a} &  \colhead{PI}}
\startdata
  Config. 1  & 2012-05-18   &20:52:50   &+44:19:32   &45    &MMT/Hectospec &IRE  &Fang\\
  Config. 2  & 2012-05-18   &20:52:50   &+44:19:32   &45    &MMT/Hectospec &IRE &Fang\\
  Config. 3  & 2012-06-14   &20:54:31   &+44:34:32   &45    &MMT/Hectospec &IRE &Fang \\
  Config. 4  & 2012-06-15   &20:57:36   & 43:58:05   &45    &MMT/Hectospec &IRE &Fang\\
  Config. 5 /MMT1 & 2012-11-04   &20:49:23   &+44:39:16   &90    &MMT/Hectospec &Var,IRE &Hillenbrand\\
  Config. 6 /MMT2 & 2012-12-04   &20:56:47   &+43:46:02   &90    &MMT/Hectospec &Var,IRE &Hillenbrand\\
  Config. 7 /MMT3 & 2012-12-06   &20:49:48   &+43:30:23   &90    &MMT/Hectospec &Var,IRE &Hillenbrand\\
  Config. 8  & 2017-06-02   &20:50:51   &+44:08:19   &45   & MMT/Hectospec &IRE,Var,Xray &Fang\\
\enddata
\tablenotetext{a}{IRE: showing infrared excess, Var: showing photometric variability, Xray: having an X-ray counterpart, H$\alpha$-R: showing H$\alpha$ emission indicated by the H$\alpha$-R color, red BVI: \rev{having late spectral type indicated by red $B-V$ and $V-I$ color.}}
\end{deluxetable*}
\normalsize

\begin{table*}
\caption{Stars Used in Deriving the Instrumental Response Functions\label{tab:table_RF}}
 \centering
\begin{tabular}{lcccccccccc}
\hline\hline
\multicolumn{1}{c}{Name} &\multicolumn{1}{c}{RA} &\multicolumn{1}{c}{DEC}&\multicolumn{1}{c}{Adopted $T_{\rm eff}$} &\multicolumn{1}{c}{Wavelength ranges} &\multicolumn{1}{c}{Instruments} \\ 
\multicolumn{1}{c}{} &\multicolumn{1}{c}{(J2000)} &\multicolumn{1}{c}{(J2000)} &\multicolumn{1}{c}{(K)} &\multicolumn{1}{c}{(\AA)}&\multicolumn{1}{c}{}\\
\hline
BD-06 1263  & 05 37 41.45  & $-$06 33 57.8   & 7,000&3700--9150 & Hectospec \\
BSD 40-279  & 20 52 07.65  & +44 03 44.4   & 18,500  & 4880--10240 & Hydra \\
Cl* IC 4665 P 150  & 17 47 33.32 & +05 31 46.0  & 5,800 &       4210--9380 & Deimos \\
TYC 3179-1399-1  & 20 57 36.23 & +43 57 27.9   & 6,500 &5120--7740 & Norris\\
 BSD 40-1004 &  20 56 10.96  & +44 39 13.0  & 10,500  & 6120--8760 & Norris\\
 \hline
\end{tabular}
\end{table*}

\scriptsize
\begin{longrotatetable}
\renewcommand{\tabcolsep}{0.03cm}
\begin{deluxetable}{rccccccccccccccccccccccccccccccccc}
\tablecaption{Young stars in North America Nebula and Pelican \label{tabe_allspt}}
\tablehead{\multicolumn{1}{c}{} &\multicolumn{1}{c}{ } &\multicolumn{1}{c}{ } &\multicolumn{5}{c}{Hectospec} & &\multicolumn{5}{c}{Hydra}  & &\multicolumn{5}{c}{Deimos}  & &\multicolumn{5}{c}{Norris}\\
\cline{4-8} \cline{10-14} \cline{16-20} \cline{22-26}
\multicolumn{1}{c}{  } &\multicolumn{1}{c}{  } &\multicolumn{1}{c}{   } &\multicolumn{2}{c}{With Veiling} & &\multicolumn{1}{c}{No Veiling}  &\multicolumn{1}{c}{ } &  &\multicolumn{2}{c}{With Veiling} & &\multicolumn{1}{c}{No Veiling}  &\multicolumn{1}{c}{ } &  &\multicolumn{2}{c}{With Veiling} & &\multicolumn{1}{c}{No Veiling}  &\multicolumn{1}{c}{ } &  &\multicolumn{2}{c}{With Veiling} & &\multicolumn{1}{c}{No Veiling}  &\multicolumn{1}{c}{ }\\
\cline{4-5} \cline{7-7} \cline{10-11} \cline{13-13} \cline{16-17} \cline{19-19} \cline{22-23} \cline{25-25}
\multicolumn{1}{c}{ID} &\multicolumn{1}{c}{RA} &\multicolumn{1}{c}{DEC} &\multicolumn{1}{c}{SPT} &\multicolumn{1}{c}{$r_{7465}$} & &\multicolumn{1}{c}{SPT}  &\multicolumn{1}{c}{M\tablenotemark{a}}  & &\multicolumn{1}{c}{SPT} &\multicolumn{1}{c}{$r_{7465}$} & &\multicolumn{1}{c}{SPT}  &\multicolumn{1}{c}{M\tablenotemark{a}}  & &\multicolumn{1}{c}{SPT} &\multicolumn{1}{c}{$r_{7465}$} & &\multicolumn{1}{c}{SPT}  &\multicolumn{1}{c}{M\tablenotemark{a}}  & &\multicolumn{1}{c}{SPT} &\multicolumn{1}{c}{$r_{7465}$} & &\multicolumn{1}{c}{SPT}  &\multicolumn{1}{c}{M\tablenotemark{a}}}
\startdata
1  & 20 48 50.71  &$+$43 49 49.7  &Cont  &High &  &\nodata  &  & &\nodata  &\nodata &  &\nodata  & &  &\nodata  &\nodata &  &\nodata  & &  &\nodata  &\nodata &  &\nodata  &\\
2  & 20 49 11.78  &$+$44 12 32.9  &K7.4  &0.14 &  &K7.0  &1 & &\nodata  &\nodata &  &\nodata  & &  &\nodata  &\nodata &  &\nodata  & &  &\nodata  &\nodata &  &\nodata  &\\
3  & 20 49 14.47  &$+$44 14 49.9  &M5.0  &0.02 &  &M4.8  &1 & &\nodata  &\nodata &  &\nodata  & &  &\nodata  &\nodata &  &\nodata  & &  &\nodata  &\nodata &  &\nodata  &\\
4  & 20 49 20.30  &$+$44 35 13.5  &M2.6  &0.04 &  &M2.4  &1 & &\nodata  &\nodata &  &\nodata  & &  &\nodata  &\nodata &  &\nodata  & &  &\nodata  &\nodata &  &\nodata  &\\
5  & 20 49 24.76  &$+$44 35 21.8  &M5.6  &0.02 &  &M5.6  &1 & &\nodata  &\nodata &  &\nodata  & &  &\nodata  &\nodata &  &\nodata  & &  &\nodata  &\nodata &  &\nodata  &\\
6  & 20 49 32.19  &$+$44 17 03.1  &M1.2  &0.18 &  &M0.2  &1 & &\nodata  &\nodata &  &\nodata  & &  &\nodata  &\nodata &  &\nodata  & &  &\nodata  &\nodata &  &\nodata  &\\
7  & 20 49 53.51  &$+$44 10 17.3  &\nodata  &\nodata &  &\nodata  & & &\nodata  &\nodata &  &\nodata  & &  &\nodata  &\nodata &  &\nodata  & &  &M3.6  &0.02 &  &M3.4  &1\\
8  & 20 50 06.05  &$+$44 17 49.0  &\nodata  &\nodata &  &\nodata  & & &M4.2  &0.08 &  &M4.0  &1 &  &\nodata  &\nodata &  &\nodata  & &  &\nodata  &\nodata &  &\nodata  &\\
\enddata
\end{deluxetable}
\tablenotetext{a}{1: Using X-Shooter PMS spectral templates, 2: Using model atmospheres.}
\end{longrotatetable}
\normalsize
\clearpage

\begin{longrotatetable}
\scriptsize
\renewcommand{\tabcolsep}{0.07cm}
\begin{deluxetable}{rccccccccccccccccccccccccccccccccc}
\tablecaption{Young stars in North America Nebula and Pelican \label{tabe_YSO}}
\tablehead{\multicolumn{1}{c}{} &\multicolumn{1}{c}{} &\multicolumn{1}{c}{} &\multicolumn{1}{c}{} &\multicolumn{1}{c}{}&\multicolumn{1}{c}{} &\multicolumn{1}{c}{} & \multicolumn{8}{c}{Model\tablenotemark{b}} & & \multicolumn{8}{c}{X-Shooter template\tablenotemark{c}} \\
\cline{8-15} \cline{17-24}
\multicolumn{1}{c}{} &\multicolumn{1}{c}{ } &\multicolumn{1}{c}{ } &\multicolumn{1}{c}{Adopted}&\multicolumn{1}{c}{ } &\multicolumn{1}{c}{ }  &\multicolumn{1}{c}{ } &\multicolumn{2}{c}{Fixing $r_{7465}$} & &\multicolumn{2}{c}{Varying $r_{7465}$}& &\multicolumn{2}{c}{Adopted} & &\multicolumn{2}{c}{Fixing $r_{7465}$} & &\multicolumn{2}{c}{Varying $r_{7465}$} & &\multicolumn{2}{c}{Adopted} & &\multicolumn{4}{c}{YSO criteria\tablenotemark{d}} \\
\cline{8-9} \cline{11-12} \cline{14-15} \cline{17-18} \cline{20-21} \cline{23-24} \cline{26-29}
\multicolumn{1}{c}{ID} &\multicolumn{1}{c}{RA} &\multicolumn{1}{c}{DEC} &\multicolumn{1}{c}{SPT} &\multicolumn{1}{c}{$r_{7465}$}&\multicolumn{1}{c}{Q\tablenotemark{a}} &\multicolumn{1}{c}{$T_{\rm eff}$} & \multicolumn{1}{c}{$A_{\rm V}$} & \multicolumn{1}{c}{$L_{\star}$} & &\multicolumn{1}{c}{$A_{\rm V}$} & \multicolumn{1}{c}{$L_{\star}$} & & \multicolumn{1}{c}{$A_{\rm V}$} & \multicolumn{1}{c}{$L_{\star}$} & & \multicolumn{1}{c}{$A_{\rm V}$} & \multicolumn{1}{c}{$L_{\star}$}& & \multicolumn{1}{c}{$A_{\rm V}$} & \multicolumn{1}{c}{$L_{\star}$} & & \multicolumn{1}{c}{$A_{\rm V}$} & \multicolumn{1}{c}{$L_{\star}$} &  &\multicolumn{1}{c}{IRE} &\multicolumn{1}{c}{Li}&\multicolumn{1}{c}{X-ray}&\multicolumn{1}{c}{var}\\
\multicolumn{1}{c}{} &\multicolumn{1}{c}{(J2000)} &\multicolumn{1}{c}{(J2000)} &\multicolumn{1}{c}{ } &\multicolumn{1}{c}{ } &\multicolumn{1}{c}{} &\multicolumn{1}{c}{(K)} &\multicolumn{1}{c}{(mag)} &\multicolumn{1}{c}{($L_{\odot}$)} & &\multicolumn{1}{c}{(mag)} &\multicolumn{1}{c}{$L_{\odot}$) } & &\multicolumn{1}{c}{(mag)}&\multicolumn{1}{c}{$L_{\odot}$) } & &\multicolumn{1}{c}{(mag)}  &\multicolumn{1}{c}{($L_{\odot}$)} & &\multicolumn{1}{c}{(mag)} &\multicolumn{1}{c}{$L_{\odot}$) } & &\multicolumn{1}{c}{(mag)}&\multicolumn{1}{c}{$L_{\odot}$) }}
\startdata
1  & 20 48 50.71  &$+$43 49 49.7  &Cont  &High  &   &  &&&  &&&  && & &&&  &&&  && & &Y&\nodata&\nodata&H\\
2  & 20 49 11.78  &$+$44 12 32.9  &K7.0  &0.00  &       1  &3970  &3.9&1.489&  &3.9&1.489&  &3.9&1.489 & &3.8&1.320&  &3.8&1.320&  &3.8&1.320
 & &Y&Y&\nodata&H\\
3  & 20 49 14.47  &$+$44 14 49.9  &M4.8  &0.00  &       1  &3015  &2.0&0.217&  &2.1&0.224&  &2.0&0.217 & &1.6&0.175&  &1.6&0.175&  &1.6&0.175
 & &Y&Y&\nodata&L\\
4  & 20 49 20.30  &$+$44 35 13.5  &M2.4  &0.00  &       1  &3437  &2.4&0.387&  &2.4&0.387&  &2.4&0.387 & &2.5&0.359&  &2.5&0.359&  &2.5&0.359
 & &Y&Y&\nodata&L\\
5  & 20 49 24.76  &$+$44 35 21.8  &M5.6  &0.00  &       1  &2908  &1.9&0.174&  &2.3&0.201&  &1.9&0.174 & &0.8&0.097&  &0.8&0.097&  &0.8&0.097
 & &Y&\nodata&\nodata&L\\
6  & 20 49 32.19  &$+$44 17 03.1  &M1.2  &0.18  &       1  &3602  &2.5&0.671&  &2.0&0.614&  &2.3&0.643 & &2.6&0.718&  &2.1&0.587&  &2.4&0.653
 & &Y&Y&\nodata&H\\
7  & 20 49 53.51  &$+$44 10 17.3  &M3.4  &0.00  &       1  &3279  &0.1&0.106&  &0.3&0.114&  &0.1&0.106 & &0.0&0.095&  &0.0&0.095&  &0.0&0.095
 & &N&Y&\nodata&L\\
8  & 20 50 06.05  &$+$44 17 49.0  &M4.0  &0.00  &       1  &3160  &1.4&0.467&  &3.3&0.713&  &1.4&0.467 & &1.0&0.344&  &1.0&0.344&  &1.0&0.344
 & &N&Y&\nodata&H\\
\enddata
\end{deluxetable}
\tablenotetext{a}{Ranking for the spectral classification: 1 for very good, 2 for fair, and 3 for poor.}
\tablenotetext{b}{Fitting the SEDs of young stars with model atmosphere. Fixing $r_{7465}$:  during the SED fitting, the used $r_{7465}$ is fixed to be the one derived from the spectral classification;  Varying $r_{7465}$: $r_{7465}$  is a free parameter during the SED fitting.}
\tablenotetext{c}{Fitting the SEDs of young stars with model atmosphere. Fixing $r_{7465}$ and Varying $r_{7465}$ are same as for $^{b}$.}
\tablenotetext{d}{IRE: showing infrared excess; Li: showing \LiI\ absorption; Xray: having a X-ray counterpart; var: H for highly variable, M for moderately variable, and  L for weakly variable (see \S~\ref{Sect:var} ). }
\end{longrotatetable}
\normalsize

\begin{table}
\caption{The median luminosity and luminosity dispersion of young stars within individual $T_{\rm eff}$ bins.  \label{tab:medHRD}}
 \centering
\begin{tabular}{lrccccccccc}
\hline\hline
\multicolumn{1}{c}{Log~$T_{\rm eff}$} &\multicolumn{1}{c}{median Log~$L_{\star}/L_{\odot}$} &\multicolumn{1}{c}{$\sigma$Log~$L_{\star}/L_{\odot}$} \\ 
\hline
3.96--3.88& 1.560  &0.325   \\
3.88--3.80& 1.350  &0.288   \\
3.80--3.72& 1.106  &0.377   \\
3.72--3.64& 0.732  &0.309   \\
3.64--3.56& 0.008  &0.348   \\
3.56--3.52&$-$0.391  &0.286   \\
3.52--3.48&$-$0.638  &0.340   \\
3.48--3.44&$-$0.913  &0.390   \\
 \hline
\end{tabular}
\end{table}


\vspace{5mm}
\facilities{Palomar:P200 (Norris), Keck:I (DEIMOS), MMT (Hectospec), Palomar:P48 (ZTF) }


\clearpage
\appendix

\startlongtable
\begin{deluxetable}{lcccccccccrccc}
\normalsize
\tablecaption{List of nonmembers of NAP with infrared excess in our sample \label{tab:IRE_nonmember}}
\tablehead{\colhead{ID}&\colhead {Name designation}& \colhead{RA}     & \colhead{DEC}    & \colhead{Gaia parallex}   &  \colhead{Spectral type}}
\startdata
1   &2MASS J20491197+4414123  & 20 49 11.98  &$+$44 14 12.3  &$0.3323\pm0.0448$  &F8.0\\
2   &2MASS J20492960+4400467  & 20 49 29.60  &$+$44 00 46.7  &$-0.1801\pm0.0866$ &Late B\\
3   &2MASS J20510271+4349318  & 20 51 02.71  &$+$43 49 31.9  &$0.4564\pm0.0269$  &Early B\\
4   &2MASS J20511071+4306142  & 20 51 10.72  &$+$43 06 14.2  &$0.3829\pm0.0694$  &Late B\\
5   &2MASS J20511500+4409006  & 20 51 15.01  &$+$44 09 00.7  &$0.5238\pm0.0236$  &A8.8\\
6   &2MASS J20512307+4426461  & 20 51 23.08  &$+$44 26 46.2  &$0.3227\pm0.0841$  &B7.4\\
7   &2MASS J20513007+4412312  & 20 51 30.07  &$+$44 12 31.3  &$0.2102\pm0.1775$  &A7.6\\
8   &2MASS J20513811+4415493  & 20 51 38.12  &$+$44 15 49.3  &$0.2225\pm0.2150$  &A9.7\\
9   &2MASS J20514755+4425106  & 20 51 47.55  &$+$44 25 10.6  &$0.3214\pm0.0370$  &A8.6\\
10  &2MASS J20523733+4305467  & 20 52 37.33  &$+$43 05 46.8  &$0.3788\pm0.0174$  &Early B\\
11  &2MASS J20525419+4321189  & 20 52 54.20  &$+$43 21 18.9  &$1.0723\pm0.0277$  &A7.7\\
12  &2MASS J20533175+4501286  & 20 53 31.76  &$+$45 01 28.6  &$0.1299\pm0.0800$  &F1.9\\
13  &2MASS J20533271+4449117  & 20 53 32.72  &$+$44 49 11.8  &$0.1521\pm0.0559$  &A9.9\\
14  &2MASS J20533370+4447091  & 20 53 33.70  &$+$44 47 09.2  &$-0.3471\pm0.2115$  &A4.9\\
15  &2MASS J20535583+4314147  & 20 53 55.83  &$+$43 14 14.8  &$0.2336\pm0.0313$  &A8.8\\
16  &2MASS J20535754+4356476  & 20 53 57.55  &$+$43 56 47.6  &$0.7018\pm0.1191$  &F7.3\\
17  &2MASS J20544690+4448197  & 20 54 46.90  &$+$44 48 19.8  &$0.0564\pm0.0876$  &Late B\\
18  &2MASS J20545128+4306226  & 20 54 51.28  &$+$43 06 22.6  &$0.3560\pm0.0598$  &Late B--early A\\
19  &2MASS J20552589+4446471  & 20 55 25.90  &$+$44 46 47.1  &$0.2746\pm0.1644$  &F1.9\\
20  &2MASS J20553187+4447004  & 20 55 31.88  &$+$44 47 00.5  &$0.2275\pm0.0859$  &Early F\\
21  &2MASS J20553791+4455205  & 20 55 37.91  &$+$44 55 20.6  &$-0.4208\pm0.3217$  &F7.4\\
22  &2MASS J20553880+4456560  & 20 55 38.81  &$+$44 56 56.1  &$-0.1639\pm0.1716$  &F0.1\\
23  &2MASS J20561239+4414182  & 20 56 12.39  &$+$44 14 18.3  &$0.0812\pm0.1894$  &A7.8\\
24  &2MASS J20562155+4445145  & 20 56 21.56  &$+$44 45 14.5  &$0.2396\pm0.0950$  &A9.2\\
25  &2MASS J20562449+4456225  & 20 56 24.50  &$+$44 56 22.5  &$0.3407\pm0.0631$  &Early B\\
26  &2MASS J20564398+4404138  & 20 56 43.99  &$+$44 04 13.9  &\nodata            &O9.0\\
27  &2MASS J20564699+4443555  & 20 56 46.99  &$+$44 43 55.6  &\nodata            &F1.8\\
28  &2MASS J20564785+4438373  & 20 56 47.85  &$+$44 38 37.3  &$0.2440\pm0.0462$  &B8.0\\
29  &2MASS J20573452+4359547  & 20 57 34.53  &$+$43 59 54.7  &$1.0597\pm0.0335$  &A7.8\\
30  &2MASS J20574034+4410465  & 20 57 40.35  &$+$44 10 46.6  &$-0.7086\pm0.6578$  &Late B--early A\\
31  &2MASS J20581722+4344091  & 20 58 17.23  &$+$43 44 09.1  &$0.2349\pm0.0812$  &A8.5\\
32  &2MASS J20593917+4353527  & 20 59 39.17  &$+$43 53 52.8  &$0.5561\pm0.0266$  &B7.7\\
33  &2MASS J20595153+4343302  & 20 59 51.53  &$+$43 43 30.3  &$0.6124\pm0.0435$  &F4.2\\
34  &2MASS J20591724+4417464  & 20 59 17.25  &$+$44 17 46.5  &$0.4070\pm0.0232$  &Early B\\
35  &2MASS J20532479+4351527  & 20 53 24.80  &$+$43 51 52.7  &$0.0361\pm0.0941$  &A7.3\\
36  &2MASS J20485426+4356172  & 20 48 54.27  &$+$43 56 17.3  &$0.3855\pm0.1053$  &A9.2\\
37  &2MASS J20545135+4515063  & 20 54 51.36  &$+$45 15 06.3  &$0.1201\pm0.0209$  &Early B\\
38  &2MASS J20553631+4517182  & 20 55 36.32  &$+$45 17 18.3  &$0.2048\pm0.0348$  &B8.0\\
39  &2MASS J20544203+4516532  & 20 54 42.03  &$+$45 16 53.2  &$0.3647\pm0.2272$  &B8.0\\
40  &2MASS J21000826+4315030  & 21 00 08.27  &$+$43 15 03.1  &$0.2417\pm0.0593$  &F1.9\\
\enddata
\end{deluxetable}



\section{Improvement in IRAC photometry}\label{Appen:IRAC}
\begin{figure*}
\begin{center}
\includegraphics[angle=0,width=2\columnwidth]{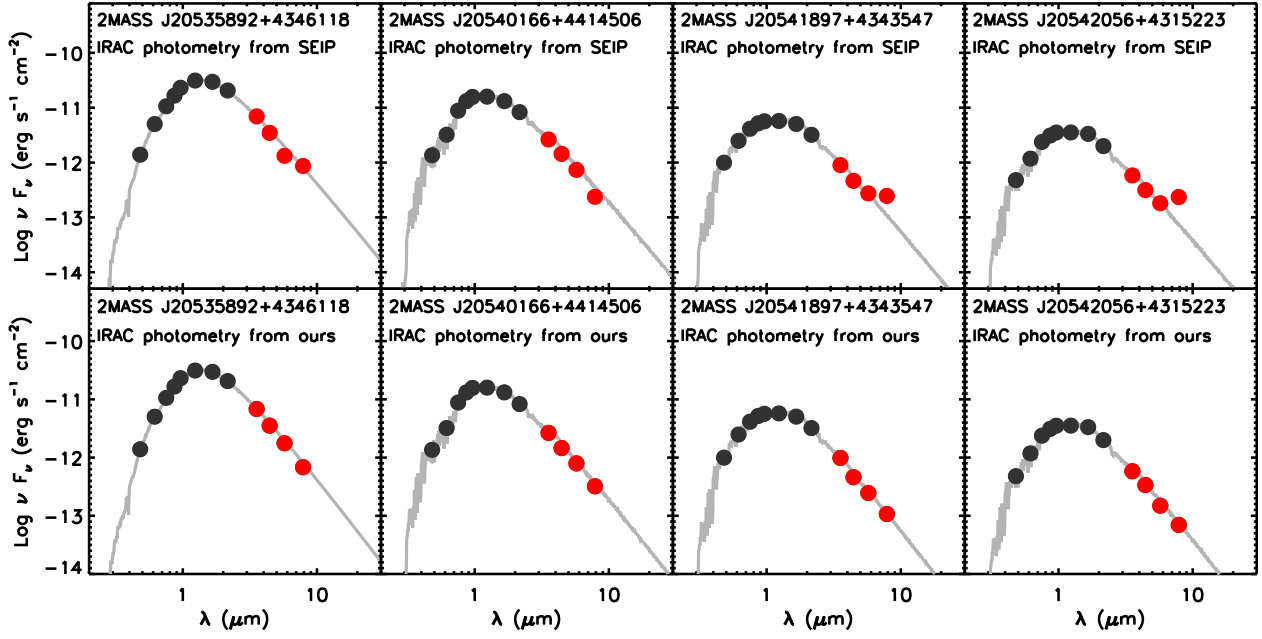}
\caption{The SEDs of four sources with bad IRAC photometry in at least one band from SEIP (Top panels) and the IRAC photometry from this works (bottom panels). In each panel, the black filled circels are the photometry from Pan-STARRS and 2MASS, and the red filled circles for photometry in  Spitzer IRAC bands. The gray line is photospheric level.}
\label{Fig:badSED}
\end{center}
\end{figure*}

In this work, we prefer to use the Spitzer photometry for our sources (mostly with infrared excess emission) collected  from \cite{2011ApJS..193...25R}. For those without photometry in \cite{2011ApJS..193...25R}, we collect the aperture photometry in the 3\farcs8 aperture from SEIP. However, we noticed some photometry from SEIP are unreliable which could be due to strong and highly variable background in the field, and some field stars could be misclassified as the ones with infrared excess emission (see Figure~\ref{Fig:badSED}). For those sources, we use our PSF photometry which is performed in the way as described in \cite{2009A&A...504..461F}. A comparison of the photometry for 4 sources from SEIP and ours are shown in Figure~\ref{Fig:badSED}.

\section{Nonmembers with infrared excess}\label{Appen:IRE_nonmember}
In our sample, 40 sources with infrared excess emission and spectral types earlier than G are excluded as nonmembers of the NAP, according to their locations in H-R diagram and/or  $Gaia$ parallax. These sources are listed in Table.~\ref{tab:IRE_nonmember}, which includes their $Gaia$ parallax and spectral types.

\section{A VY Scl Star}\label{Appen:vyscl}

Among the sources identified with X-ray emission,  we note one interesting object LkH$\alpha$ 170 = HBC 297. This star is B8.5 according to our classification and A0 in \cite{2004AJ....127.1682H}.  It is also an emission-line star, and thus a potential Herbig Ae/Be object.  Given its parallax (1.2912	$\pm$0.0224), the star should be a member of the NAP.   However, in the H-R diagram it appears about one order of magnitude fainter than main sequence stars with the same spectral type.   The source is also referred to in the literature in the context of VY Scl stars \citep{2001PASP..113...72P}. It is thus beyond the scope of this work.







\end{document}